\documentclass[tightenlines,twocolumn,superscriptaddress,preprintnumbers,amssymb,nofootinbib]{revtex4-1}
\usepackage{amsmath, amsthm, amssymb, mathrsfs, bm, mathtools} 
\usepackage{graphics}
\usepackage{epsfig}
\usepackage{graphicx}
\usepackage{wrapfig}
\usepackage{color}
\usepackage{tikz}
\usepackage[hidelinks,colorlinks=true,linkcolor=blue,citecolor=blue]{hyperref}

\RequirePackage{subfigure}

\newcommand{\beq}{\begin{equation}}
\newcommand{\eeq}{\end{equation}}
\newcommand{\bqa}{\begin{eqnarray}}
\newcommand{\eqa}{\end{eqnarray}}

\newcommand{\os}{\text{\tiny OS}}
\newcommand{\ms}{\overline{\text{\tiny MS}}}

\newcommand{\Tr}{{\mathrm{Tr}}}
\newcommand{\cep}{\text{\tiny CEP}}
\newcommand{\tcp}{\text{\tiny TCP}}

\newcommand{\x}{x}

\begin{document}

\title{Comparing critical regions in Polyakov-quark-meson models: Effects of flavor (two versus 2+1) under on-shell renormalization, and on-shell versus curvature-mass parameter fixing for the two-flavor case.}

\author{Pooja Kumari}
\email{pooja65028@gmail.com}
\affiliation{Department of Physics, University of Allahabad, Prayagraj, India-211002}
\author{Suraj Kumar Rai}
\email{surajrai050@gmail.com}
\affiliation{Department of Physics, Acharya Narendra Deo Kisan P.G. College,
Babhnan Gonda-271313; Maa Pateswari University Balrampur,India-271201}
\author{Vivek Kumar Tiwari}
\email{vivekkrt@gmail.com}
\affiliation{Department of Physics, University of Allahabad, Prayagraj, India-211002}
\date{\today}

\begin{abstract}

We compute contours of the enhanced quark number susceptibility ratio $R_q$, normalized by the free quark gas susceptibility, around the critical end point (CEP) using the on-shell renormalized two-flavor quark-meson (RQM) model and its two Polyakov-loop-extended variants: the Log RPQM and PolyLog-glue RPQM models, employing logarithmic and PolyLog-glue forms of the Polyakov-loop potential without and with quark back-reaction, respectively. These contours are compared with those from the curvature-mass parametrized quark-meson-with-vacuum-term (QMVT) model and its logarithmic Polyakov-loop extension (Log PQMVT)~\cite{vkkr12}, to assess how different treatments of quark one-loop vacuum fluctuations affect the critical region. Although the critical fluctuation regions near the RQM/Log RPQM CEP are reduced in $T$-$\mu$ extent relative to the QMVT/Log PQMVT case, the higher-$T$, lower-$\mu$ CEP location shifts its enhanced-susceptibility domain toward higher $T$ and lower $\mu$. Since the $\sigma$ and $\pi$ curvature and pole masses differ in the RQM/Log RPQM model but coincide in QMVT/Log PQMVT, we also compare contours of the enhanced scalar susceptibility ($\sim 1/m_\sigma^2$) around the respective CEPs. To isolate the roles of strange flavor and the $U_A(1)$ anomaly, we compare $R_q = 2, 3, 5$ contours from our two-flavor RQM/Log RPQM/PolyLog-glue RPQM model calculations with the corresponding 2+1 flavor RQM/RPQM model very recent results in Ref.~\cite{Akakn}. The critical region is narrower and less extended in $T$ and $\mu$ for the 2+1 flavor case. The third flavor further compresses the contour widths in the Log RPQM model, an effect moderated by quark back-reaction in the PolyLog-glue RPQM model, while $T$-direction stretching remains comparable to the two-flavor models. Chiral-limit ($m_\pi = 0$) phase diagrams show that the tricritical point (TCP) lies well inside the $R_q = 2$ contour for all the two-flavor scenarios, indicating its influence on the critical fluctuations near the CEP; with the third flavor, however, the TCP moves outside the $R_q = 2$ contour in the 2+1 flavor RQM model and sits near its boundary in the Log RPQM and PolyLog-glue RPQM models. The critical exponents for the divergence of the quark number susceptibility approaching the CEP in the RQM/RPQM models and the TCP in the RQM model are also computed.

\end{abstract}
\maketitle
\section{Introduction}

The strong interaction theory called Quantum Chromodynamics (QCD) mandates that its basic units, quarks and gluons, are weakly coupled at high energies, whereas they get confined within hadrons in its low-energy vacuum. Under extreme conditions of temperature or density, a phase transition occurs, dissolving hadrons into a quark-gluon plasma \cite{Cabibbo75,SveLer,Mull,Ortms,Riske}. The study of such a phase transition is highly relevant for understanding the physics of the early universe, neutron stars, and relativistic heavy-ion collision experiments. Since first principle Lattice QCD simulations \cite{AliKhan:2001ek,Digal:01,Fodor:03,Allton:05,Karsch:05,Aoki:06,Cheng:06,Cheng:08,JLange,ABAZ,Ejiri.Lat} fail at non zero baryon densities due to the fermion sign problem \cite{Karsch:02}, one often relies on QCD-like effective theory models \cite{Alf,Fukhat,FejosI,FejosII,FbRenk,Fejos3,Fejos4} built upon two very important properties : chiral symmetry and color charge confinement, through which QCD reveals itself.

The axial $SU_{A}(2)$ (A = L-R) part of the  $SU_{L}(2) \times SU_{R}(2)$ symmetry known as the chiral symmetry of the QCD Lagrangian for two flavor of massless quarks,~gets spontaneously broken by formation of a quark condensate and one gets three massless pions as Goldstone modes.~Having small masses, the real pions are pseudo Goldstone modes, because this symmetry gets broken explicitly also due to very small mass of  u and d quarks. The spontaneous breaking of $SU_{A}(3)$ chiral symmetry when strange s quark is also considered as massless, gives two chiral condensates; non-strange and strange. Since s quarks are about 20 to 30 times heavier than the light quarks,~one observes octet of  pseudo-Goldstone modes as three light pions,~four heavy kaons and a heavier  $\eta$ meson due to large  explicit breaking of chiral symmetry in the s direction.~The QCD would have had $U_{A}(3)$ chiral symmetry for three massless quarks, but for the $U_A(1)$ axial anomaly, due to which the would be $U_{A}(1)$ axial symmetry of the classical Lagrangian is explicitly broken at the quantum level by instanton effects, leaving the $\eta'$ roughly $1\text{ GeV}$ heavy \cite{tHooft:76prl} as a consequence.~The restoration (breaking) of this symmetry at high (low) temperatures is studied in the two and three linear sigma model where meson Lagrangian is written in powers of chiral invariants $\sigma^{2}+\vec\pi^2$ with  a scalar $\sigma$ and a pseudoscalar $\vec\pi$ as chiral partners for two flavor.~The chiral invariants for three flavor are $\text{Tr}(\cal{M}^\dagger \cal{M})$ and $\text{Tr}[({\cal{M}^\dagger {M}})^2]$ where ${\cal{M}}=(\lambda_{a}/2)(\sigma_{a}+i\pi_{a})$ and ${\cal{M}}^\dagger$ are $3\times 3$ matrices containing nine scalar ($\sigma_a$) and pseudo-scalar $\pi_a$ meson fields with eight Gelmann matrices $\lambda_a$ ($a=1 \cdots 8$) and $\lambda_0=\sqrt{\frac{2}{3}}{\mathbb I}_{3\times3}$ \cite{Ortman,Rischke:00,Lenagh,Roder}. Coupling the meson systems to two (three) flavor  of quarks via a Yukawa interaction gives the very useful QCD-like effective theory framework of two (2+1) flavor quark-meson (QM) model~\cite{scav,mocsy,bj,Schaefer:2006ds,SchaPQM2F,Schaefer:09,SchaPQM3F,Mao,TiPQM3F}.

The two (three) flavor  Nambu-Jona-Lasinio (NJL) model~\cite{costaA,costaB,fuku08} is an alternative framework where quarks are the primary fields in  the Lagrangian and the multi-quark point interactions generate the mesons. The  confinement of QCD color charges, is accounted for in a statistical sense when the chiral models are coupled to a constant background of $SU_{c}(N)$ gauge field ~\cite{SveLer,Polyakov:78plb,benji,BankUka,Pisarski:00prd,fuku,Vkt:06}.~Both the  color confinement-deconfinement  and chiral symmetry breaking-restoring phase transitions can be studied in the integrated framework of Polyakov loop enhanced quark meson model (PQM)~\cite{SchaPQM2F,SchaPQM3F,Mao,TiPQM3F} when the free energy density from gluons in the form of Polyakov-loop potential~\cite{ratti,Roesnr} is added to the QM model effective potential. The form of Polyakov-loop potential has also been improved from a pure gauge potential to a unquenched glue potential after including the effect of back reaction of quarks \cite{Haas,Redlo,BielichP,Herbst,THerbst2}.~The confinement-deconfinement and chiral phase transitions get linked to each other even at small temperatures and large chemical potentials ~\cite{BielichP} due to the above improvement.

The QCD transition at $\mu = 0$ for physical quark masses is a crossover according to Lattice QCD \cite{Wupertal2010,WB2014,HotQCD2014}, whereas effective theories \cite{rob,hjss,Asak,Bard,Berg} predict a first-order transition at lower $T$ and higher $\mu$. Therefore the $\mu - T$ plane phase diagram features a critical end point (CEP) belonging to the $3\text{D}$ Ising $Z(2)$ universality class \cite{Fuji,Hatta,redlsaki,KFuku}, where the chiral crossover line ends to become first order. In the light chiral limit ($m_{ud} = 0=m_{\pi}$), the temperature driven transition at $\mu = 0$ is second-order in the $3\text{D}$ $O(4)$ universality class. Consequently, the physical point CEP implies a tricritical Point (TCP) in massless two-flavor QCD, where the $O(4)$ second-order line terminates into a first-order region. An analogous TCP appears at $\mu = 0$ in $2+1$ flavor QCD-like effective models (with physical strange quark mass and $m_\pi = 0$) \cite{vktChpt1,vktChpt2}.~Finding signals of CEP \cite{Son,Misha3,Kraja2,Jeon,Ejik} are very important for confirming the formation of QGP matter. Experiments dedicated to the CEP  search were launched at Relativistic Heavy Ion Collider (RHIC), Brookhaven National Laboratory (BNL) USA, beam energy scan (BES) program (BES-I: 2010–2014, BES-II: 2018–2021) \cite{AdamAg,Odyniec:2019XZ,Cliu}. Experimental signatures rely on measuring event-by-event fluctuations and higher-order moments of conserved charges (such as net baryon number, charge, and strangeness) connected to thermodynamic susceptibilities \cite{MishaPRL,Krajamisha,bedang,Misha11,APandav,MishaConf}. Comprehensive review of theoretical aspects of CEP physics can be found in Refs.~\cite{Bzda,Ldu}. Tracking non-Gaussian fluctuations near the CEP requires accounting for non-equilibrium dynamics \cite{xin,Akam,vovchen,xan,Parot,Pradeep1,Pradeep2}, including maximum entropy approaches for freeze-out \cite{Pradeep1,Pradeep2}.~The fututre Compressed Baryonic Matter experiments at the Facility for Antiproton and Ion Research (GSI-Darmstadt, Germany) and the Nuclotron-based Ion Collider Facility at the Joint Institute for Nuclear Research  (Dubna, Russia) will focus on finding a CEP at high chemical potentials.

Chiral models \cite{Ortman,Rischke:00,Lenagh,Roder,jakobi,Herpay:05,Herpay:06,Herpay:07,Kovacs:2006ym,kahara,Bowman:2008kc,Fejos,Jakovac:2010uy,koch,marko,fuku11,grahl,GFejo}, two and 2+1 flavor QM/PQM models \cite{scav,mocsy,SchaPQM2F,Schaefer:09,SchaPQM3F,Mao,TiPQM3F,bj,Schaefer:2006ds} are widely used to unravel different aspects of  QCD phase transition.~The quark one-loop vacuum fluctuations used to be neglected in the QM model under the standard mean field approximation (s-MFA).~The  chiral phase transition for the chiral limit  turn out to first order in these studies which is at odds with the theoretical arguments~\cite{rob,hjss}.~The above pathology was cured by including the quark one-loop vacuum correction in the two flavor QM model~\cite{vac}.~Earlier works~\cite{Fraga1,Fraga2,Fraga3} studied the effect of fermionic vacuum corrections  in the context of finite temperature and density Yukawa theory.~In the modified framework of quark meson/Polyakov loop augmented quark meson with the vacuum term (QMVT/PQMVT) model,~diverse studies ~\cite{vac,Gatto,Anna,lars,guptiw,chatmoh1,schafwag12,TranAnd,vkkr12,Dima,vkkt13,Weise1,
Herbst,Weyrich,kovacs,zacchi1,zacchi2,Rai,Weise3} found that the QCD thermodynamics and its phase structure get strongly influenced by the effect of the quark one-loop vacuum correction.~After regularizing the one-loop divergence in the minimal subtraction scheme,~the above studies fixed the model parameters using the curvature masses of the mesons which are defined by the double derivatives of the effective potential at its minimum.~The above parameter fixing method looks inconsistent when one notes that the effective potential is the generator of the n-point functions of the theory at zero external momenta.~The contributions of quark one-loop vacuum corrections in meson self energies are included at zero momentum in the curvature masses.

Studies in Refs. \cite{Kobes,Rebhan,laine,BubaCar,fix1,Naylor,Adhiand1} emphasize that meson pole masses are physical and gauge invariant. Because radiative corrections alter the tree-level relations between physical quantities and Lagrangian parameters, calculating the effective potential using uncorrected tree-level parameters is inconsistent. The renormalization scale $\Lambda$ dependence of running parameters is accounted for in the (modified) minimal subtraction ($\overline{\text{MS}}$) scheme, whereas on-shell parameters retain their tree-level values. After calculating counter-terms in both the $\overline{\text{MS}}$ and on-shell schemes following standard renormalization prescriptions, the renormalized parameters of the two schemes are connected. The effective potential of the on-shell renormalized quark-meson ($\text{RQM}$) and Polyakov-loop enhanced quark-meson ($\text{RPQM}$) models is ultimately calculated using the $\overline{\text{MS}}$ procedure, taking the relations between running and on-shell parameters (physical quantities) as input \cite{Adhiand1,Adhiand2,Adhiand3,asmuAnd}. Following a consistent treatment of quark one-loop vacuum fluctuations, the  effective potentials for two-flavor and $2+1$ flavor $\text{RQM}/\text{RPQM}$ models were recently calculated in Refs. \cite{Adhiand1,Adhiand2,Adhiand3,asmuAnd,RaiTiw22,raiti23} and Refs. \cite{vkkr23,skrvkt24,vktChpt1,vktChpt2,Joand,Gholami}, respectively.

The impact of treating quark one-loop vacuum divergences differently on effective potentials, phase diagrams, the CEP positions, and thermodynamic properties of the 
two-flavor QM model was extensively investigated in Refs.\cite{RaiTiw22,raiti23}. In these works, the on-shell renormalized RQM/RPQM model was compared against the QMVT/PQMVT model with curvature mass based parameter fixing. In the QMVT/PQMVT model, quark one-loop vacuum fluctuations strongly smooth the chiral transition, causing the first-order line to shrink and the CEP to shift toward the bottom right (high $\mu$, low $T$) of the phase diagram. In contrast, in the RQM/RPQM model, the moderate softening effect of these corrections shifts the CEP toward higher temperatures and lower chemical potentials.
The regions of critical fluctuations around the CEP of the QMVT and Log PQMVT models were studied in Ref. \cite{vkkr12}. In this work, our first goal is to make a comprehensive comparison of the size and shape of the critical regions surrounding the CEP across the RQM, Log RPQM, QMVT, and Log PQMVT models. Such a comparison will reveal how different treatments of the quark one-loop vacuum fluctuations in the two-flavor QM/PQM models influence the extent of the critical regions. In the Log RPQM and Log PQMVT models, we utilize the logarithmic form of the Polyakov loop potential. The phase diagrams in the chiral limit ($m_\pi = 0$), together with the location of the tricritical point (TCP), will be computed for all the aforementioned model scenarios. The proximity of the CEP to the TCP will be quantified by comparing the light chiral limit phase diagrams with the corresponding phase diagrams computed using physical point parameters ($m_\sigma = 500$ MeV). The temperature dependence of $m_\sigma$ and $m_\pi$, along with the quark number and scalar susceptibilities around both the CEP and the TCP, will be plotted and compared. Because the CEP is a thermodynamic singularity, the quark number and scalar susceptibilities are significantly enhanced in its vicinity. The spread of the critical regions enveloping the CEP will be mapped by drawing contours of constant enhanced normalized quark number susceptibility. Furthermore, the scalar $\sigma$ mass approaches zero ($m_\sigma \to 0$) at the CEP in the RQM and QMVT models as a function of $T$ and $\mu$. Noting that the curvature and pole masses of the $\sigma$ and $\pi$ mesons differ in the RQM model but remain the same in the QMVT model, it is also interesting to compute and compare the contours of enhanced scalar susceptibilities ($\sim 1/m_\sigma^2$) around the CEP for both the cases.

A long standing question concerns the role of the strange quark and the axial anomaly in shaping the nature and location of the chiral transition. In the two-flavor chiral limit, the $O(4)$ symmetry of the light-quark sector forbids a cubic invariant in the effective potential, so the transition at $\mu=0$ is expected to belong to the O(4) universality class and remain continuous~\cite{rob}. Extending to 2+1 flavors introduces the $U(1)_A$-anomalous 't Hooft determinant term~\cite{tHooft:76prl}, which couples the light and strange condensates and, upon expanding around the strange condensate, generates an explicit cubic term in the light-sector potential - a term forbidden at $N_f=2$ but symmetry-allowed once the anomaly dynamically links the two sectors~\cite{rob,Pelis,pisarski24}. This cubic coupling drives the system toward a first-order transition already at the mean-field level, with its strength growing as the light quark mass is reduced toward the chiral limit $m_\pi \to 0$, where it terminates at a tricritical point (TCP). Away from this limit, the same coupling controls the location and reach of the CEP at the physical point, whose position depends on the strange quark  (equivalently, kaon) mass~\cite{Pelis,pisarski24,bernhardt23}.

In the specific context of the 2+1 flavor RQM model, the 't Hooft coupling $c$ is significantly enhanced because the condensate-dependent part of the $U(1)_A$ anomaly term is modified when the meson self-energies from quark loops are evaluated using the pole meson masses. The explicit symmetry-breaking strengths $h_x$ and $h_y$ are also weakened by on-shell renormalization -- by a small amount for $h_x$ and by a relatively larger amount for $h_y$~\cite{vkkr23}. These features of the RQM model shift the CEP to higher temperature ($T$) and lower chemical potential ($\mu$) relative to the curvature-mass-based, parameterized 2+1 flavor QMVT model~\cite{vkkr23,skrvkt24}.

It is relevant to note that the CEP in the RPQM model~\cite{skrvkt24,Akakn} lies closer to the range of critical-point locations predicted by the most recent results from independent theoretical approaches. Padé resummations of the lattice-computed Taylor expansion in $\mu_{B} \,(=3\mu_f)$~\cite{Clark}, conformal maps~\cite{Basar}, hybrid lattice-plus-gauge/gravity approaches~\cite{Hipp}, and the functional renormalization group~\cite{Lugao} all point to a common region of the QCD phase diagram, $T_c \sim 100$--$110$ MeV and $\mu_{B} \sim 420$--$650$ MeV (corresponding to $\mu_{\text{CEP}} \sim 140$--$217$ MeV)~\cite{MishaConf}. The size and shape of the critical region surrounding the CEP, and the proximity of the TCP (located in the light chiral limit, $m_\pi=0$, phase diagrams) to the CEP in the 2+1 flavor RQM and RPQM models, have been computed very recently in Ref.~\cite{Akakn}.

To definitively isolate the role of third (strange) flavor and axial $U_{A}(1)$ anomaly in governing the topography of critical fluctuations enveloping the CEP, our second objective is to perform a comprehensive, side-by-side comparison of the normalized quark number susceptibility contours from our present two-flavor RQM/RPQM calculations with the corresponding 2+1 flavor results of Ref.~\cite{Akakn}. For a systematic comparison, we employ logarithmic (Log) and PolyLog-glue forms of the Polyakov-loop potential -- without and with quark back-reaction, respectively -- in our present two-flavor RPQM model calculations, as in Ref.~\cite{Akakn} for the 2+1 flavor RPQM model. Such a comparative analysis is needed to establish essential theoretical boundaries for interpreting experimental signatures across different collision energies at RHIC-BES, FAIR, and NICA.

The third goal of the present work is to compute the critical exponents for finding the power law divergence of the quark number susceptibility when the $\mu$ approaches the $\mu_{\cep}$ at $T_{\cep}$  in the RQM, Log RPQM and PolyLog-glue RPQM models.~The RQM model critical exponents will be compared with QM model exponents in Ref.~\cite{Schaefer:2006ds,vkkr12} whereas the Log RPQM and PolyLog-glue RPQM model critical exponents will be compared with the Log PQMVT model in Ref.~\cite{vkkr12}.

The paper is organized as follows. Section~\ref{sec:II} presents a brief description of the QM/PQM, RQM/RPQM, and QMVT/PQMVT models and their effective potentials. Section~\ref{sec:IIA} contains specifically the QMVT model. Section~\ref{sec:IIB} details the various Polyakov-loop potentials, and section~\ref{sec:IIC} provides the grand thermodynamic potentials for the RPQM and PQMVT models. The results and discussion are presented in section~\ref{sec:III}. Here, the two-flavor RQM and Log-RPQM model phase diagrams for the physical-point and chiral-limit parameters are compared with the corresponding QMVT and Log-PQMVT phase diagrams in Section~\ref{sec:IIIA}. This section also displays the normalized quark number susceptibility contours ($R_q=2$) enveloping the CEP and TCP. Section~\ref{sec:IIIA}-1 compares the morphology of the critical regions around the CEP across the RQM/Log-RPQM and QMVT/Log-PQMVT models. In Section~\ref{subsec:IIIB}, the phase diagrams for the two-flavor RQM, Log-RPQM, and PolyLog-glue-RPQM models—along with their CEPs, TCPs, and normalized susceptibility contours ($R_q=2$)—are comprehensively compared with those of the corresponding 2+1-flavor RQM and RPQM model scenarios. The morphology of the critical regions for the two-flavor versus 2+1-flavor models is compared in Section~\ref{subsec:IIIB}-1, where the critical exponent of the TCP in the RQM model is also calculated. Section~\ref{secIIIC} presents the critical exponents for the power-law divergence of the quark number susceptibility at the CEP in the RQM, Log-RPQM, and PolyLog-glue-RPQM models. Finally, Section~\ref{sec:IV} provides a summary of the work. The calculation of the on-shell renormalized RQM model vacuum effective potential and its parameters is presented in Appendix~\ref{appendA}. The curvature mass calculations for the $\sigma$ and $\pi$ mesons are detailed in Appendix~\ref{appendB}, while Appendix~\ref{appendC} contains the relevant integrals.

\section{Model Formulation}
\label{sec:II}
The framework of Polyakov-loop extended quark meson (PQM) model \cite{Schaefer:2006ds,SchaPQM2F,vac,guptiw,vkkr12,raiti23}  is constructed by coupling the two flavor of quarks with   the scalar $\sigma$ and pseudo-scalar pion $\vec{\pi}$ fields where color degrees of freedom of quarks are also coupled to the temporal component of gauge field represented by the Polyakov loop potential.~The thermal expectation value of color trace of Wilson loop in temporal direction defines the Polyakov loop field $\Phi$ as :
\begin{equation}
\Phi(\vec{x}) = \frac{1}{N_c} \langle \Tr_c L(\vec{x})\rangle, \qquad \qquad  \bar\Phi(\vec{x}) =
\frac{1}{N_c} \langle \Tr_c L^{\dagger}(\vec{x}) \rangle
\end{equation}
here $L(\vec{x})$ is a $SU_c(3)$ color gauge group matrix in its fundamental representation.
\begin{equation}
\label{eq:Ploop}
L(\vec{x})=\mathcal{P}\mathrm{exp}\left[i\int_0^{\beta}d \tau
A_0(\vec{x},\tau)\right]
\end{equation}
Here $\mathcal{P}$ denotes the path ordering, $\beta = T^{-1}$, and $A_0$ is the temporal component of vector field \cite{Polyakov:78plb}. The homogeneous Polyakov loop fields $\Phi(\vec{x})=\Phi$=constant and $\bar\Phi(\vec{x})=\bar\Phi$=constant as in Ref. \cite{ratti,Roesnr}.
 
The Lagrangian of quark meson model together with the Polyakov loop potential ${\cal U} \left( \Phi, \bar\Phi, T \right)$ constitutes the Lagrangian of Polyakov quark-meson (PQM) model as:
\bqa
{\cal L_{PQM}}&=&{\cal L_{QM}}-{\cal U} \big( \Phi , \bar\Phi , T \big).
\label{lag:PQM}
\eqa
The Lagrangian  of QM model \cite{Rischke:00,Schaefer:09,TiPQM3F} is the following:
\bqa
\label{lag}
{\cal L_{QM}}&=&\bar{\psi}[i\gamma^\mu D_\mu- g\; T_a\big( \sigma_a 
+ i\gamma_5 \pi_a\big) ] \psi+\cal{L_M}\;.
\eqa
~Quarks couple with the uniform temporal background gauge field through the co-variant derivative $D_{\mu} = \partial_{\mu} -i A_{\mu}$. The $A_{\mu} = \delta_{\mu 0} A_0$ (Polyakov gauge) and $A_{\mu} = g_s A^{a}_{\mu} \lambda^{a}/2$ with vector potential $A^{a}_{\mu}$ for color gauge fields.~The $\lambda_a$ ($a=1 \cdots 8$) are Gell-Mann matrices in the color space and $g_s$ is the $SU_c(3)$ gauge coupling.~In the pure QM model,~the gauge coupling is zero and $D_{\mu} \equiv \partial_{\mu}$.~The  $O(4)$ sigma model Lagrangian  is :
\bqa
\nonumber
\label{lagM}
\cal{L_M}&=&\frac{1}{2}(\partial_\mu \sigma)^{2}+\frac{1}{2}(\partial_\mu \bm{\pi})^{2}-U(\sigma,\bm{\pi}).\
\eqa

\bqa
U(\sigma,\bm{\pi})=\frac{1}{2}m^2(\sigma^2+\bm{\pi}^2)+\frac{\lambda}{4}(\sigma^2+\bm{\pi}^2)^2- h\sigma
\eqa
The doublet of quark fields $\psi$ (color $N_c$-plet  Dirac spinor) is coupled to the isospin singlet scalar $\sigma$ and isospin triplet pseudo-scalar mesons $\bm{\pi}$ fields 
through the flavor blind Yukawa coupling $g$. The $\sigma$ field  develops the non-zero vacuum expectation value $\overline{\sigma}$. The $SU_L(2) \times SU_R(2)$ chiral symmetry is broken spontaneously  by the chiral condensate $\overline{\sigma}=x$ and  and it is broken explicitly also by the external field $h  \neq 0$.~Incorporating the thermal and quantum fluctuations of the quarks/anti-quarks and treating the mesons as mean fields,~the QM model grand potential \cite{Schaefer:2006ds,SchaPQM2F,vac,guptiw,vkkr12},~is obtained after adding the vacuum effective potential $ U(\overline{\sigma}=x) $ with the quark/anti-quark  contribution $\Omega_{q\bar{q}}$ at finite temperature $T$ and quark chemical potential $\mu_{f} (f=u,d)$.
\bqa
\label{Grandpxy}
&&\Omega_{\rm MF }(T,\mu)=U(\x)+\Omega_{q\bar{q}} (T,\mu;\x)\;. \\ \nonumber \\ 
&&\Omega_{q\bar{q}}(T,\mu;\x)=\Omega_{q\bar{q}}^{vac}+\Omega_{q\bar{q}}^{T,\mu}(\x). \\ \label{eq:mesop}
&&U(\x)=\frac{m^{2}}{2}\x^{2}+\frac{\lambda}{4} \x^{4} -h \x.  \\
\label{vac1}
&&\Omega_{q\bar{q}}^{vac} =- 2 N_c\sum_f  \int \frac{d^3 p}{(2\pi)^3} \ E_q \  \theta( \Lambda_c^2 - \vec{p}^{2})\;.\\ \nonumber \\
\label{vac2}
&&\Omega_{q\bar{q}}^{T,\mu}(\x)=- 2 N_c \sum_{f} \int \frac{d^3 p}{(2\pi)^3} T \left[ \ln g_f^{+}+\ln g_f^{-}\right].\;  \\
\label{GrandQM}
&&\Omega_{\rm QM }(T,\mu,\x)=U(\x)+\Omega_{q\bar{q}}^{T,\mu}(\x)\;. 
\eqa
The $ g^{\pm}_f = \left[1+e^{-E_{f}^{\pm}/T}\right] $ where $E_{f}^{\pm} =E_f \mp \mu_{f}$ and $E_f=\sqrt{p^2 + m{_f}{^2}}$ is the quark/anti-quark energy.~The quark mass  $m_{u(d)}=g\x$ and $\mu_{u}=\mu_{d}=\mu$.~The quark one-loop vacuum term with ultraviolet cut-off $\Lambda_c$ in Eq.~(\ref{vac1}), which accounts for the Dirac sea contribution, is omitted from the grand effective potential of the quark-meson (QM) model in Eq.~(\ref{GrandQM}) under the standard mean-field approximation (s-MFA).~Several studies have used the minimal subtraction scheme to regularize quark one-loop vacuum divergences after including vacuum fluctuations within the extended mean-field approximation (e-MFA) \cite{vac, lars,guptiw, vkkr12, chatmoh1, schafwag12, vkkt13, Rai}. However, their treatment of the vacuum effective potential $\Omega_{vac} (\x)=U(\x)+\Omega_{q\bar{q}}^{vac}$ introduces an inconsistency because they fix the model parameters using the curvature masses of mesons, which are obtained by taking the second derivatives of the effective potential with respect to different fields at its minimum. This inconsistency becomes apparent when one notes that calculating curvature masses involves evaluating meson self-energies at zero momentum, as the effective potential is the generator of the $n$-point functions of the theory at vanishing external momenta \cite{laine,Adhiand1,Adhiand2,Adhiand3,asmuAnd,RaiTiw22,raiti23,vkkr23,skrvkt24}. Furthermore, Refs.~\cite{Kobes, Rebhan} have emphasized that the pole definition of the meson mass is the genuinely physical and gauge-invariant one.

In the present work, alongside the curvature-mass-parameterized effective potential of the QMVT model, we primarily utilize the consistent e-MFA RQM model effective potential. This RQM model effective potential is derived by matching the counter-terms in the $\overline{\text{MS}}$ scheme to those in the on-shell (OS) scheme \cite{Adhiand1,Adhiand2,Adhiand3,asmuAnd,RaiTiw22,raiti23}. The relationships between the renormalized parameters of the two schemes are established by relating physical quantities—specifically the on-shell pole masses $m_{\pi}$ and $m_{\sigma}$, along with the pion decay constant $f_{\pi}$—to the expressions for the $\overline{\text{MS}}$ running couplings and mass parameters. These matched relations subsequently serve as inputs for evaluating the effective potential via the modified minimal subtraction procedure. Following the cancellation of the $1/\epsilon$ divergences, the vacuum effective potential $\Omega_{vac}=U(\x_{\ms})+\Omega^{q,vac}_{\ms}+\delta U(\x_{\ms})$ in the $\overline{\text{MS}}$ scheme is expressed in Refs.~\cite{Adhiand1,Adhiand2,Adhiand3,asmuAnd} in terms of the scale-$\Lambda$-independent constituent quark mass parameter $\Delta=g_{\ms}\x_{\ms}$ as follows:

\begin{align}
\label{OmegDelx}
\Omega_{\text{vac}}(\Delta) &= \frac{m^2_0 \Delta^2}{2g^2_0} + \frac{\lambda_0 \Delta^4}{4g^4_0} - \frac{h_0 \Delta}{g_0} + \frac{2N_c\Delta^4}{(4\pi)^2}\left[\frac{3}{2} + \ln\frac{\Lambda^2}{\Delta^2}\right]
\end{align}
The scale $\Lambda_0$ gets fixed by requiring that the minimum of the RQM model effective potential does not shift from that of the QM model~\cite{Adhiand1,Adhiand2,Adhiand3,asmuAnd,RaiTiw22,raiti23,vkkr23,skrvkt24} as :
\bqa
\label{Sclcond}
&\ln\left(\frac{\Lambda^2_0}{m_u^2}\right)+\mathcal{F}(m^2_\pi)+m^2_\pi \mathcal{F}^{\prime}(m^2_\pi)=0\;. 
\eqa
The terms $\mathcal{F}(m^2_\pi) $ and $ \mathcal{F}^{\prime}(m^2_\pi) $ and the derivations of the expressions of the renormalized parameters  $m^{2}_{0}=(m^{2}+m^2_{\text{\tiny{FIN}}})$,~$h_{0}=(h+h_{\text{\tiny{FIN}}})$, and $\lambda_0=(\lambda+\lambda_{\text{\tiny{FIN}}})$, are presented in the appendix.~The $m^2_{\text{\tiny{FIN}}}$,
$ h_{\text{\tiny{FIN}}} $, and $\lambda_{\text{\tiny{FIN}}} $,  
are the finite on-shell corrections in the parameters at the scale $\Lambda_0$. The experimental values of the pseudo-scalar meson mass $m_{\pi}$, the scalar $\sigma$ mass $m_{\sigma}$, and the pion decay constant $f_{\pi}$ serve as inputs to determine the tree-level QM model quartic coupling $\lambda=(m_{\sigma}^2-m_{\pi}^2)/2f_{\pi}^2$, the mass parameter $m^2=(3m_{\pi}^2-m_{\sigma}^2)/2$, and $h=f_{\pi}m_{\pi}^2$. The expressions for the renormalized parameters and the  effective potential as originally derived in Refs.~\citep{Adhiand1,Adhiand2,Adhiand3} are presented in Appendix~\ref{appendA}.

The dressing of the meson propagator in the on-shell scheme leads to the renormalization of $f_{\pi}$ and $g$. However, these quantities remain unchanged since $g_{\ms}=g_{ren}=g_{0}=g$ and $\x_{\ms}=\x$ at $\Lambda_0$. Furthermore, at the minimum, $\x_{\ms}=f_{\pi,ren}=f_\pi$. Using $\Delta=g \x$, the vacuum effective potential in Eq.~(\ref{OmegDelx}) can be expressed in terms of $\x$ as :
{\small
\begin{align}
\label{vacRQM2f}
\Omega_{vac}^{\rm RQM}(\x) &= \frac{(m^{2}+m^2_{\text{\tiny{FIN}}})}{2} \ \x^{2}
- (h+h_{\text{\tiny{FIN}}}) \ \x
\nonumber \\
&+ \frac{(\lambda+\lambda_{\text{\tiny{FIN}}})}{4} \ \x^{4}
- \frac{2N_c g^4}{(4\pi)^2} \ \x^4 \ln\left(\frac{\x^2}{f_{\pi}^2}\right)
\nonumber \\
&+ \frac{2N_c g^4 \, \x^4}{(4\pi)^2}
\left[\frac{3}{2}-\mathcal{C}(m^2_\pi)-m^2_\pi \mathcal{C}^{\prime}(m^2_\pi)\right].
\end{align}
}
The  $\mu \text{ and } T$ dependent grand effective potential of the RQM model is obtained after adding the finite temperature quark-antiquark contribution $ \Omega_{q\bar{q}}^{T,\mu}(\x)$ to the renormalized vacuum  $\Omega_{vac}^{\rm RQM}(\x) $ contribution.
\bqa
\label{GrandRQM}
&&\Omega_{\rm RQM }(T,\mu,\x)=\Omega_{vac}^{\rm RQM}(\x)+\Omega_{q\bar{q}}^{T,\mu}(\x)\;. 
\eqa 

By minimizing the grand effective potential in Eq.~(\ref{GrandRQM}) via the condition $\frac{\partial \Omega_{\rm RQM}(T,\mu,x)}{\partial \x}= 0$, one obtains the $T$ and $\mu$ dependence of the order parameter $\x$.~The expressions of  curvature masses  of the scalar and pseudo-scalar mesons for the RQM model are derived in Refs.~\cite{vkkr23,vktChpt1}.~The renormalized chiral symmetry breaking strength $h_{0}$ is determined from Eq.~(\ref{vacRQM2f}) by imposing the equation of motion, $\frac{\partial \Omega_{\rm vac}^{\rm RQM}(x)}{\partial \x}=0$.
\bqa
\label{hx0}
&h_{0}=m_{\pi,c}^2 \ f_{\pi}.
\eqa 
The pion  curvature mass $m_{\pi,c}$ expression is the following.
\begin{align}
\label{mpicr}
&m_{\pi,c}^2=m_{\pi}^2\biggl\{ 1-\frac{N_{c}g^2}{4\pi^2} \ m_{\pi}^2 \ \mathcal{C}^{\prime}(m^2_\pi,m_u)  \biggr\}.
\end{align}
The $\pi$ curvature mass  $m_{\pi;c}= 135.95$ MeV is smaller by 2.05 MeV from its pole mass $m_{\pi}= 138$ MeV.~In all the computations, we have taken pion decay constant  $f_{\pi}=92.4$ MeV and the Yukawa coupling $g=3.25$.

\subsection{The QMVT model and curvature mass based parameter fixing}
\label{sec:IIA}
In this subsection, we describe the calculation of the effective potential as derived in Refs.~\cite{guptiw,vkkr12}, where the quark one-loop vacuum divergence in Eq.~(\ref{vac1}) is regularized via the minimal subtraction scheme. In this approach, the model parameters are fixed using the $\sigma$ and $\pi$ meson curvature masses (screening masses), which are defined by the second derivatives of the thermodynamic potential evaluated at its minimum. The zero-temperature quark one-loop vacuum contribution can be expressed as :

\begin{align}
\Omega_{qq}^{\mathrm{vac}}
&= -2N_fN_c \int \frac{d^3p}{(2\pi)^3}\,E_q \\
&= -2N_fN_c \int \frac{d^4p}{(2\pi)^4}\,\ln\left(p_0^2+E_q^2\right) + \mathcal{K}.
\end{align}
Since the infinite constant $\mathcal{K}$ does not depend on the fermion mass, it is safely discarded. Applying dimensional regularization to Eq.~(13) in $d=3-2\epsilon$ dimensions yields the potential up to zeroth order in $\epsilon$, expressed as:

\begin{equation}
\Omega_{q\bar{q}}^{\rm vac} =  \frac{N_c N_f}{16 \pi^2} m_q^4 \left\{
  \frac{1}{\epsilon} - \frac{1}{2} 
\left[  -3 + 2 \gamma_E + 4 \ln\left(\frac{m_q}{2\sqrt{\pi} M}\right)
\right] 
\right\},
\label{Omega_DR}
\end{equation}
where M denotes the arbitrary renormalization scale. Introducing a counterterm $\delta \mathcal{L}$ into the Lagrangian of the QM or PQM model.

\begin{equation}
\delta \mathcal{L} = \frac{N_c N_f}{16 \pi^2} g^4{\sigma}^{4} \left\{ \frac{1}{\epsilon} - \frac{1}{2}
\left[  -3 + 2 \gamma_E - 4 \ln\left(2\sqrt{\pi}\right)  \right] \right\},
\label{counter}
\end{equation}

gives the renormalized fermion vacuum loop contribution as 
\begin{equation}
\Omega_{q\bar{q}}^{\rm reg} =  -\frac{N_c N_f}{8 \pi^2} m_q^4  \ln\left(\frac{m_q}{ M}\right).
\label{Omega_reg}
\end{equation}

The vacuum contribution in the Eq.~(\ref{vac1}) will be replaced by appropriately renormalized fermion vacuum loop contribution in Eq.~(\ref{Omega_reg})

The quartic coupling $\lambda$ and mass parameter $m^2$ are determined in the vacuum using the effective potential where the purely $\sigma$-dependent mesonic potential $U(\sigma)$ is combined with the renormalized vacuum term of Eq.~(\ref{Omega_reg})

\bqa
\nonumber
\Omega^\text{M}(\sigma) &=& \Omega_{q\bar{q}}^{\rm reg}+U(\sigma) 
= -\frac{N_c N_f}{8 \pi^2} g^4{\sigma}^4 \ln\left(\frac{g\sigma}{ M}\right) \\
&&+ \frac{m^2}{2}{\sigma}^2+\frac{\lambda}{4}{\sigma}^4-h\sigma, 
\label{OmegaZEROT}
\eqa

The first derivative of $\Omega(\sigma)$ with respect to $\sigma$ at 
$\sigma=f_\pi$ in the vacuum is put to zero

\begin{equation}
\frac{\partial \Omega^\text{M}_{\rm MF}(\sigma)}{\partial \sigma}\bigg|_{\sigma=f_\pi} =0
\label{DERIZERO}
\end{equation}
The second derivative of $\Omega^\text{M}(\sigma)$ with respect to $\sigma$ at 
$\sigma=f_\pi$ in the vacuum gives the mass of $\sigma$

\begin{equation}
m_\sigma ^2 = \frac{\partial^2 \Omega^\text{M}_{\rm MF}(\sigma)}{\partial\sigma^2}
	    = \frac{\partial^2 \Omega^\text{M}(\sigma)}{\partial \sigma^2}\bigg|_{\sigma=f_\pi}
\label{msigma2}
\end{equation}

Solving the equations (\ref{DERIZERO}) and (\ref{msigma2}), we obtain 

\begin{equation}
\lambda = \lambda_s+\frac{N_c N_f}{8 \pi^2} g^4\left[3+4\ln\left(\frac{g f_\pi}{ M}\right)\right]
\label{lambd}
\end{equation}

and 

\begin{equation}
m^2 = m^2_{s}-\frac{N_c N_f}{4 \pi^2} g^4\ f_\pi^2 
\label{lambdv2}
\end{equation}

where $\lambda_s$  and  $m_{s}^2$ are the values of the parameters 
in the pure sigma model 

\bqa
\lambda_s =\frac{m_\sigma^2-m_\pi^2}{2 f_\pi^2} \text{ and } \quad \quad  m^2_s  =\frac{3m_\pi^2-m_\sigma^2}{2}.
\label{lambs}
\eqa

Inserting the values of $\lambda$ and $m^2$ into Eq.~(\ref{OmegaZEROT}), the $M$-dependence cancels out neatly upon rearranging the terms and we find the following vacuum effective potential of quark meson with vacuum term (QMVT) model.

\begin{equation}
\Omega^{\text {QMVT}}_{vac}(\sigma) =  -\frac{N_c N_f}{8 \pi^2} g^4{\sigma}^4 \ln\left(\frac{\sigma}{f_\pi}\right)
+ \frac{m_r^2}{2}\sigma^2+\frac{\lambda_r}{4}\sigma^4-h\sigma .
\label{OmegasigF}
\end{equation}

Here, $\lambda_r$ and $m_r^2$ denote the renormalized values of these parameters, obtained after properly accounting for the fermion vacuum contribution.

\begin{equation}
\lambda_r = \lambda_s+\frac{3 N_c N_f}{8 \pi^2} g^4
\label{lambdR}
\end{equation}
and 
\begin{equation}
m_r^2 = m^2_s-\frac{N_c N_f}{4 \pi^2} g^4\ f_\pi^2 
\label{Rlambdv2}
\end{equation}

The expression of effective potential in terms $x$ (where $\sigma=\overline{\sigma}=x$) is the following. 

\begin{equation}
\Omega^{\text {QMVT}}_{vac}(x) =  -\frac{N_c N_f}{8 \pi^2} g^4{x}^4 \ln\left(\frac{x}{f_\pi}\right)
+ \frac{m_r^2}{2}x^2+\frac{\lambda_r}{4}x^4-hx .
\label{VacQMVT}
\end{equation}
Adding the finite temperature quark-antiquark contribution $ \Omega_{q\bar{q}}^{T,\mu}(\x)$ to the vacuum effective potential $\Omega_{vac}^{\rm QMVT}(\x) $ of quark meson with vacuum term (QMVT) model, we find the  $\mu \text{ and } T$ dependent grand effective potential of the QMVT model.
\bqa
\label{GrandQMVT}
&&\Omega_{\rm QMVT }(T,\mu,\x)=\Omega_{vac}^{\rm QMVT}(\x)+\Omega_{q\bar{q}}^{T,\mu}(\x)\;. 
\eqa 

By minimizing the grand effective potential in Eq.~(\ref{GrandQMVT}) via the condition $\frac{\partial \Omega_{\rm QMVT}(T,\mu,x)}{\partial \x}= 0$, one obtains the $T$ and $\mu$ dependence of the order parameter $\x$.~The expressions of  curvature masses  of the scalar and pseudo-scalar mesons for the QMVT model are derived in  \ref{appendC}.

\subsection{Different forms of Polyakov-loop potential}
\label{sec:IIB}

Various forms of the Polyakov-loop effective potential $\mathcal{U}(\Phi,\bar{\Phi},T)$ have been employed in the literature to investigate the confinement-deconfinement phase transition. The simplest of these is formulated by constructing a potential that preserves all underlying symmetries while capturing the spontaneous breaking of the $Z(3)$ symmetry in the deconfined phase~\cite{SveLer,benji,BankUka}. The minimal structure for this Polyakov-loop potential is given by the following polynomial expression:

\bqa
\label{plykov_poly}
\hspace{-0.5 cm}\frac{\mathcal{U_{\rm Poly}}}{T^4}&=&-\frac{b_2(T)}{2}\Phi\bar{\Phi}-\frac{b_3}{6}(\Phi^3+\bar{\Phi}^3)+\frac{b_4}{4}(\Phi\bar{\Phi})^2\;,
\eqa
the coefficients of the Eq.~(\ref{plykov_poly}) are given by
\bqa
b_2(T)=a_0+a_1\left(\frac{T_0}{T}\right)+a_2\left(\frac{T_0}{T}\right)^2+a_3\left(\frac{T_0}{T}\right)^3\;,
\eqa
where $a_0=6.75$, $a_1=-1.95$, $a_2=2.625$, $a_3=-7.44$, $b_3=0.75$ and $b_4=7.5$ .\\

To improve upon the polynomial form, one includes the contribution arising from the integration over the $SU(3)$ group volume in the generating functional for the Euclidean action. Evaluated via the Haar measure, this integration introduces a Jacobian determinant whose logarithm is added to the effective potential. A key advantage of this formulation is that the positive coefficient of the logarithm ensures that the potential remains bounded from below for large $\Phi$ and $\bar{\Phi}$. This logarithmic version of the Polyakov-loop potential~\cite{fuku,Roesnr} takes the form given below.
\bqa
\label{plykov_log}
\nonumber
\hspace{-0.5 cm}\frac{\mathcal{U_{\rm Log}}}{T^4}&=&b(T)\ln[1-6\Phi\bar{\Phi}+4(\Phi^3+\bar{\Phi}^3)-3(\Phi\bar{\Phi})^2]\;\\&&-\frac{1}{2}a(T)\Phi\bar{\Phi}\;.
\eqa
"By fitting pure-gauge lattice data for the pressure, entropy density, energy density, and the evolution of the Polyakov loop $\langle\Phi\rangle$, the parameters for both the polynomial and logarithmic potentials were determined~\cite{ratti,Roesnr}. The resulting coefficients for Eq.~(\ref{plykov_log}) are given by~\cite{Roesnr},
\bqa
&&a(T)=a_0+a_1\left(\frac{T_0}{T}\right)+a_2\left(\frac{T_0}{T}\right)^2\;,\\
&&b(T)=b_3\left(\frac{T_0}{T}\right)^3\;,
\eqa
where $a_0=3.51$, $a_1=-2.47$, $a_2=15.2$, $b_3=-1.75$.
The logarithmic potential exhibits qualitative consistency with leading-order results from the strong-coupling expansion~\cite{JLange}. Moreover, since the potential diverges as $\Phi, \bar{\Phi} \to 1$, the expectation value of the Polyakov loop is bounded below 1, reaching unity only asymptotically as $T \to \infty$. 

A refined Polyakov-loop effective potential was constructed in Ref.~\cite{Redlo} by incorporating Polyakov-loop fluctuations. Its parameters are tuned to reproduce both the longitudinal and transverse susceptibilities, in addition to other existing lattice data. By adding a logarithmic term to the polynomial form, the expression for the new PolyLog Polyakov-loop potential is obtained below.
\bqa
\label{plykov_polylog}
\nonumber
\hspace{-0.5 cm}\frac{\mathcal{U_{\rm PolyLog}}}{T^4}&=&b(T)\ln[1-6\Phi\bar{\Phi}+4(\Phi^3+\bar{\Phi}^3)-3(\Phi\bar{\Phi})^2]\;\\ \nonumber
&&+a_2(T)\Phi\bar{\Phi}+a_3(T)(\Phi^3+\bar{\Phi}^3)+a_4(T)(\Phi\bar{\Phi})^2.\\
\eqa
In the PolyLog parametrization,~the coefficients in the Eq.~(\ref{plykov_polylog}) are defined as :
\bqa
&&a_i(T)=\frac{a^{(i)}_0+a^{(i)}_1\left(\frac{T_0}{T}\right)+a^{(i)}_2\left(\frac{T_0}{T}\right)^2}{1+a^{(i)}_3\left(\frac{T_0}{T}\right)+a_4^{(i)}\left(\frac{T_0}{T}\right)^2}\;\\
&&b(T)=b_0\left(\frac{T_0}{T}\right)^{b_1}\left[1-e^{b_2\left(\frac{T_0}{T}\right)^{b_3}}\right].
\eqa
\\
The  summary of the parameters is given in the Table~\ref{tab:plglg}.
\begin{table}[!htbp]
    \caption{Parameters of the PolyLog Polyakov-loop potential have been taken from the Ref.~\cite{Redlo}.}
    \label{tab:plglg}
    \resizebox{0.48\textwidth}{!}{
    \begin{tabular}{p{1.5cm} p{1.5cm} p{1.5cm} p{1.5cm} p{1.5cm} p{1.5cm} p{0.5 cm}}
      \toprule 
      PolyLog& $a^{(2)}_0$ & $a^{(2)}_1$ & $a^{(2)}_2$ & $a^{(2)}_3$ & $a^{(2)}_4$&\\
      & 22.07 & -75.7 & 45.03385 & 2.77173 & 3.56403&\\
      & $a^{(3)}_0$ & $a^{(3)}_1$ & $a^{(3)}_2$ & $a^{(3)}_3$ & $a^{(3)}_4$&\\
      &-25.39805&57.019&-44.7298&3.08718&6.72812\\
      & $a^{(4)}_0$ & $a^{(4)}_1$ & $a^{(4)}_2$ & $a^{(4)}_3$ & $a^{(4)}_4$&\\
      &27.0885&-56.0859&71.2225&2.9715&6.61433&\\
      &$b_0$&$b_1$&$b_2$&$b_3$& &\\
      &-0.32665&5.8559&-82.9823&3.0& &\\
      \hline      
      \hline 
    \end{tabular}}
\end{table}
In pure-gauge Yang-Mills theory, the deconfinement phase transition is first-order with a critical temperature $T_c^{\rm YM} = T_0 = 270$~MeV. This first-order transition becomes a crossover in the presence of dynamical quarks. In full dynamical QCD, the parameter $T_0$ depends on the number of quark flavors and the chemical potential~\cite{SchaPQM2F,Haas,kovacs,BielichP,THerbst2}. This dependence arises because $T_0$ is linked to the QCD mass scale, $\Lambda_{\rm QCD}$, which is modified by the fermionic matter fields. This flavor and chemical potential dependence $T_0 \to T_0(N_f,\mu)$ takes the following form :
\bqa
\label{t0_mu}
T_0(N_f,\mu)=\hat{T} \ e^{-1/(\alpha_0 b(N_f,\mu))}\;,
\eqa
with 
\bqa
b(N_f,\mu)=\frac{1}{6\pi}(11N_c-2N_f)-b_\mu\frac{\mu^2}{(\hat{\gamma}\hat{T})^2}.
\eqa
Here the parameter $\hat{T}$  is fixed at the scale $\tau$,  $\hat{T}=T_\tau=1.77$ GeV and $\alpha_0=\alpha(\Lambda)$ at a UV scale $\Lambda$.~The $T_0(N_f=0)$ = 270 MeV gives  $\alpha_0$ = 0.304 and $b_\mu\simeq\frac{16}{\pi}N_f$. The curvature of $T_0(\mu)$ is governed by the parameter $\hat{\gamma}$  with the systematic error estimation range $0.7\lesssim\hat{\gamma}\lesssim1$ \cite{SchaPQM2F,THerbst2}.~Massive flavors lead to suppression factors of the order $T_0^2/(T^2_0 +m^2)$ in the $\beta$-function.

When the backreaction of quarks is included in full dynamical QCD, the Polyakov-loop potential is replaced by the QCD glue potential. In Ref.~\cite{Haas}, Functional Renormalization Group (FRG) equations were applied to QCD to compare the pure-gauge potential $\mathcal{U}_{\rm YM}$ with the ``glue'' potential $\mathcal{U}_{\rm glue}$, which incorporates quark polarization in the gluon propagator. Although significant differences were found between the two, it was observed that they share the same shape and can be mapped onto one another by relating the temperatures of the two systems, $T_{\rm YM}$ and $T_{\rm glue}$. Identifying the previously discussed Polyakov-loop potential as $\mathcal{U}_{\rm YM}$, the improved potential $\mathcal{U}_{\rm glue}$ is constructed as follows~\cite{Haas}:
\bqa
\frac{\mathcal{U}_{\rm glue}}{T^4_{\rm glue}}(\Phi, \bar{\Phi}, T_{\rm glue})&=&\frac{\mathcal{U}_{\rm YM}}{T^4_{\rm YM}}(\Phi, \bar{\Phi}, T_{\rm YM})
\eqa
here the temperature $T_{\rm glue}$ is related to $T_{\rm YM}$ as 
\bqa
\frac{T_{\rm YM}-T^{\rm YM}_{\rm c}}{T^{\rm YM}_{\rm c}}=0.57 \
\frac{T_{\rm glue}-T^{\rm glue}_{\rm c}}{T^{\rm glue}_{\rm c}}
\eqa
The $T^{\rm glue}_{\rm c}$ is the transition temperature for the unquenched case.~The coefficient 0.57 comes from the comparison of the two effective potentials.~$T^{\rm glue}_{\rm c}$ lies within a range $T^{\rm glue}_{\rm c} \in [180,270]$.~In practice,~one  uses the replacement $T  \longrightarrow T^{\rm YM}_{\rm c}(1+0.57(\frac{T}{T^{\rm glue}_{\rm c}}-1))$ in the right-hand side of the Polyakov-loop potentials where $T_0$ means $T^{\rm YM}_{\rm c}$ and ($T\sim T_{\rm YM}$) on the left side of the arrow while ($T\sim T_{\rm glue}$) on the right side \cite{skrvkt24}.~In our computations for two-flavor, we take both $T_{c}^{glue}$ and $T_0$ as 208 MeV.


\subsection{Ployakov-loop potential augmented RPQM and PQMVT Model}
\label{sec:IIC}
Because the Polyakov-loop potential contributes only at non-zero temperatures, it does not affect the vacuum parameters of the chiral part of the effective potential. Consequently, the Polyakov-loop potential modifies only the finite-temperature part $\Omega_{q\bar{q}}^{T,\mu}$ of the quark-antiquark contribution to the total effective potential, given by Eq.~(\ref{vac2}).
\bqa
\label{FTpoly}
\Omega_{q\bar{q}}^{T,\mu}(\x,\Phi,\bar{\Phi})=- 2 \sum_q \int \frac{d^3 p}{(2\pi)^3} T \left[ \ln g_f^{+}+\ln g_f^{-}\right].\;
\eqa
The $\Omega_{q\bar{q}}^{T,\mu}(\x,\Phi,\bar{\Phi})=\Omega_{q\bar{q}}(T,\mu;\x,\Phi,\bar{\Phi}) $ and new $ g_{f}^{\pm}$ being different from their earlier values in the  Eq.~(\ref{vac2}),~are the following :
\bqa
\hspace{-1 cm}g_{f}^{+}&=&\left[1+3\Phi e^{-E_{q}^{+}/T}+3\bar{\Phi}e^{-2E_{q}^{+}/T}+e^{-3E_{q}^{+}/T}\right]\;,\qquad\\
g_{f}^{-}&=&\left[1+3\bar{\Phi} e^{-E_{q}^{-}/T}+3\Phi e^{-2E_{q}^{-}/T}+e^{-3E_{q}^{-}/T}\right]\;.
\eqa
$E_{f}^{\pm} =E_f \mp \mu_{f} $ and $E_f=\sqrt{p^2 + m{_f}{^2}}$ with $f=u,d$ represent the quark/antiquark energy.~Here $m_{u}=m_{d}=g\x$ is the mass of the light quarks $u$ and $d$.~In this work we have considered  $\mu_{u}=\mu_{d}=\mu$.

We construct the grand thermodynamic potential for the renormalized Polyakov-loop-enhanced quark-meson (RPQM) model~\cite{raiti23}, using the PQM model Lagrangian in Eq.~(\ref{lag:PQM}). The total potential is formulated by adding the RQM vacuum effective potential $\Omega_{\rm vac}^{\rm RQM}(x)$ from Eq.~(\ref{vacRQM2f}) to the thermal quark-antiquark contributions in the presence of the Polyakov-loop potential.
\bqa
\label{rpqmomega}
\nonumber
\Omega_{\rm RPQM}(T,\mu;\x,\Phi,\bar{\Phi})&=&\Omega_{vac}^{\rm RQM}(\x)+\mathcal{U}(\Phi,\bar{\Phi}) \\
&&+ \Omega_{q\bar{q}} (T,\mu;\x,\Phi,\bar{\Phi}) 
\eqa
By adding the vacuum effective potential in Eq.~(\ref{VacQMVT}) for the curvature-mass-parameterized QMVT model to the thermal quark-antiquark contributions in the presence of the Polyakov-loop potential, one obtains the grand thermodynamic potential of the PQMVT model, given below :
\bqa
\label{pqmvtomega}
\nonumber
\Omega_{\rm PQMVT}(T,\mu;\x,\Phi,\bar{\Phi})&=&\Omega_{vac}^{\rm QMVT}(\x)+\mathcal{U}(\Phi,\bar{\Phi}) \\
&&+ \Omega_{q\bar{q}} (T,\mu;\x,\Phi,\bar{\Phi}). 
\eqa

\begin{table*}[!htbp]
    \caption{Different Model Parameters at the physical point and in the light chiral limit study.}
    \label{tab:table2}
    \begin{tabular}{p{0.1\textwidth} p{0.1\textwidth} p{0.1\textwidth} p{0.1\textwidth} p{0.1\textwidth} p{0.1\textwidth} }
    \toprule 
      Model& $m_{\sigma}$&$m_{\pi}(\text{MeV})$&$\lambda_r$&$m^2_r(\text{MeV})$&$h(\text{MeV}^3)$ \\
      \hline 
       QMVT& $500$&  $138$&$38.96$ &$-(491.12)^2$ &$(120.73)^3$ \\
       &$500$&$\phantom{00}0$&$40.07$&$-(519.39)^2$ &$\phantom{(000.00)^3}0$ \\
      \hline
      \phantom{text} & \phantom{text}&\phantom{text}&$\lambda_0$ &$m_0^2(\text{MeV}^2)$& $h(\text{MeV}^3)$\\
      \hline
      RQM&$500$&$138$&$10.341$&${-(451.72)^2}$ &$(119.53)^3$  \\
       &$500$&$\phantom{00}0$&$11.045$&${-(488.95)^2}$ &$\phantom{(000.00)^3}0$ \\
      \hline 
    \end{tabular}
\end{table*}
The temperature $T$ and chemical potential $\mu$ dependence of the chiral condensate $x$, $\Phi$, and $\bar{\Phi}$ are obtained in the RPQM and PQMVT models by determining the global minimum of the grand potentials given in Eq.~(\ref{rpqmomega}) and Eq.~(\ref{pqmvtomega}), respectively.
\bqa
\nonumber
\frac{\partial{\Omega_{\rm RPQM/PQMVT}}}{\partial
      {\x}}&=&\frac{\partial \Omega_{\rm RPQM/PQMVT}}{\partial\Phi}\\ &&=\frac{\partial \Omega_{\rm RPQM/PQMVT}}{\partial\bar{\Phi}} =0
\label{EoMMF3}
\eqa

The effective potentials and phase diagrams of the two-flavor RQM and QMVT models were computed and compared in Ref.~\cite{RaiTiw22}, whereas the phase diagrams, various thermodynamic quantities, and the critical endpoints (CEPs) of the RPQM and PQMVT models were previously calculated in Ref.~\cite{raiti23} using both the logarithmic and PolyLog-glue parameterizations of the Polyakov-loop potential. In those works, two flavors of quarks were coupled to the four scalar ($\sigma$ and $\vec{a}_0$) and four pseudoscalar ($\eta$ and $\vec{\pi}$) mesons of the $SU_L(2) \times SU_R(2)$ symmetric linear sigma model. In the present work, we employ the QM model, utilizing the $O(4)$ linear sigma model as the minimal effective theory coupled to the two quark flavors. This approach integrates out (or freezes) the heavier $\eta$ and $\vec{a}_0$ fields, restricting the meson field space to a single four-component real vector $\Phi = (\sigma, \vec{\pi})$.We utilize this $O(4)$ framework of the two-flavor RQM/RPQM and QMVT/PQMVT models to calculate the phase diagrams using both physical-point and chiral-limit parameters. Furthermore, we compute the quark number susceptibilities to analyze the critical region around the CEP across these different model scenarios. Additionally, we will compute the critical region around the TCP in the RQM model. We will also evaluate the critical exponents characterizing the divergence of the quark number susceptibility at this TCP, as well as at the CEP in the RQM, Log-RPQM, and PolyLog-glue models.

\begin{figure*}[htb]
\subfigure[]{
\label{fig1a} 
\begin{minipage}[b]{0.48\textwidth}
\centering \includegraphics[width=\linewidth]{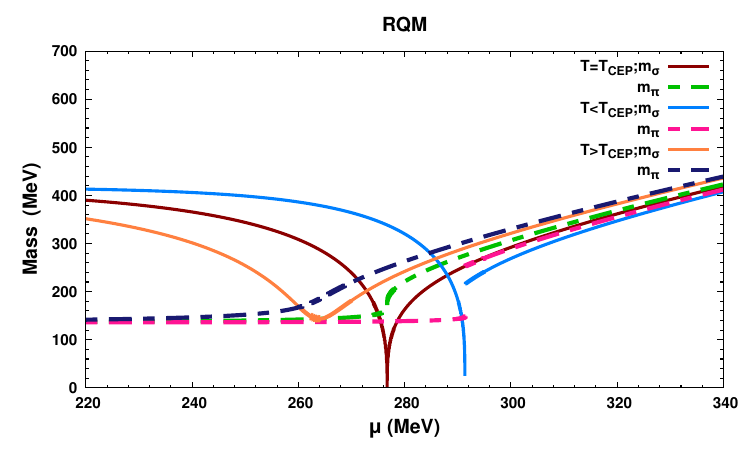}
\end{minipage}}
\hfill
\subfigure[]{
\label{fig1b} 
\begin{minipage}[b]{0.48\textwidth}
\centering \includegraphics[width=\linewidth]{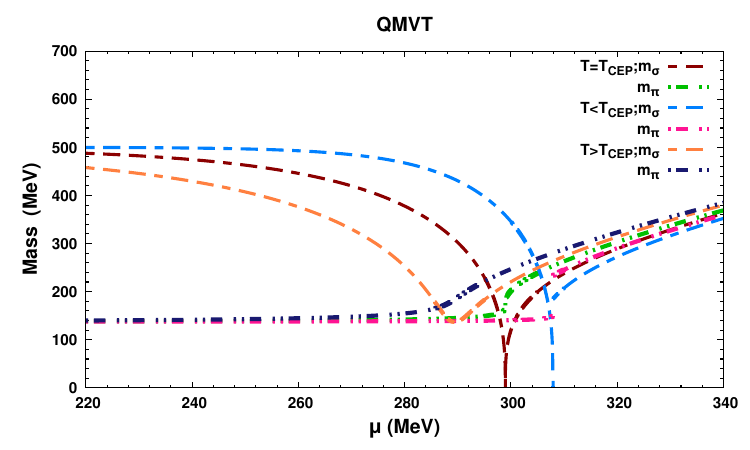}
\end{minipage}}
\caption{The variation of the pion and sigma masses with chemical potential is illustrated at three temperatures: at the critical endpoint temperature ($T_{\cep}$), as well as 15~MeV above and below it. The results for the two-flavor RQM ($T_{\cep} = 35.68$~MeV) and QMVT ($T_{\cep} = 28.36$~MeV) models are shown in the left panel (a) and right panel (b), respectively.}
\label{fig1}
\end{figure*}
\begin{table*}[!htbp]
    \caption{Pseudo-critical temperature, Critical temperature for $m_{\pi}=0$, Critical end points and Tricritical points.}
    \label{tab:table3}
    \begin{tabular}{p{1.8cm}| p{3.3cm}|p{3cm} |p{3cm}|p{2.6cm} |p{2.6cm}}
      \toprule
       Flavor & \hphantom{textt} Models & $T^\chi_c(\mu=0,m_\pi=138)$ \hphantom{textext} MeV & $T_c(\mu=0,m_\pi=0)$ \hphantom{textext} MeV & $(\mu_{\cep},T_{\cep})$ MeV & $(\mu_{\tcp},T_{\tcp})$ MeV \\
      \hline

      2F & RQM
      &\qquad$145.20$&\qquad$150.47$ & $\quad(276.69,35.68)$ & $\quad(243.19,68.25)$ \\
       & Log RPQM &\qquad$168.77$&\qquad$184.86$&\quad$(270.58,92.45)$& $\quad(221.93,136.46)$\\
       & PolyLog-glue RPQM  &\qquad$176.39$&\qquad$181.42$&\quad$(268.29,68.26)$& $\quad(228.19,112.17)$ \\
       & QMVT Ref.\cite{schafwag12} &\qquad$157.00$&\qquad$159.81$&$\quad(298.91,28.36)$&$\quad(263.16,67.00)$ \\
       & PQMVT-Log Ref.\cite{schafwag12}&\qquad$186.69$&\qquad$191.79$&$\quad(296.10,74.51)$&$\quad(242.12,134.71)$ \\
      \hline

      3F Ref.~\cite{Akakn} & RQM &\qquad$133.60$&\qquad$136.80$ & $\quad(265.42,38.71)$ & $\quad(220.15,71.52)$ \\
       & Log RPQM &\qquad$148.90$&\qquad$164.50$&\quad$(252.00,94.60)$& $\quad(194.30,126.53)$\\
       & PolyLog-glue RPQM &\qquad$157.30$&\qquad$160.90$&\quad$(252.70,70.90)$& $\quad(202.20,108.14)$ \\
      \hline
    \end{tabular}
\end{table*}

\section{Results and Discussion}
\label{sec:III}

The effective potential of chiral models becomes flat as the critical end point (CEP) is approached in the chemical potential ($\mu$) and temperature ($T$) plane. Consequently, the $\sigma$ meson mass approaches zero ($m_{\sigma} \rightarrow 0$) at the CEP for both the RQM and QMVT models, as it is defined by the second derivative (curvature) of the effective potential in the $\sigma$ direction. The critical fluctuations around the CEP are governed by the divergence of the correlation length, $\xi \propto 1/m_{\sigma}$. Figs.~\ref{fig1a} and \ref{fig1b} depict the variation of $m_{\sigma}$ and $m_{\pi}$ with respect to $\mu$ in the RQM and QMVT models, respectively, for three fixed temperatures: $T=T_{\cep}+15.0 \text{ MeV}$, $T=T_{\cep}$, and $T=T_{\cep}-15.0 \text{ MeV}$. It is worth pointing out that because the $\sigma$ and $\pi$ mesons are chiral partners, their masses ($m_{\sigma}$ and $m_{\pi}$) become degenerate after the chiral symmetry-restoring phase transition has taken place. The chiral transitions depicted in these figures are smooth crossovers at $T=T_{\cep}+15.0 \text{ MeV}$ and first-order transitions at $T=T_{\cep}-15.0 \text{ MeV}$.

\begin{figure*}[htb]
\subfigure[]{
\label{fig2a} 
\begin{minipage}[b]{0.48\textwidth}
\centering \includegraphics[width=\linewidth]{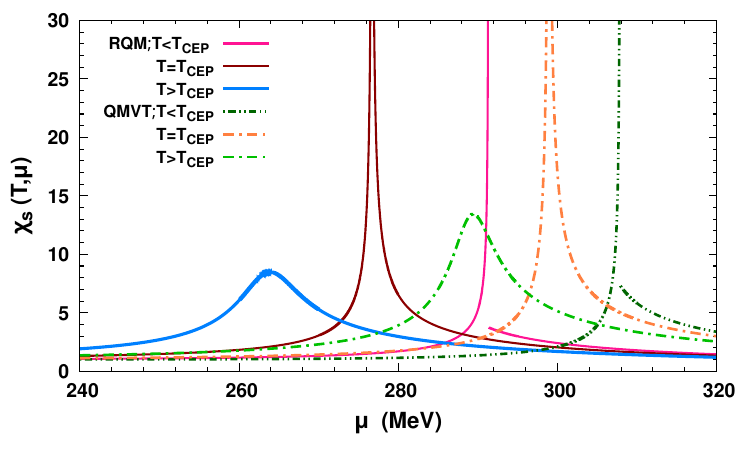}
\end{minipage}}
\hfill
\subfigure[]{
\label{fig2b} 
\begin{minipage}[b]{0.48\textwidth}
\centering \includegraphics[width=\linewidth]{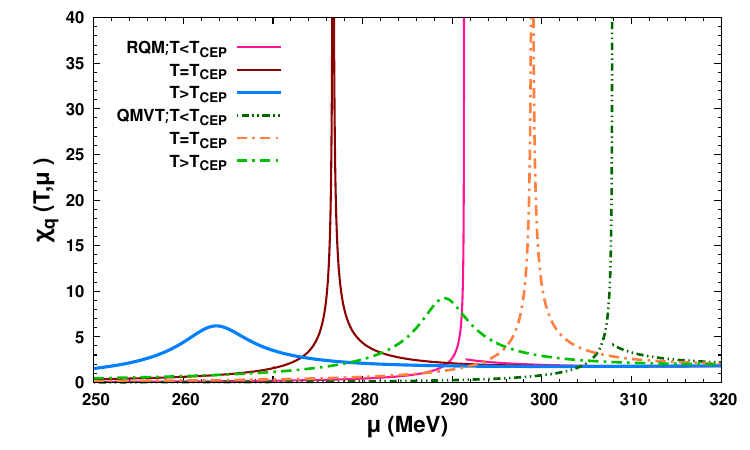}
\end{minipage}}
\caption{The left panel (a) and right panel (b) present the scalar and quark number susceptibilities, respectively, as functions of the chemical potential for the RQM and QMVT models. Both quantities are evaluated near the CEP at three temperatures: exactly at the critical endpoint temperature ($T_{\rm CEP}$), as well as 15~MeV above and below it.}
\label{fig:mini:fig2}
\end{figure*}

It is important to note that when the parameters of the RQM model are computed by matching the counter-terms in the on-shell and $\overline{\text{MS}}$ schemes \cite{Adhiand1,asmuAnd,RaiTiw22,raiti23}, the vacuum ($T=\mu=0$) curvature and pole masses of the $\sigma$ meson differ. Specifically, we find a curvature mass of $m_{\sigma,c}=417 \text{ MeV}$, corresponding to a pole mass of $m_{\sigma}=500 \text{ MeV}$. As shown in Fig.~\ref{fig1a}, $m_{\sigma,c}$ in the RQM model starts from this smaller value of $417 \text{ MeV}$ and becomes zero at the CEP, located at $(\mu_{\cep},T_{\cep})=(276.69, 35.68) \text{ MeV}$. In contrast, for the QMVT model, where the pole and curvature masses are identical, $m_{\sigma}$ starts its variation from $500 \text{ MeV}$ in Fig.~\ref{fig1b} and goes to zero at a CEP positioned at noticeably higher $\mu$ and lower $T$, specifically at $(\mu_{\cep},T_{\cep})=(298.91, 28.36) \text{ MeV}$, as detailed in Table~\ref{tab:table3}. This relative shift in the CEP location arises because quark one-loop vacuum fluctuations moderately soften the strength of the chiral transition in the RQM model, where the explicit symmetry-breaking strength, $h$, is also reduced after renormalization because the $\pi$ meson curvature mass gets reduced. In contrast, the chiral transition in the QMVT model is strongly smoothed. As a result, the RQM model exhibits a more extended first-order line, with its CEP located at a higher $T$ and lower $\mu$, whereas the QMVT model's CEP is pushed to the lower-right corner of the phase diagram (shown later) at a noticeably smaller $T$ and higher $\mu$ with a smaller first-order line \cite{RaiTiw22,raiti23}.

The scalar susceptibility is defined as $\chi_{s}(\mu,T) = \frac{1}{m_{\sigma}^{2}(\mu,T)}$. It diverges at the CEP when varying $\mu$ ($T$) while keeping $T_{\cep}$ ($\mu_{\cep}$) fixed.~The  variations of the scalar susceptibility,$\chi_{s}$, with respect to the chemical potential corresponding to $T=T_{\cep}+15.0 \text{ MeV}$, $T=T_{\cep}$, and $T=T_{\cep}-15.0 \text{ MeV}$ in the RQM and QMVT  models are shown in Fig.~\ref{fig2a}.~Finding the quark number susceptibility as a function of $(\mu,T)$ requires evaluating the second-order derivative of the grand potential with respect to the chemical potential. Numerical differentiation via the finite-difference method gives consistent results only up to first order; the second-order derivatives obtained this way show numerous spurious fluctuations, which grow more pronounced as the number of $\mu$-dependent implicit variables increases. To avoid these fluctuations, we compute the second-order derivatives of the grand potential using a semi-analytic method \citep{Ghosh:2014,Gholami:2025}. Minimization of the grand potential with respect to the order parameters $X_i = \{x, \Phi, \bar\Phi\}$ gives the gap equations
\begin{equation}
\label{gap}
\frac{\partial \Omega(T, \mu, X_{i})}{\partial X_i}=0,
\end{equation}
whose solutions give the condensates $\bar{X}_i(T,\mu)$. The quark number density is then
\begin{align}
n_{q}=-\frac{d \Omega}{d \mu}=-\frac{\partial \Omega }{\partial \mu}-\sum_i \frac{\partial \Omega}{\partial X_i}\frac{d X_i}{d \mu}\Big|_{X_i=\bar{X}_i}=-\frac{\partial \Omega }{\partial \mu},
\end{align}
using Eq.~(\ref{gap}). The quark number susceptibility follows as
\begin{align}
 \chi_{q}=\frac{d n_q}{d\mu}=-\frac{d^2\Omega}{d^2\mu}=-\frac{\partial^2\Omega}{\partial^2 \mu}+\sum_i \frac{\partial^2\Omega}{\partial \mu \partial X_i}\frac{d X_i}{d \mu}\Big|_{X_i=\bar{X}_i}.
\end{align}
In the RQM/QMVT model, the quark number density depends on $\mu$ explicitly and implicitly through the single order parameter $\bar{x}$, so that
\begin{equation}
    \chi_q = \frac{d n_q}{d \mu}=-\frac{d^2\Omega}{d^2\mu} = -\left( \frac{\partial^2 \Omega}{\partial \mu^2} + \frac{\partial^2 \Omega}{\partial \mu \partial x} \frac{\partial \bar{x}}{\partial \mu} \right).
    \label{eq:chi_q_chain}
\end{equation}
The implicit derivative $\partial\bar x/\partial\mu$ follows from the total derivative of the gap equation, Eq.~(\ref{gap}), with respect to $\mu$:
\begin{equation}
    \frac{d}{d \mu} \left( \frac{\partial \Omega}{\partial x} \right) = \frac{\partial^2 \Omega}{\partial \mu \partial x} + \frac{\partial^2 \Omega}{\partial x^2} \frac{\partial \bar{x}}{\partial \mu} = 0.
\end{equation}
Defining the $\sigma$ curvature mass as $m_{\sigma,c}^2 = \partial^2\Omega/\partial x^2$, and using $\partial^2\Omega/\partial\mu\partial x = -\partial n_q/\partial x$, this gives
\begin{equation}
   \frac{\partial \bar{x}}{\partial \mu} = \frac{-\partial^2 \Omega/\partial \mu \partial x}{\partial^2 \Omega/\partial x^2}=\frac{\partial n_q/\partial x}{m_{\sigma,c}^2}.
    \label{eq:implicit_deriv}
\end{equation}
Substituting back, the $\chi_{q}$ takes the form
\begin{equation}
 \chi_q = -\frac{\partial^2 \Omega}{\partial \mu^2} + \frac{\left( \partial n_q/\partial x \right)^2}{m_{\sigma,c}^2}=\chi_q^{\text{reg}} + \frac{\left( \partial n_q/\partial x \right)^2}{m_{\sigma,c}^2},
\end{equation}
where $\chi_q^{\text{reg}} = -\partial^2\Omega/\partial\mu^2$ is regular and non-divergent.

\begin{figure*}[htb]
\subfigure[]{
\label{fig3a} 
\begin{minipage}[b]{0.48\textwidth}
\centering \includegraphics[width=\linewidth]{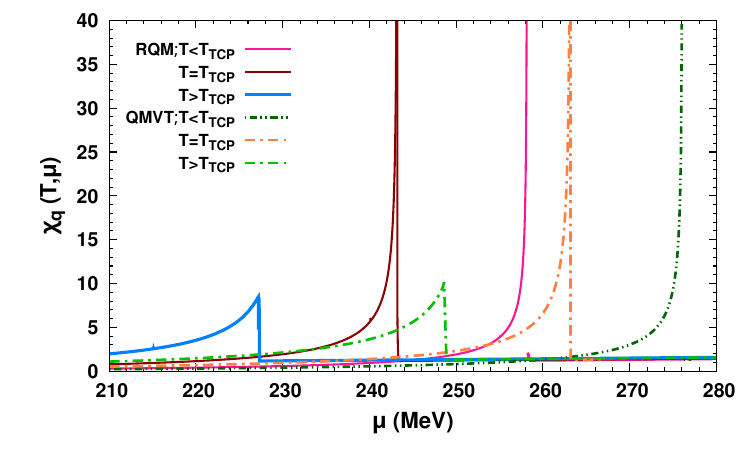}
\end{minipage}}
\hfill
\subfigure[]{
\label{fig3b} 
\begin{minipage}[b]{0.48\textwidth}
\centering \includegraphics[width=\linewidth]{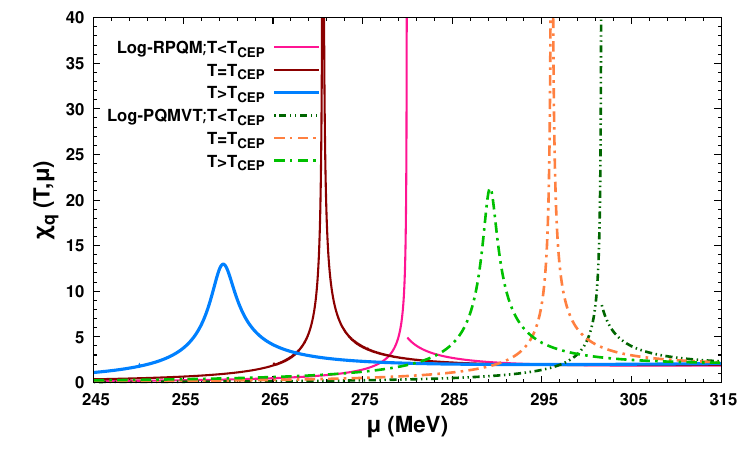}
\end{minipage}}
\caption{The quark number susceptibility as a function of chemical potential is presented in both panels. The left panel (a) shows the results for the RQM and QMVT models near the tricritical point (TCP), while the right panel (b) displays the results for the Log-RPQM and Log-PQMVT models near the critical endpoint (CEP). In both panels, the curves are evaluated at three temperatures: exactly at the respective critical temperature ($T_{\rm TCP}$ or $T_{\rm CEP}$), as well as 15~MeV above and below it.}
\label{fig:mini:fig3} 
\end{figure*}

Physically, $\chi_q$ characterizes the static fluctuations of the net quark number: $\chi_q = \frac{1}{VT}\langle(\delta N_q)^2\rangle$. Its behavior is tied to the vanishing of the $\sigma$ meson curvature mass and changes qualitatively depending on whether the system approaches the $Z(2)$ critical end point (CEP) at physical quark masses or the $O(4)$ tricritical point (TCP) in the chiral limit. Charge conjugation symmetry is explicitly broken at finite chemical potential ($\mu>0$), mixing the chiral condensate ($x=\langle\sigma\rangle$) with the net quark number density $n_q$. Because the quark density depends on the chiral order parameter, the total derivative of $n_q$ with respect to $\mu$ generates a singular contribution proportional to the scalar susceptibility \citep{Hatta,redlsaki}, so that $\chi_q \sim (\partial n_q/\partial x)^2/m_{\sigma,c}^2$. Thus, whenever the effective potential flattens and $m_{\sigma,c}\to 0$, apart from the scalar susceptibility, the quark number susceptibility also diverges.

In the exact chiral limit, the explicit symmetry-breaking strength vanishes ($h=0$) and the second-order chiral transition line terminates at the $O(4)$ tricritical point. At the TCP, both the quadratic curvature $m_{\sigma,c}^2$ and the quartic coupling of the effective potential vanish simultaneously. The singular behavior of $\chi_q$ is thus governed by $Z(2)$ universality near the CEP and by $O(4)$ universality near the TCP.

The  quark number susceptibility, $\chi_{q}(\mu,T)$, variations on the $\mu$ aixs corresponding to $T=T_{\cep}+15.0 \text{ MeV}$, $T=T_{\cep}$, and $T=T_{\cep}-15.0 \text{ MeV}$ in the RQM and QMVT  models are depicted in Fig.~\ref{fig2b}.~The chiral transition is a smooth crossover for $T=T_{\cep}+15.0 \text{ MeV}$ and first order for 
$T=T_{\cep}-15.0 \text{ MeV}$ whereas the $\chi_{q}$ diverges at the Z(2) CEP as discussed above.~The $\chi_{q}(\mu,T)$ variations in QMVT model are seen as shifted to higher chemical potentials because its $\mu_{\cep}$ has large value compared to that of the RQM model.~The chemical potential variations of $\chi_{q}(\mu,T)$ computed for the chiral limit ($m_{\pi}=0$)  parameters of the RQM and QMVT model are shown in Fig.~\ref{fig3a}  at the tricritical point $T=T_{\tcp}$  and at  $T=T_{\tcp}+15.0 \text{ MeV}$ as well as $T=T_{\tcp}-15.0 \text{ MeV}$. The TCP lies at $(\mu_{\tcp},T_{\tcp})=(243.19, 68.25) \text{ MeV}$ in the RQM model whereas its location in the QMVT model is at noticeably higher $\mu$ at $(\mu_{\tcp},T_{\tcp})=(263.16, 67.00) \text{ MeV}$.~Since the chiral transition at $T=T_{\tcp}+15.0 \text{ MeV}$ in both the RQM and QMVT models is second order, we notice that the susceptibility gets to  its peaks and then drops vertically, a behavior that corresponds to the $O(4)$ critical point.~The susceptibilities diverge at the TCP and show the first order discontinuity for  $T=T_{\tcp}-15.0 \text{ MeV}$ in Fig.~\ref{fig3a}.

When the logarithmic form of the Polyakov-loop potential representing the physics of confinement-deconfinement phase transition is integrated with the RQM model, the CEP shifts to significantly higher temperature at $(\mu_{\cep},T_{\cep})=(270.58, 92.45) \text{ MeV}$ in the Log-RPQM model.~Since the curvature mass based parameter fixing generates strong softening effect on the strength of chiral transition,~the CEP in the Log PQMVT model gets located at the $(\mu_{\cep},T_{\cep})=(296.10, 74.51) \text{ MeV}$ which lies at noticeably smaller temperature (and higher $\mu$) than that of the Log RPQM model CEP as shown in Table~\ref{tab:table3}.~The variations of susceptibilities, $\chi_{q}(\mu,T)$, with respect to $\mu$ for three fixed temperatures  $T=T_{\cep}+15.0 \text{ MeV}$, $T=T_{\cep}$, and $T=T_{\cep}-15.0 \text{ MeV}$ in the Log RPQM and Log PQMVT model have been depicted in Fig.~\ref{fig3b}.~Due to the effect of Polyakov-loop potential, the susceptibility peaks for $T=T_{\cep}+15.0 \text{ MeV}$ of the crossover transitions in the Log RPQM and Log PQMVT models in Fig.~\ref{fig3b} are larger and sharper than the corresponding RQM and QMVT model  $\chi_{q}(\mu,T)$ peaks in Fig.~\ref{fig2b}. The model susceptibilities diverge at $T=T_{\cep}$  and show first order discontinuity at $T=T_{\cep}-15.0 \text{ MeV}$ in  Fig.~\ref{fig3b}.

\begin{figure*}[htb]
\subfigure[]{
\label{fig4a} 
\begin{minipage}[b]{0.48\textwidth}
\centering \includegraphics[width=\linewidth]{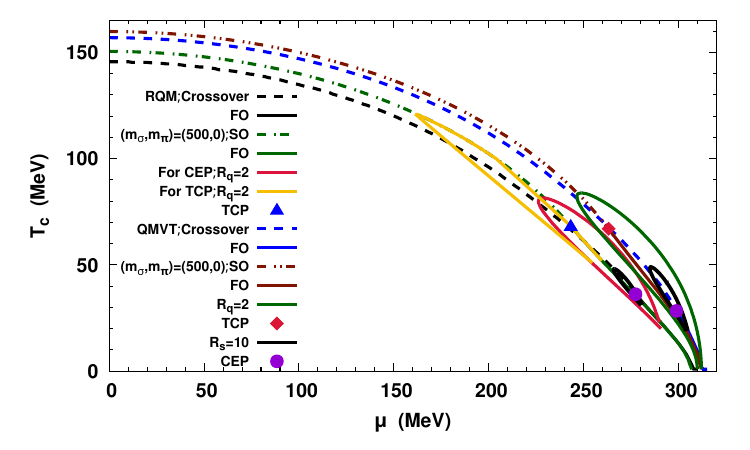}
\end{minipage}}
\hfill
\subfigure[]{
\label{fig4b} 
\begin{minipage}[b]{0.48\textwidth}
\centering \includegraphics[width=\linewidth]{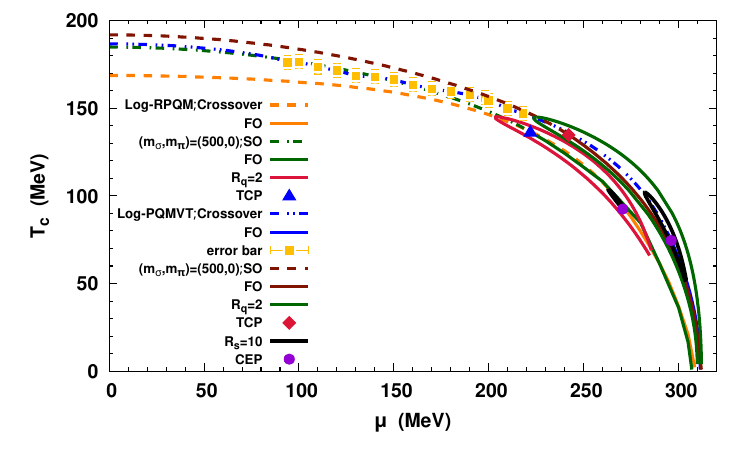}
\end{minipage}}
\caption{The phase diagrams computed for $m_{\sigma}=500$~MeV, using the physical-point and chiral limit parameter sets, are shown in panel (a) for the RQM and QMVT models, and in panel (b) for the Log-RPQM and Log-PQMVT models. The line styles and the positions of the critical endpoints (CEPs) and tricritical points (TCPs) are indicated accordingly. In each phase diagram, the CEP is enclosed by a contour line of constant quark-number susceptibility ratio ($R_q=2$) and a contour line of constant scalar susceptibility ratio ($R_s=10$). The $R_q=2$ contour which is computed for the chiral limit parameters around the TCP of RQM model is also plotted in the left panel (a).}
\label{fig:mini:fig4}
\end{figure*}
\subsection{ Two-Flavor RQM and Log-RPQM versus QMVT and Log-PQMVT Model Phase Diagrams : CEP and TCP}
\label{sec:IIIA}
The CEP and TCP are located by identifying the $(\mu,T)$ coordinates where the quark number susceptibility $\chi_q$ diverges. Furthermore, the critical regions enveloping the CEP \citep{Asak,Bard,Berg,Hatta,Son,Misha3,Jeon,Schaefer:2006ds} are mapped by drawing contours corresponding to the locus of $(\mu,T)$ points where the scalar susceptibility $\chi_s(\mu,T)$ (relative to its vacuum value) and the quark number susceptibility $\chi_q(\mu,T)$ (relative to its free-gas value) exhibit strong enhancement. The normalized quark number susceptibility, $R_q$, is defined as the ratio of the numerically computed susceptibility $\chi_q(\mu,T)$ to its free quark gas value:
\begin{equation}
R_{q} = \frac{\chi_q(\mu,T)}{\chi_{q}^{\text{free}}(\mu,T)}.
\end{equation}
In the chiral limit ($m_q \to 0$), the free quark gas susceptibility is given by
\begin{equation}
\chi^{\text{free}}_{q}(T,\mu) = \frac{\nu_q}{6} \left( T^2 + \frac{3\mu^2}{\pi^2} \right),
\end{equation}
where $\nu_q = 2N_cN_f$ with $N_c = 3$ colors and $N_f = 2$ flavors. Similarly, the normalized scalar susceptibility, $R_s$, is defined as:
\begin{equation}
R_{s} = \frac{\chi_s(\mu,T)}{\chi_{s}(0,0)} = \frac{m_{\sigma}^{2}(0,0)}{m_{\sigma}^{2}(\mu,T)}.
\end{equation}

\begin{figure*}[htb]
\subfigure[]{
\label{fig5a} 
\begin{minipage}[b]{0.48\textwidth}
\centering \includegraphics[width=\linewidth]{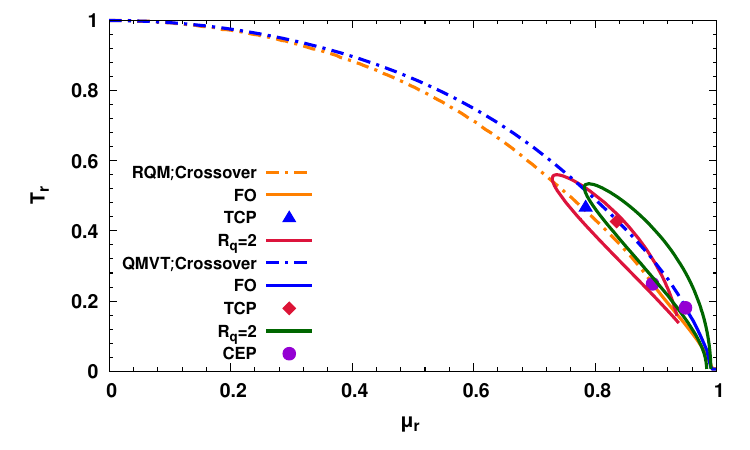}
\end{minipage}}
\hfill
\subfigure[]{
\label{fig5b} 
\begin{minipage}[b]{0.48\textwidth}
\centering \includegraphics[width=\linewidth]{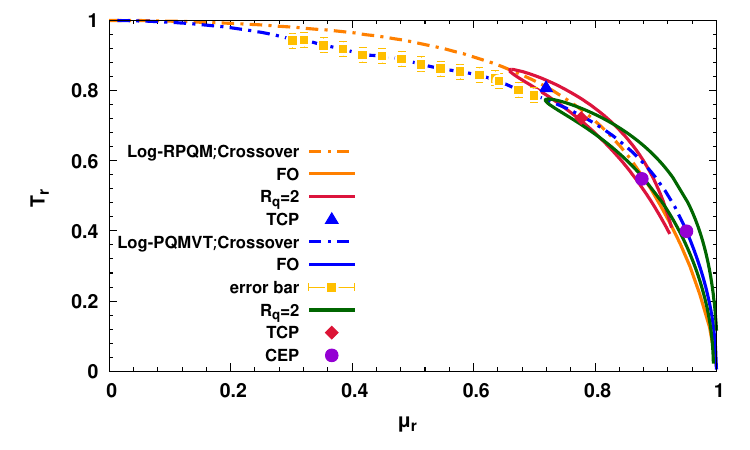}
\end{minipage}}
\caption{Phase diagrams computed using the physical-point parameters ($m_{\pi} = 138$~MeV and $m_{\sigma}=500$~MeV) are shown in the reduced-coordinate $T_r$--$\mu_r$ plane for (a) the RQM and QMVT models, and (b) the Log-RPQM and Log-PQMVT models. The reduced temperature, $T_r = T/T_c^\chi$, is defined by scaling the temperature by the pseudo-critical chiral crossover temperature ($T_c^\chi$) at $\mu = 0$. Similarly, the reduced chemical potential, $\mu_r = \mu/\mu_0$, is defined as the ratio of the chemical potential to the first-order transition chemical potential ($\mu_0$) at $T = 0$.}
\label{fig:mini:fig5}
\end{figure*}

Figs.~\ref{fig4a} and \ref{fig4b} compare the phase diagrams—including the positions of the critical end point (CEP) and tricritical point (TCP), along with their enveloping critical region contours for $R_{q}=2$ and $R_{s}=10$—for the two-flavor RQM and QMVT models and the Log RPQM and Log PQMVT models, respectively. The calculations are performed for both the physical pion mass ($m_\pi = 138$~MeV) and the chiral limit ($m_\pi = 0$) at a fixed scalar meson mass of $m_\sigma = 500$~MeV. One finds large, broad, and well-developed critical regions around the CEPs of the RQM and QMVT models, located at $(\mu_{\cep}, T_{\cep}) = (276.69, 35.68)\text{ MeV}$ and $(298.91, 28.36)\text{ MeV}$, respectively, but the widths of the critical regions in the QMVT model are somewhat larger. The total spreads along the $T$ and $\mu$ axes—82.9 and 65.6 MeV, respectively, in the QMVT model for the constant $R_{q}=2$ quark number susceptibility contour in Fig.~\ref{fig4a}—are larger than the corresponding spreads of 61.6 and 63.7 MeV in the RQM model.~The spread and width of the constant $R_{s}=10$ scalar susceptibility contour in the RQM model are noticeably smaller than those of the QMVT model. This feature is a consequence of the fact that the $\mu$ variation of the RQM model $\sigma$ curvature mass begins from its vacuum value of $m_{\sigma,c}(\mu=0,T=0)=417$ MeV, whereas the corresponding variation in the QMVT model starts from $m_{\sigma}(\mu=0,T=0)=500$ MeV, where the curvature mass and pole mass of the $\sigma$ meson are equal. Fig.\ref{fig:mini:fig4} illustrates two-dimensional $\mu-T$ planes of the phase diagram for $m_{\pi}=0$ and $m_{\pi}=138$ MeV, whereas the true physical description requires a three-dimensional $\mu-T-m_{\pi}$ parameter space. The $R_{q}=2$ susceptibility contour in the $\mu-T$ plane for $m_{\pi}=138$ MeV casts a projection onto the $\mu-T$ plane at $m_{\pi}=0$.~Because the TCPs of the RQM and QMVT models, located at $(\mu_{\tcp}, T_{\tcp}) = (243.19, 68.25)\text{ MeV}$ and $(263.16, 67.0)\text{ MeV}$, respectively, sit comfortably within these projected shadow contours on the $\mu-T$ plane at $m_{\pi}=0$, they would influence the critical fluctuations surrounding the CEPs in both models \cite{Hatta,xiong}. The $R_q =2$ contour computed for the chiral limit around the TCP of RQM model has also been plotted in 
Fig.~\ref{fig4a}.~In the RQM model, the $R_{q}=2$ critical region contour around the TCP extends 121 MeV above and 51 MeV below the TCP on the temperature axis, and spans from 161.17 MeV on the left to 259.51 MeV on the right along the $\mu$ axis. As a result, the $R_{q}=2$ critical regions of the TCP and CEP show a noticeable overlap in Fig.~\ref{fig4a}. This feature is discussed in detail at the end of  Section~\ref{subsec:IIIB}.

The logarithmic form of the Polyakov-loop potential in the Log-PQMVT model causes a very large stretching of 140.9 MeV along the $T$ axis in the $R_{q}=2$ quark number susceptibility contour around its CEP at $(\mu_{\cep}, T_{\cep}) = (296.10, 74.51)\text{ MeV}$ in Fig.~\ref{fig4b}, whereas the elongation in the corresponding Log-RPQM model contour around its CEP, located noticeably higher up in the phase diagram at $(\mu_{\cep}, T_{\cep}) = (270.58, 92.45)\text{ MeV}$, is 79.0 MeV. In contrast, the corresponding $\mu$-axis spreads are quite comparable, being 88.2 MeV for the Log-PQMVT model and 82.1 MeV for the Log-RPQM model. Because the phase boundary of the Log-PQMVT model exhibits significant curvature at its CEP, which is located at a very high chemical potential, 70.5 MeV of the $R_{q}=2$ contour's total 140.9 MeV span along the $T$ axis lies below $T_{\cep}$. Consequently, the contour develops a neck-like structure below the CEP, as detailed in subsection \ref{sec:IIIA}-1. The Polyakov-loop potential exerts a strong thermal effect, shifting the CEP positions significantly higher along the $T$ axis, whereas the corresponding shifts along the $\mu$ axis toward lower values are significantly smaller relative to the RQM and QMVT models. In comparison to the QMVT model, the critical fluctuations in the Log-PQMVT model exhibit a large relative enhancement of 58.0 MeV along the $T$ axis, while the $\mu$-axis relative increase is a moderate 22.6 MeV. In contrast, the $T$-axis enhancement in the Log-RPQM model relative to the RQM model is only 17.4 MeV, whereas the $\mu$-axis relative enhancement is 18.4 MeV, as shown in Table~\ref{tab:table4}. The widths of the contours are noticeably compressed due to the logarithmic form of the Polyakov-loop potential with no quark back-reaction, yielding a distinctively pinched contour shape similar to the findings in Refs.~\cite{schafwag12,vkkr12,Akakn}. The tricritical points (TCPs)—located at $(\mu_{\tcp}, T_{\tcp}) = (221.93, 136.46)\,\text{MeV}$ and $(242.12, 134.71)\,\text{MeV}$ in the Log-RPQM and Log-PQMVT models, respectively—lie well inside the $R_{q}=2$ contour in Fig.~\ref{fig4b}. These TCPs will influence the critical fluctuations around their respective CEPs. Furthermore, the $R_{s}=10$ scalar susceptibility contour for the Log-RPQM model in Fig.~\ref{fig4b} is significantly narrower than the corresponding contour in the RQM model. In contrast, the thinning in the Log-PQMVT model is noticeably less pronounced than in the Log-RPQM model: its $R_{s}=10$ contour is only somewhat thinner than that of the QMVT model, though its overall spread turns out to be larger.

We are analyzing how, in the presence of quark one-loop vacuum fluctuations in the two-flavor QM model, on-shell versus curvature-mass-based parameter fixing and the Polyakov-loop potential (without quark back-reaction) affect the extent of critical fluctuations around the CEP, while also quantifying its proximity to the TCP through phase diagram calculations in the light chiral limit ($m_\pi = 0$). Differently treated quark one-loop vacuum fluctuations and Polyakov-loop potentials not only affect the pseudo-critical temperature $T_{c}^{\chi}$ of the $\mu=0$ chiral crossover transition (see Table~\ref{tab:table3}) but also shift the overall phase diagrams up (for curvature mass parametrized QMVT case) or down (for on-shell parametrization in the RQM scenario) in the $\mu$-$T$ plane, necessitating the definition of reduced quantities for a rigorous comparison. Furthermore, vacuum corrections in the RQM model induce a significant shift of the critical end point (CEP) toward higher $\mu$ and lower $T$ compared to the s-MFA QM model, whereas the curvature-mass-parametrized QMVT model drives the CEP toward an even larger $\mu$ and considerably lower $T$. In contrast, the Polyakov-loop potentials cause a differential lifting of the CEP toward higher temperatures and lower chemical  otentials~\cite{RaiTiw22,raiti23}. The reduced temperature is defined as $T_{r} = T / T_{c}^{\chi}$, where the temperature $T$ is scaled by the pseudo-critical temperature $T_{c}^{\chi}$ at $\mu = 0$, whereas the reduced chemical potential is given by $\mu_{r} = \mu / \mu_{0}$, with $\mu_{0}$ denoting the value at which a first-order phase transition occurs on the $\mu$-axis at $T = 0$~\cite{Akakn}.~The phase diagrams in the reduced chemical potential and temperature ($\mu_r-T_r$) plane are plotted in Fig.~\ref{fig:mini:fig5}.

\begin{table*}[!htbp]
    \caption{The $T$ and $\mu$ axis spreads of the  normalized quark number susceptibility contours for the $R_{q}=2,\bf {3 \text{ and } 5}$ in the different 2-flavor and 2+1-flavor model scenarios are presented in the upper part of the Table.~The size of the contours above and below $T_{\cep}$ are denoted respectively by the $\Delta T_{a}$ and $\Delta T_{b}$ while the total contour size on $T$ axis is $\Delta T$.~The size of the contours on the lower side and higher side of the $\mu_{\cep}$ are denoted respectively by the $\Delta \mu_{l}$ and $\Delta \mu_{h}$ while the total contour size on $\mu$ axis is $\Delta \mu$. The lower part of the table summarizes the $T$ and $\mu$ axis extents of the normalized scalar susceptibility contours for the $R_{s}=10,\bf {15 \text{ and } 25}$ in the different two-flavor models.}
    \label{tab:table4}
    \begin{tabular}{p{2.965cm} |p{2.4cm}|p{2.36cm}|p{2.51cm}|p{2.46cm}|p{2.1cm}|p{2.45cm}}
      \toprule
       \hphantom{texttext} Models &  \multicolumn{6}{|c}{Entries below the $\Delta T_{a}$, $\Delta T_{b}$, $\Delta T$, $\Delta\mu_{l}$, $\Delta\mu_{h}$ and $\Delta\mu$ correspond to ${\bf {R_q}}= 2\textbf{(3)(5)}$} \ \\
        & $\Delta T_{a}$ & $\Delta T_{b}$ & $ \Delta T=\Delta T_{a}+\Delta T_{b}$ &  $\Delta\mu_{l}$ &  $\Delta \mu_{h}$& $\Delta \mu=\Delta \mu_{l}+\Delta \mu_{h}$ \\
 \hline

       RQM {(2F)} &$45.9\textbf{(27.4)(15.2)}$  &$15.7 \textbf{(9.7)(5.7)}$ &$61.6\textbf{(37.1)(20.9)}$ & $50.6\textbf{(26.8)(14.3)}$ & $13.1\textbf{(7.8)(3.9)}$& $63.7\textbf{(34.6)(18.2)}$  \\
       Log-RPQM &$52.6\textbf{(37.7)(23.2)} $ &$26.4\textbf{(17.4)(10.4)}$&$79.0\textbf{(55.1)(33.6)}$ &$67.0\textbf{(37.7)(19.1)}$& $15.1\textbf{(9.1)(5.1)}$& $82.1\textbf{(46.8)(24.2)}$\\
      
       QMVT  &  $55.5\textbf{(34.2)(19.8)}$ & $27.4\textbf{(16.4)(8.4)}$ & $82.9\textbf{(50.6)(28.2)}$ & $52.5\textbf{(27.3)(13.8)}$ &$13.1\textbf{(9.6)(5.2)}$ & $65.6\textbf{(36.9)(19.0)}$ \\
       PQMVT  & $70.4\textbf{(52.8)(34.0)}$ &$70.5\textbf{(31.5)(15.5)} $& $140.9\textbf{(84.3)(49.5)} $ & $72.3\textbf{(40.4)(19.9)}$ & $15.9\textbf{(10.1)(5.8)} $ & $88.2\textbf{(50.5)(25.7)}$\\
       
       PolyLog-glue-RPQM  & $58.4\textbf{(41.9)(26.1)}$&  $ 28.3\textbf{(11.3)(9.3)}$ & $86.7\textbf{(53.2)(35.4)} $& $ 62.5\textbf{(38.4)(21.1)}$ &$ 18.3\textbf{(7.8)(4.7)}$ &$ 80.8\textbf{(46.2)(25.8)}$  \\
 \hline

       RQM {(2+1)F}~\cite{Akakn} &$29.3\textbf{(17.6)(9.8)}$&$9.1\textbf{(5.7)(3.3)}$ &$38.4\textbf{(23.3)(13.1)}$ &$35.7\textbf{(19.9)(10.5)}$& $ 9.3\textbf{(5.6)(3.2)}$&$45.0\textbf{(25.5)(13.7)}$ \\
       Log-RPQM &$32.3\textbf{(23.2)(14.2)}$& $14.6\textbf{(6.6)(4.6)}$&$46.9\textbf{(29.8)(18.8)}$& $56.7\textbf{(35.5)(19.1)}$ &$14.7\textbf{(7.1)(4.7)}$ &$ 71.4\textbf{(42.6)(23.8)}$\\
       PolyLog-glue-RPQM &$38.5\textbf{(26.3)(15.8)}$&$15.0\textbf{(9.6)(5.9)}$ &$53.5\textbf{(35.9)(21.7)}$   &$50.8\textbf{(31.0)(17.3)}$ &$ 14.2\textbf{(9.2)(5.5)}$&    $65.0\textbf{(40.2)(22.8) } $ \\
  \hline
  
  \toprule
       \hphantom{texttext} Models &  \multicolumn{6}{|c}{Entries below the $\Delta T_{a}$, $\Delta T_{b}$, $\Delta T$, $\Delta\mu_{l}$, $\Delta\mu_{h}$ and $\Delta\mu$ correspond to ${\bf {R_s}}= 10\textbf{(15)(25)}$} \ \\
        & $\Delta T_{a}$ & $\Delta T_{b}$ & $ \Delta T=\Delta T_{a}+\Delta T_{b}$ &  $\Delta\mu_{l}$ &  $\Delta \mu_{h}$& $\Delta \mu=\Delta \mu_{l}+\Delta \mu_{h}$ \\
 \hline
 
       RQM {(2F)} &$12.5\textbf{(8.1)(4.7)}$  &$4.6 \textbf{(2.7)(1.4)}$ &$17.1\textbf{(10.8)(6.1)}$ & $10.6\textbf{(6.9)(4.0)}$ & $4.2\textbf{(1.6)(1.2)}$& $14.8\textbf{(8.5)(5.2)}$  \\
       Log-RPQM &$11.2\textbf{(4.9)(...)} $ &$1.4\textbf{(0.4)(...)}$&$12.6\textbf{(5.3)(...)}$ &$7.7\textbf{(3.2)(...)}$& $0.8\textbf{(0.4)(...)}$& $8.5\textbf{(3.6)(...)}$\\
      
       QMVT  &  $20.7\textbf{(13.3)(7.8)}$ & $9.4\textbf{(6.4)(2.4)}$ & $30.1\textbf{(19.7)(10.2)}$ & $14.1\textbf{(8.6)(4.9)}$ &$6.0\textbf{(4.0)(1.7)}$ & $20.1\textbf{(12.6)(6.6)}$ \\
       PQMVT  & $27.5\textbf{(17.1)(8.3)}$ &$23.5\textbf{(13.5)(5.5)} $& $51.0\textbf{(30.6)(13.8)} $ & $13.9\textbf{(7.7)(3.4)}$ & $7.9\textbf{(4.8)(2.2)} $ & $21.8\textbf{(12.5)(5.6)}$\\
  \hline  
    \end{tabular}
\end{table*}

In terms of reduced coordinates, the CEP lies at $(\mu_{r \cep}, T_{r \cep}) = (0.892, 0.245)$ for the RQM model in Fig.~\ref{fig5a}, with the TCP located at $(\mu_{r\tcp}, T_{r\tcp}) = (0.784, 0.468)$. In contrast, the reduced coordinates in the QMVT model are $(\mu_{r\cep}, T_{r\cep}) = (0.948, 0.180)$ for the CEP and $(\mu_{r\tcp}, T_{r\tcp}) = (0.835, 0.426)$ for the TCP. Relative to the QMVT model, both the CEP and TCP in the RQM model are positioned toward the upper left in the phase diagram. Although the extents of the $R_{q}=2$ quark number susceptibility contour along the $T_r$ and $\mu_r$ axes are smaller in the RQM model than those in the QMVT model, the critical region contour in the RQM model reaches higher (on $T$ axis) to close at $(\mu_{r},T_{r})=(0.729,0.560)$. In contrast, the $R_{q}=2$ critical region contour in the QMVT model closes at a lower-right position of $(\mu_{r},T_{r})=(0.782,0.534)$. This feature is a consequence of the fact that the RQM model CEP lies at a noticeably higher position compared to that of the QMVT model. The curvatures of the crossover and first-order transition lines for the Log-RPQM and Log-PQMVT models in Fig.~\ref{fig5b} change significantly when compared with those of the RQM and QMVT models in Fig.~\ref{fig5a} due to the effect of the Polyakov loop potential (where the quark back-reaction is absent). Driven by the logarithmic Polyakov-loop potential, the critical end points (CEPs) of the RQM and QMVT models undergo relative uplifts of $(\Delta \mu_{r \cep}, \Delta T_{r \cep}) = (-0.015, 0.303)$ and $(-0.009, 0.219)$, shifting to $(\mu_{r \cep}, T_{r \cep}) = (0.877, 0.548)$ and $(0.950, 0.399)$ in the Log-RPQM and Log-PQMVT models, respectively, as illustrated in Fig. \ref{fig5b}. We point out that the CEP displacement is noticeably larger for the Log-RPQM model than that for the Log-PQMVT model. Analogously, experiencing relative uplifts of $(\Delta \mu_{r \tcp}, \Delta T_{r \tcp}) = (-0.065, 0.341)$ and $(-0.058, 0.295)$, the tricritical points (TCPs) of the RQM and QMVT models are relocated to $(\mu_{r \tcp}, T_{r \tcp}) = (0.719, 0.809)$ and $(0.777, 0.721)$ in the Log-RPQM and Log-PQMVT models, respectively. Because the critical endpoint (CEP) of the Log-RPQM model is located at a significantly higher position than that of the Log-PQMVT model, the $R_q = 2$ critical region contour in the Log-RPQM model reaches markedly higher, closing at $(\mu_r, T_r) = (0.659, 0.860)$. In contrast, the corresponding contour in the Log-PQMVT model closes at a lower value of $T_r$, $(\mu_r, T_r) = (0.719, 0.776)$, even though the overall extents of its $R_q = 2$ quark number susceptibility contour along the $T_r$ and $\mu_r$ axes are larger.

\begin{figure*}[htb]
\subfigure[]{
\label{fig6a} 
\begin{minipage}[b]{0.48\textwidth}
\centering \includegraphics[width=\linewidth]{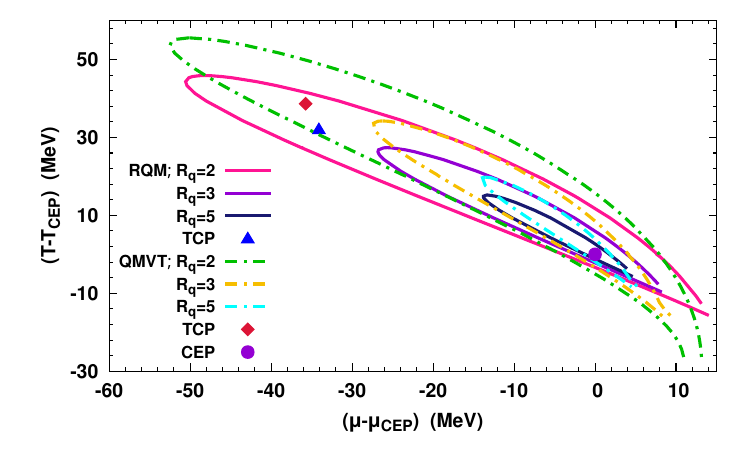}
\end{minipage}}
\hfill
\subfigure[]{
\label{fig6b}
\begin{minipage}[b]{0.48\textwidth}
\centering \includegraphics[width=\linewidth]{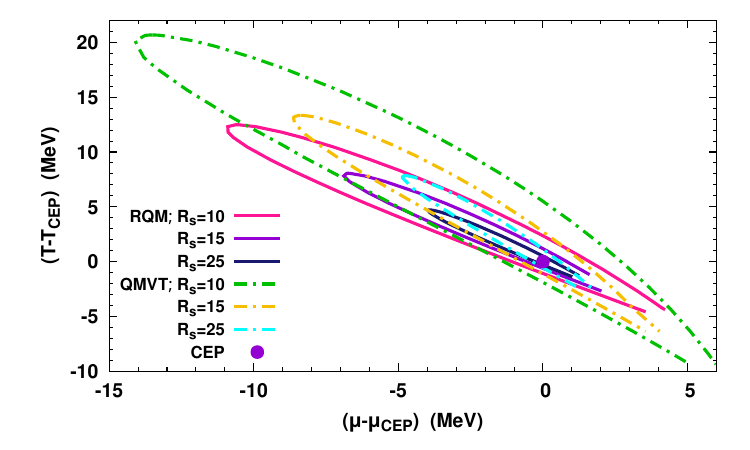}
\end{minipage}}
\caption{The left panel (a) presents normalized quark number susceptibility contours for $R_q=2$, $3$, and $5$ in the RQM and QMVT models. The right panel (b) displays the corresponding scalar susceptibility contours for $R_s=10$, $15$, and $25$ for both models.}
\label{fig:mini:fig6} 
\end{figure*}

\subsection*{1. Morphology of the critical regions of the CEP}
Previous studies utilizing the standard mean-field approximation (s-MFA)—where quark one-loop fluctuations in the QM model are neglected—found that the susceptibility contours of the critical region are relatively small \cite{Schaefer:2006ds,vkkr12,schafwag12,Akakn}. In these scenarios, the contours extend asymptotically parallel to the first-order transition line but exhibit narrow widths perpendicular to it. Furthermore, these works demonstrated that incorporating a logarithmic Polyakov-loop potential (creating the PQM model) shifts the CEP to a much higher temperature and lower chemical potential ($\mu$). As a result, the critical region around this PQM model CEP becomes highly compressed and significantly narrower (pinched) than in the standard QM model. Building on this foundation, we investigate in this subsection how the morphologies of the critical fluctuation regions around the CEP are altered when quark one-loop vacuum fluctuations are explicitly included. We systematically compare the effects of two different procedures for determining the renormalized effective potentials and model parameters when quark one-loop vacuum corrections are accounted for in the QM model: the on-shell scheme (yielding the RQM model) and the curvature-mass-based scheme (yielding the QMVT model). Finally, we examine how the critical region contours are further modified when the logarithmic Polyakov-loop potential with no quark back-reaction is integrated into both of these frameworks, resulting in the Log-RPQM and Log-PQMVT models, respectively.

\begin{figure*}[htb]
\subfigure[]{
\label{fig7a} 
\begin{minipage}[b]{0.48\textwidth}
\centering \includegraphics[width=\linewidth]{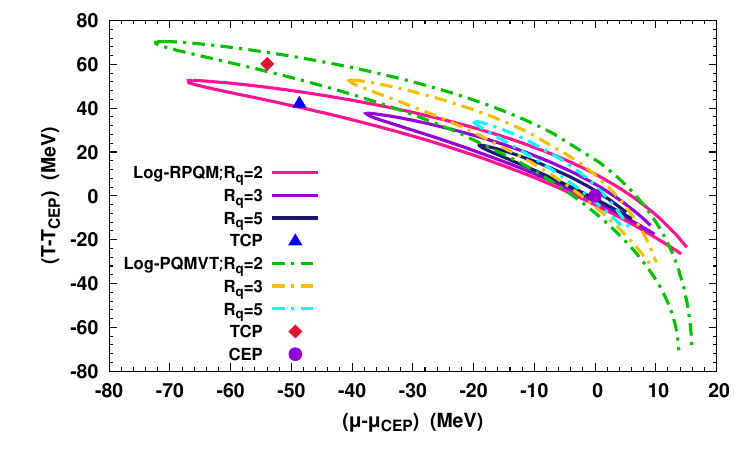}
\end{minipage}}
\hfill
\subfigure[]{
\label{fig7b} 
\begin{minipage}[b]{0.48\textwidth}
\centering \includegraphics[width=\linewidth]{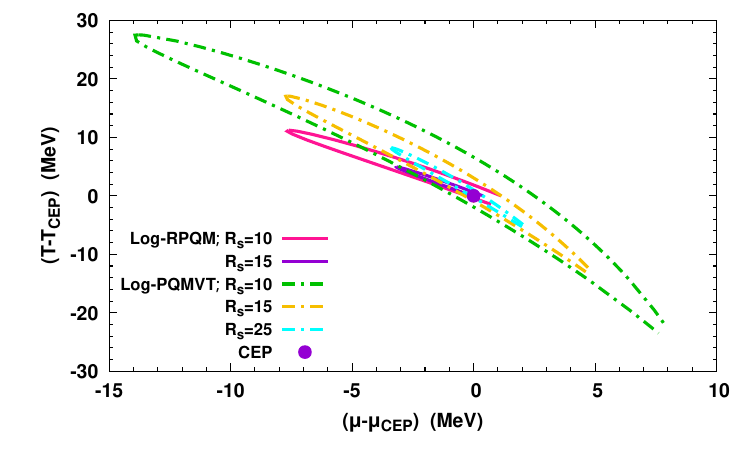}
\end{minipage}}
\caption{The left panel (a) presents contour plots of the normalized quark number susceptibility for $R_q=2$, $3$, and $5$ around the CEP of the Log-RPQM and Log-PQMVT models. The right panel (b) plots the corresponding scalar susceptibility contours for $R_s=10$, $15$, and $25$ for both models.}
\label{fig:mini:fig7}
\end{figure*}

To analyze the size and shape of the critical regions enveloping the CEP in detail, susceptibility contours for three constant ratios ($R_q = 2, 3,$ and $5$) are plotted in Fig.~\ref{fig:mini:fig6} for the RQM and QMVT models, and in Fig.~\ref{fig:mini:fig7} for the Log-RPQM and Log-PQMVT models. In Fig.~\ref{fig6a}, the $R_q = 2, 3,$ and $5$ contours for the RQM model exhibit total spreads of 61.6, 37.1, and 20.9 MeV along the $T$-axis, with respective $\mu$-axis total extents of 63.7, 34.6, and 18.2 MeV. In contrast, the corresponding contours for the QMVT model show total spreads of 82.9, 50.6, and 28.2 MeV along the $T$-axis, with $\mu$-axis total extents of 65.6, 36.9, and 19.0 MeV, respectively. The detailed distributions of these spreads relative to $T_{\cep}$ and $\mu_{\cep}$ are summarized in Table~\ref{tab:table4}. Note that the $T$-axis spreads of the contours are noticeably larger in the QMVT model than in the RQM model, whereas the $\mu$-axis spreads remain comparable in both models. The $R_{q}=2, 3,$ and $5$ contours of the QMVT model are systematically broader and larger, yet they display a distinct geometry below the CEP. Because the phase boundary bends sharply at the very high chemical potential where the CEP is located, a significant portion of the $R_{q}=2$ contour (27.4 MeV out of its total 82.9 MeV span along the $T$ axis) is forced below $T_{\cep}$, creating a neck-like structure.~As shown in Fig.~\ref{fig6b}, the scalar susceptibility contours for the ratios $R_s = 10, 15,$ and $25$ in the RQM model exhibit total extensions along the $T$-axis of 17.1, 10.8, and 6.1 MeV, respectively, whereas their respective $\mu$-axis extensions are 14.8, 8.5, and 5.2 MeV. In contrast, the QMVT model scalar susceptibility contours are significantly larger and broader than those of the RQM model. Specifically, the $R_s = 10, 15,$ and $25$ contours in the QMVT model have total extensions of 30.1, 19.7, and 10.2 MeV along the $T$-axis, while the corresponding $\mu$-axis spreads are 20.1, 12.6, and 6.6 MeV, respectively. This large difference occurs because the variation of the scalar $\sigma$ meson mass—whether with chemical potential (at fixed $T$) or temperature (at fixed $\mu$)—originates from a vacuum curvature mass (pole mass) of $m_{\sigma}(\mu=0,T=0)=500$ MeV in the QMVT model, compared to a starting mass of $m_{\sigma,c}(\mu=0,T=0)=417$ MeV in the RQM model, where the curvature mass of the $\sigma$ meson is quite different from its pole mass. The detailed distributions of the sizes of scalar susceptibility contours for $R_s = 10, 15,$ and $25$ relative to $T_{\cep}$ and $\mu_{\cep}$ are also summarized in Table~\ref{tab:table4}.

Compared to the RQM and QMVT models, incorporating the logarithmic Polyakov-loop potential in the Log-RPQM and Log-PQMVT models significantly alters the shape of the critical region contours around their respective CEPs. As shown in Fig.~\ref{fig7a}, the $R_q = 2, 3,$ and $5$ contours of the Log-PQMVT model experience the largest total stretching along the $T$-axis, reaching 140.9, 84.3, and 49.5 MeV, respectively, while their corresponding $\mu$-axis extents are 88.2, 50.5, and 25.7 MeV. In contrast, the $R_q = 2, 3,$ and $5$ contours of the Log-RPQM model exhibit smaller total spreads of 79.0, 55.1, and 33.6 MeV along the $T$-axis, with respective $\mu$-axis extents of 82.1, 46.8, and 24.2 MeV. Compared to the on-shell parameterized Log-RPQM model, the CEP for the curvature-mass parameterized Log-PQMVT model is situated significantly further to the lower-right in the $\mu$-$T$ phase diagram. As a result, the Polyakov-loop potential generates the most expansive contour extensions in the Log-PQMVT model. This large extension is driven by the pronounced curvature of the first-order and crossover transition lines converging at a considerably high $\mu_{\cep}$. Furthermore, this steep phase boundary is directly responsible for the characteristic neck-like structure seen in the $R_{q}=2, 3,$ and $5$ contours beneath the CEP, as it drags exactly half of the $R_{q}=2$ contour's total temperature span (70.5 MeV out of 140.9 MeV) completely below $T_{\cep}$. In the RQM and QMVT models, quark one-loop vacuum fluctuations create a smoothing effect that generates a broader and larger critical region perpendicular to the crossover transition line. However, this smoothing effect is significantly compromised because the logarithmic Polyakov-loop potential modifies the fermionic determinant term. This modification induces differential geometric effects on the critical fluctuation contours across different model scenarios, depending on the local curvature of the phase boundary at the CEP. Consequently, the contours near the CEPs of the Log-RPQM and Log-PQMVT models become narrower due to compression in the temperature direction, consistent with findings in Refs.~\cite{schafwag12,Akakn}.Furthermore, the narrower and relatively smaller susceptibility contours in the Log-RPQM model do not exhibit a neck-like structure below the CEP. This is because the phase boundary near the Log-RPQM CEP (which is located at a higher $T$ and lower $\mu$) possesses a gentler curvature, causing its $R_{q}=2, 3,$ and $5$ contours to lie vertically below those of the Log-PQMVT model in Fig.~\ref{fig7a}. Recall that the quark one-loop vacuum correction generates only a moderate softening effect on the strength of the chiral transition in the Log-RPQM model, whereas this softening is substantially larger in the Log-PQMVT model.

The widths of the scalar susceptibility contours for $R_s = 10, 15,$ and $25$ are significantly reduced in the Log-PQMVT model compared to the QMVT model. This reduction is due to the logarithmic Polyakov-loop potential, which causes the drop in the $\sigma$ meson mass (starting from $m_{\sigma}(\mu=0,T=0)=500$ MeV) to occur over narrower $\mu$ and $T$ intervals. However, the total extensions of the $R_s = 10, 15,$ and $25$ contours along the $T$-axis in Fig.~\ref{fig7b} for the Log-PQMVT model are 51.0, 30.6, and 13.8 MeV, respectively, which are noticeably larger than those in the QMVT model. In contrast, the respective $\mu$-axis spreads in the Log-PQMVT model are 21.8, 12.5, and 5.6 MeV, remaining comparable to those of the QMVT model. This behavior is explained by the fact that the Polyakov-loop potential exerts a particularly strong effect in the temperature direction. Furthermore, because the smoothing effect of the quark one-loop vacuum correction on the strength of the chiral transition is moderate in the Log-RPQM model compared to that of the Log-PQMVT model, the Polyakov-loop potential drastically modifies the $R_s = 10$ and $15$ contours, while the $R_s = 25$ contour ceases to exist. This disappearance occurs because the $\sigma$ curvature mass (beginning from a smaller initial value of $m_{\sigma,c}(\mu=0,T=0)=417$ MeV) drops very sharply to its minimum over an extremely small $\mu$ and $T$ interval. Consequently, the $R_s = 10$ and $15$ contours for the Log-RPQM model in Fig.~\ref{fig7b} have quite small extensions of 12.6 and 5.3 MeV along the $T$-axis, whereas the corresponding $\mu$-axis spreads are even smaller at 8.5 and 3.6 MeV. Therefore, the scalar susceptibility contours in the Log-RPQM model appear very small, thin, and pinched.

\subsection{Phase diagrams, CEPs, and TCPs of two-flavor versus (2+1)-flavor RQM and RPQM models.}
\label{subsec:IIIB}

\begin{figure*}[htb]
\subfigure[]{
\label{fig8a} 
\begin{minipage}[b]{0.48\textwidth}
\centering \includegraphics[width=\linewidth]{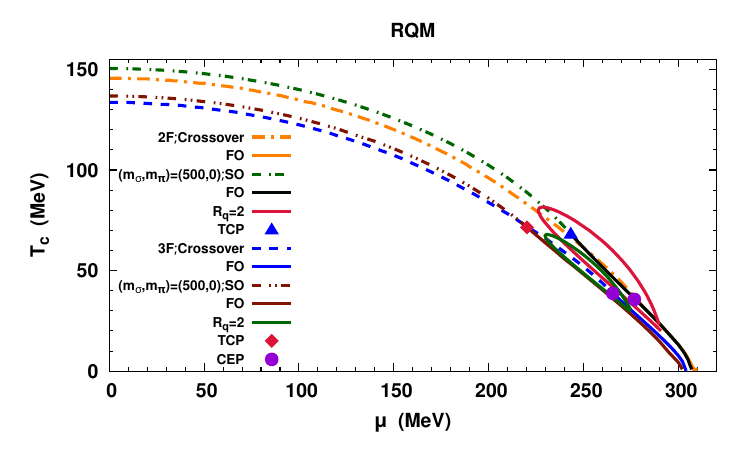}
\end{minipage}}
\hfill
\subfigure[]{
\label{fig8b}
\begin{minipage}[b]{0.48\textwidth}
\centering \includegraphics[width=\linewidth]{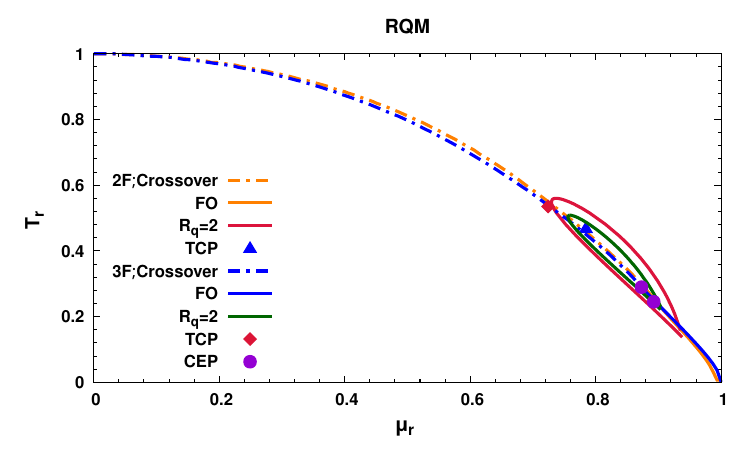}
\end{minipage}}

\caption{The phase diagrams computed for $m_{\sigma}=500$~MeV, using the physical-point and light-chiral-limit parameter sets, are compared in panel (a) for the two-flavor and 2+1-flavor RQM models. The right panel (b) presents the phase diagrams for the physical-point parameters in the reduced $\mu_r$--$T_r$ plane, comparing the two-flavor and 2+1-flavor RQM models. The line styles and the positions of the critical endpoints (CEPs) and tricritical points (TCPs) are indicated accordingly. In each phase diagram, the CEP is enclosed by a contour line of a constant quark-number susceptibility ratio ($R_q=2$).}

\label{fig:mini:fig8} 
\end{figure*}

Figure~\ref{fig8a} compares the two-flavor and (2+1)-flavor RQM model phase diagrams in the $\mu-T$ plane for the physical point and light chiral limit ($m_\pi=0$) parameters. The diagrams for the physical point, which include the critical end points (CEPs), tricritical points (TCPs), and their surrounding critical region contours for $R_{q}=2$, are also plotted in the reduced chemical potential and temperature ($\mu_r-T_r$) plane for comparison in Fig.~\ref{fig8b}. Located at $(\mu_{\cep}, T_{\cep})=(265.42, 38.71)\text{ MeV}$ and $(\mu_{\tcp}, T_{\tcp})=(220.15, 71.52)\text{ MeV}$, the CEP and TCP of the (2+1)-flavor RQM model are shifted to higher temperatures and lower chemical potentials in the $\mu-T$ plane relative to the positions of their two-flavor counterparts, which lie at $(\mu_{\cep}, T_{\cep}) = (276.69, 35.68)\text{ MeV}$ and $(\mu_{\tcp}, T_{\tcp}) = (243.19, 68.25)\text{ MeV}$, respectively. As shown in Fig.~\ref{fig8a}, the total sizes of the constant $R_{q}=2$ quark number susceptibility contour for the two-flavor RQM model along the $T$ and $\mu$ axes are 61.6 and 63.7 MeV, respectively. These are significantly larger than the corresponding spreads of 38.4 and 45.0 MeV on the $T$ and $\mu$ axes, respectively, reported for the (2+1)-flavor RQM model in Ref.~\cite{Akakn}. Furthermore, the fluctuation regions have noticeably larger widths in the two-flavor case, as its $R_{q}=2$ contour is significantly broader than that of the (2+1)-flavor RQM model. The TCP lies well inside the $R_{q}=2$ contour of the two-flavor case, whereas it is located completely outside the boundary of the corresponding contour in the (2+1)-flavor RQM model. Thus, the critical fluctuations around the CEP will be affected by the presence of the nearby TCP in the two-flavor RQM model; however, such an effect is absent in the (2+1)-flavor case. Note that the overall (2+1)-flavor phase diagrams at the physical point and the light chiral limit ($m_{\pi}=0$) lie below those of the two-flavor RQM model. Specifically, at $\mu=0$, the pseudo-critical (critical for the chiral limit) temperature $T_{c}^{\chi}(T_{c})=145.2 (150.47)$ MeV for the crossover (second-order) chiral transition in the two-flavor model is noticeably larger than the corresponding values of $T_{c}^{\chi}(T_{c})=133.6 (136.8)$ MeV computed in the (2+1)-flavor scenario. Therefore, it is necessary to compare the two-flavor and (2+1)-flavor RQM model phase diagrams in the reduced-variable ($\mu_r-T_r$) plane.

In Fig.~\ref{fig8b}, the reduced coordinates of the CEP and TCP for the two-flavor RQM model are located at $(\mu_{r\cep}, T_{r\cep}) = (0.892, 0.245)$ and $(\mu_{r\tcp}, T_{r\tcp}) = (0.784, 0.468)$, respectively. In contrast, the CEP and TCP of the (2+1)-flavor RQM model are positioned at higher $T_r$ and lower $\mu_r$, lying at $(\mu_{r\cep}, T_{r\cep}) = (0.87, 0.289)$ and $(\mu_{r\tcp}, T_{r\tcp}) = (0.724, 0.535)$, respectively. The extents of the $R_{q}=2$ quark number susceptibility contour along the $T_r$ and $\mu_r$ axes in the two-flavor RQM model are significantly larger than the corresponding $T_r$ and $\mu_r$ spreads for the (2+1)-flavor scenario. The two-flavor RQM model contour extends higher and further to the left, closing at $(\mu_{r},T_{r})=(0.729,0.560)$ in the phase diagram, whereas the corresponding (2+1)-flavor contour closes to the right of this point at $(\mu_{r},T_{r})=(0.7556,0.5089)$. Having a significantly broader width, the two-flavor RQM model contour contains the (2+1)-flavor $R_{q}=2$ contour well within its boundary.

\begin{figure*}[htb]
\subfigure[]{
\label{fig9a} 
\begin{minipage}[b]{0.48\textwidth}
\centering \includegraphics[width=\linewidth]{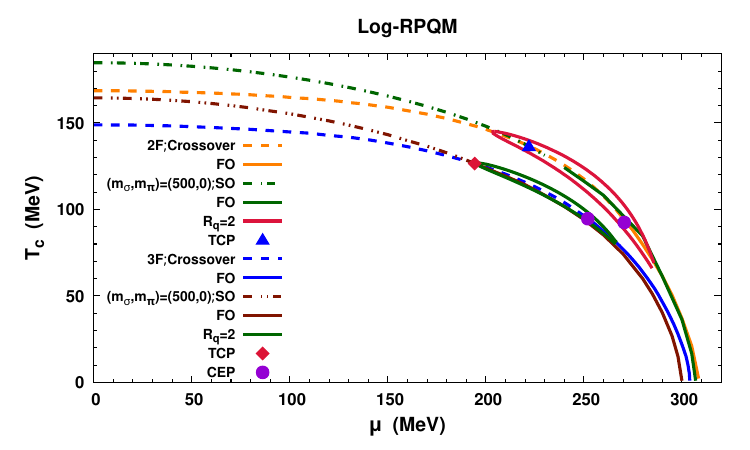}
\end{minipage}}
\hfill
\subfigure[]{
\label{fig9b}
\begin{minipage}[b]{0.48\textwidth}
\centering \includegraphics[width=\linewidth]{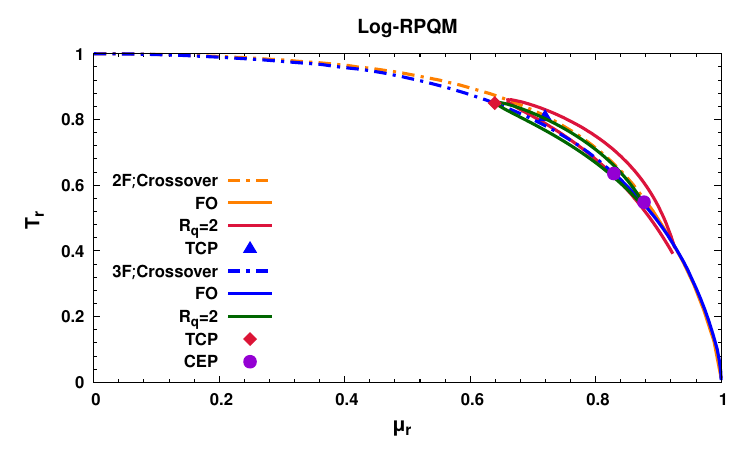}
\end{minipage}}

\caption{The phase diagrams computed for $m_{\sigma}=500$~MeV, using the physical-point and light-chiral-limit parameter sets, are compared in panel (a) for the two-flavor and 2+1-flavor Log-RPQM models. The right panel (b) presents the phase diagrams for the physical-point parameters in the reduced $\mu_r$--$T_r$ plane, comparing the two-flavor and 2+1-flavor Log-RPQM models. The line styles and the positions of the critical endpoints (CEPs) and tricritical points (TCPs) are indicated accordingly. In each phase diagram, the CEP is enclosed by a contour line of a constant quark-number susceptibility ratio ($R_q=2$).}

\label{fig:mini:fig9} 
\end{figure*}

Figure~\ref{fig9a} compares the two-flavor and (2+1)-flavor Log-RPQM model phase diagrams in the $\mu-T$ plane for the physical point and light chiral limit ($m_\pi=0$) parameters, including their critical end points (CEPs), tricritical points (TCPs), and the $R_{q}=2$ critical region contours enveloping the CEPs. For comparison, the phase diagrams for the physical point are also plotted in the reduced chemical potential and temperature ($\mu_r-T_r$) plane in Fig.~\ref{fig9b}. In the two-flavor Log-RPQM model depicted in Fig.~\ref{fig9a}, the constant $R_{q}=2$ quark number susceptibility contour spans 79.0 and 82.1 MeV along the $T$ and $\mu$ axes, respectively. These are significantly larger than the corresponding spreads of 46.9 and 71.4 MeV on the $T$ and $\mu$ axes, respectively, reported for the (2+1)-flavor RQM model in Ref.~\cite{Akakn}. Since the logarithmic Polyakov-loop potential with no quark back-reaction causes the CEP and TCP to shift to significantly higher locations in both the two-flavor and (2+1)-flavor Log-RPQM model phase diagrams compared to the base RQM model, the $R_{q}=2$ contours expand accordingly in overall size, while their widths become compressed relative to the corresponding contours in the RQM model. It is worth pointing out that the $R_{q}=2$ contour for the two-flavor Log-RPQM model is not only larger in overall extent but also broader than the compressed $R_{q}=2$ contour of the (2+1)-flavor Log-RPQM model. Thus, both the size and width of the critical fluctuation regions around the CEP decrease due to the effect of the third flavor, regardless of whether the base RQM model or the Polyakov-loop enhanced RPQM model is considered. Furthermore, the TCP will influence the critical fluctuations around the CEP of the two-flavor Log-RPQM model, as the TCP lies well inside the $R_{q}=2$ contour. An analogous effect is quite weak in the (2+1)-flavor Log-RPQM model because the TCP lies at the boundary of its $R_q=2$ contour.

In the $\mu-T$ plane, the CEP and TCP of the (2+1)-flavor Log-RPQM model are located at $(\mu_{\cep}, T_{\cep}) = (252.00, 94.60)\text{ MeV}$ and $(\mu_{\tcp}, T_{\tcp}) = (194.30, 126.53)\text{ MeV}$, respectively. Compared to their two-flavor counterparts—which lie at $(\mu_{\cep}, T_{\cep})=(270.58, 92.45)\text{ MeV}$ and $(\mu_{\tcp}, T_{\tcp})=(221.93, 136.46)\text{ MeV}$—both critical points in the (2+1)-flavor scenario shift to lower chemical potentials, while their temperature shifts are different: the CEP moves slightly higher, whereas the TCP drops moderately lower. The context for this result becomes clear by noting that the pseudo-critical (critical) temperature for the $\mu=0$ chiral crossover (second-order) transition is $T_{c}^{\chi}(T_{c})=168.77 (184.86)\text{ MeV}$ in the two-flavor case, whereas it is $148.90 (164.50)\text{ MeV}$ in the (2+1)-flavor Log-RPQM model, whose phase diagrams lie significantly below those of the two-flavor case. Therefore, one needs to contrast the two-flavor and (2+1)-flavor Log-RPQM model phase diagrams in Fig.~\ref{fig9b} in terms of reduced variables ($\mu_{r}-T_{r}$). In this reduced parameter space, the CEP and TCP coordinates of the (2+1)-flavor Log-RPQM model are located at higher $T_r$ and lower $\mu_r$, specifically at $(\mu_{r\cep}, T_{r\cep}) = (0.828, 0.635)$ and $(\mu_{r\tcp}, T_{r\tcp}) = (0.639, 0.849)$, respectively. In contrast, the CEP and TCP for the two-flavor Log-RPQM model are found at $(\mu_{r\cep}, T_{r\cep}) = (0.877, 0.548)$ and $(\mu_{r\tcp}, T_{r\tcp}) = (0.719, 0.809)$, respectively. For the two-flavor Log-RPQM model, the extents of the $R_{q}=2$ quark number susceptibility contour along the $T_r$ and $\mu_r$ axes—which lie to the left of and above the CEP—are somewhat larger than the corresponding $T_r$ and $\mu_r$ spreads for the (2+1)-flavor Log-RPQM model. The two-flavor Log-RPQM model contour extends slightly higher and marginally further to the right, closing at $(\mu_{r},T_{r})=(0.659,0.860)$ in the phase diagram, while the corresponding (2+1)-flavor contour closes slightly lower and to the left of this point at $(\mu_{r},T_{r})=(0.6443,0.852)$. Although the two-flavor contour has a large overlap, it does not completely contain the (2+1)-flavor Log-RPQM model $R_{q}=2$ contour within its boundary, as is the case with the base RQM model.

\begin{figure*}[htb]
\subfigure[]{
\label{fig10a} 
\begin{minipage}[b]{0.48\textwidth}
\centering \includegraphics[width=\linewidth]{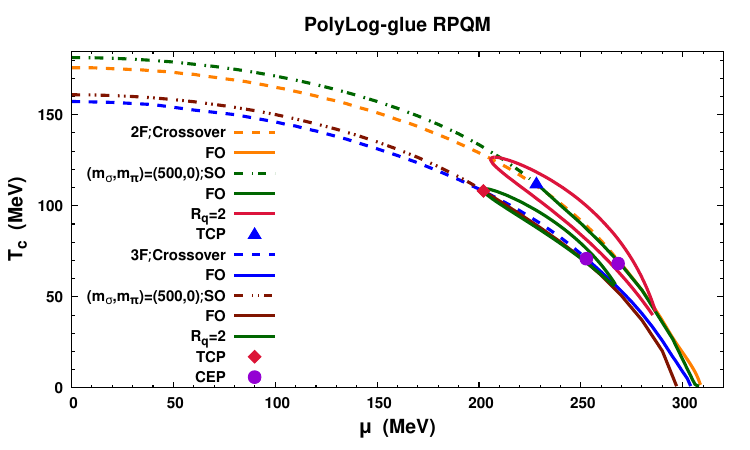}
\end{minipage}}
\hfill
\subfigure[]{
\label{fig10b}
\begin{minipage}[b]{0.48\textwidth}
\centering \includegraphics[width=\linewidth]{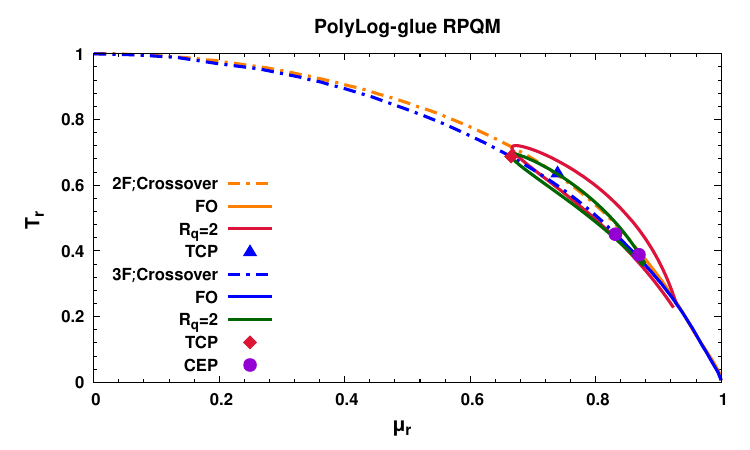}
\end{minipage}}
\caption{The phase diagrams computed for $m_{\sigma}=500$~MeV, using the physical-point and light-chiral-limit parameter sets, are compared in panel (a) for the two-flavor and 2+1-flavor PolyLog-glue-RPQM models. The right panel (b) presents the phase diagrams for the physical-point parameters in the reduced $\mu_r$--$T_r$ plane, comparing the two-flavor and 2+1-flavor PolyLog-glue-RPQM models. The line styles and the positions of the critical endpoints (CEPs) and tricritical points (TCPs) are indicated accordingly. In each phase diagram, the CEP is enclosed by a contour line of a constant quark-number susceptibility ratio ($R_q=2$).}
\label{fig:mini:fig10} 
\end{figure*}

The phase diagrams, CEPs, TCPs, and $R_{q}=2$ susceptibility contours for the PolyLog-glue-RPQM model are presented in the $\mu-T$ plane of Fig.~\ref{fig10a} and the reduced-variable $\mu_{r}-T_{r}$ plane of Fig.~\ref{fig10b}. In the physical $\mu-T$ plane, the $R_{q}=2$ contour for the two-flavor case extends 86.7 and 80.8 MeV along the $T$ and $\mu$ axes, respectively. These extents significantly exceed the corresponding spreads of 53.5 and 65 MeV reported for the (2+1)-flavor RQM model in Ref.~\cite{Akakn}. Furthermore, due to the inclusion of quark back-reaction in the PolyLog-glue-RPQM model, the spread along the $T$ axis moderately increases while the $\mu$-axis spread somewhat decreases in both the two-flavor and (2+1)-flavor scenarios, relative to the Log-RPQM model results in Fig.~\ref{fig9a} where such back-reaction is absent. In both the two-flavor and (2+1)-flavor scenarios, the CEP and TCP of the PolyLog-glue-RPQM model shift to higher locations in the phase diagram relative to the base RQM model. However, the inclusion of quark back-reaction moderates this effect, making these upward shifts smaller than those observed in the Log-RPQM model, where such back-reaction is absent. This moderating influence also extends to the $R_{q}=2$ susceptibility contours: although their widths are somewhat narrower than those in the base RQM model, they remain noticeably broader than the significantly compressed contours of the Log-RPQM model. Finally, consistent with the behavior of the previous models, the introduction of the third flavor in the PolyLog-glue-RPQM framework universally decreases both the overall size and width of the critical fluctuation regions around the CEP. Just as observed in the previous models, the TCP remains situated well inside the $R_{q}=2$ contour of the two-flavor PolyLog-glue-RPQM model, meaning it continues to influence the critical fluctuations near the CEP. In contrast, this interplay is significantly weakened in the (2+1)-flavor PolyLog-glue-RPQM model, where the TCP sits directly on the contour's boundary.

The structural shifts in the PolyLog-glue-RPQM phase diagrams are fundamentally anchored by the $\mu=0$ chiral crossover (second-order) transitions. In the two-flavor case, this pseudo-critical (critical) temperature is $T_{c}^{\chi}(T_{c})=176.39 (181.42)\text{ MeV}$, but it drops significantly to $157.30 (160.90)\text{ MeV}$ upon introducing the third flavor. Because the entire (2+1)-flavor phase diagram lies at noticeably lower temperatures, its CEP and TCP in the physical $\mu-T$ plane correspondingly shift to lower chemical potentials compared to the two-flavor case. Specifically, the (2+1)-flavor CEP and TCP reside at $(\mu_{\cep}, T_{\cep}) = (252.70, 70.90)\text{ MeV}$ and $(\mu_{\tcp}, T_{\tcp}) = (202.20, 108.14)\text{ MeV}$, whereas their two-flavor counterparts are located at $(268.29, 68.26)\text{ MeV}$ and $(228.19, 112.17)\text{ MeV}$. Notably, the temperature shifts diverge: the addition of the third flavor slightly raises the CEP temperature but lowers the TCP temperature. To properly compare these regions given their disparate absolute scales, Fig.~\ref{fig10b} maps these phase diagrams into the dimensionless $\mu_r-T_r$ plane. Evaluated in reduced variables, the (2+1)-flavor critical points migrate to higher $T_r$ and lower $\mu_r$ domains relative to the two-flavor model. The (2+1)-flavor CEP and TCP land at $(\mu_{r\cep}, T_{r\cep}) = (0.831, 0.450)$ and $(\mu_{r\tcp}, T_{r\tcp}) = (0.665, 0.687)$, respectively, contrasting with the two-flavor coordinates of $(0.869, 0.388)$ and $(0.739, 0.638)$. Analyzing the $R_q=2$ critical regions—which project to the left of and above their respective CEPs—reveals that the two-flavor contour spans somewhat larger reduced coordinate distances than the (2+1)-flavor version. Furthermore, the two-flavor contour terminates at a slightly higher and more leftward coordinate of $(\mu_{r},T_{r})=(0.666,0.720)$, whereas the (2+1)-flavor boundary closes lower and to the right at $(0.6797,0.6956)$. While the two susceptibility regions overlap substantially, the two-flavor boundary fails to fully encompass the (2+1)-flavor contour, unlike the behavior observed in the base RQM model, where the two-flavor boundary completely contains the (2+1)-flavor contour.

\begin{figure*}[htb]
\subfigure[]{
\label{fig11a} 
\begin{minipage}[b]{0.48\textwidth}
\centering \includegraphics[width=\linewidth]{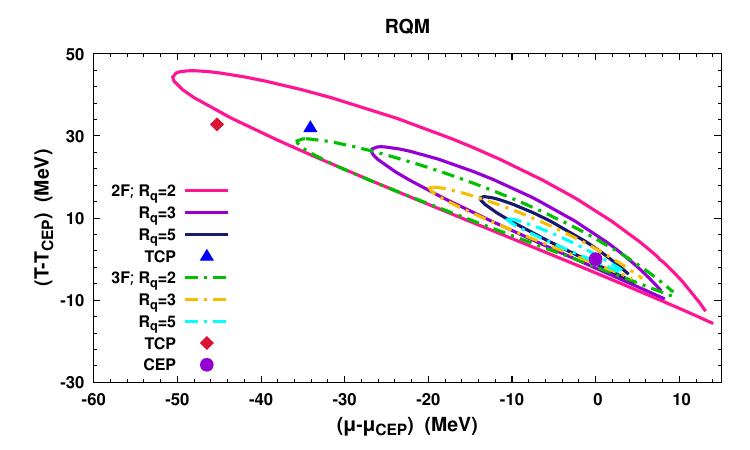}
\end{minipage}}
\hfill
\subfigure[]{
\label{fig11b}
\begin{minipage}[b]{0.48\textwidth}
\centering \includegraphics[width=\linewidth]{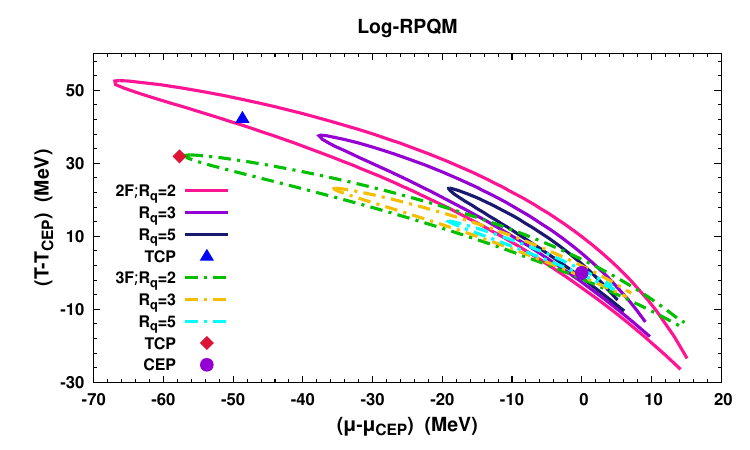}
\end{minipage}}

\caption{The left panel(a) displays the contour plots of the normalized quark number susceptibility ratios $(R_q = 2, 3, 5)$ for the two-flavor and $2+1$-flavor RQM models, while the right panel(b) shows the corresponding contours for the Log-RPQM models.}

\label{fig:mini:fig11} 
\end{figure*} 

\subsection*{1. Two-flavor versus 2+1-flavor models and the morphology of critical regions.}

\begin{figure*}[htb]
\subfigure[]{
\label{fig12a} 
\begin{minipage}[b]{0.48\textwidth}
\centering \includegraphics[width=\linewidth]{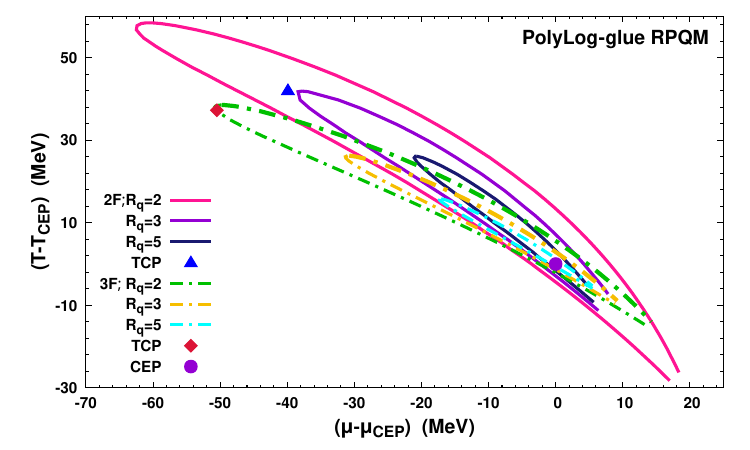}
\end{minipage}}
\hfill
\subfigure[]{
\label{fig12b}
\begin{minipage}[b]{0.48\textwidth}
\centering \includegraphics[width=\linewidth]{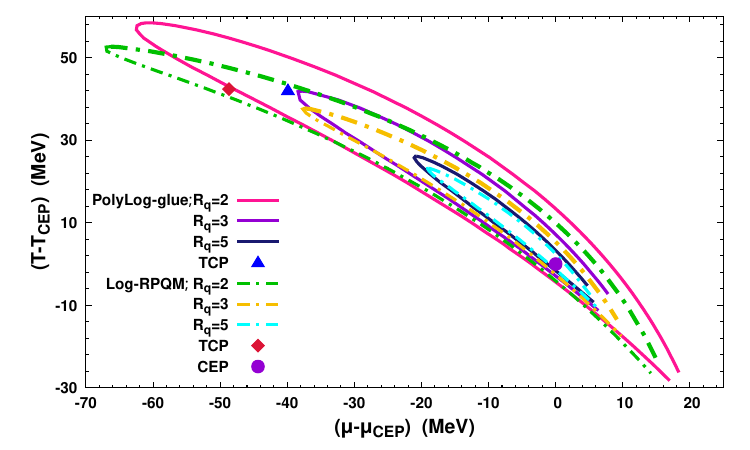}
\end{minipage}}

\caption{Contour plots of the normalized quark number susceptibility ratios $(R_q = 2, 3, 5)$ are mapped in the left panel (a) for the two and \(2+1\)-flavor PolyLog-glue RPQM models while the right panel (b) highlights the differences in the two-flavor contours between the Log-RPQM and PolyLog-glue RPQM models.} 
 
\label{fig:mini:fig12} 
\end{figure*}

To conduct a detailed geometric analysis of the critical regions enveloping the CEP, susceptibility contours for three constant fluctuation ratios ($R_q = 2, 3,$ and $5$) in both the two-flavor and (2+1)-flavor scenarios are mapped across three figures: Fig.~\ref{fig11a} for the base RQM models, Fig.~\ref{fig11b} for the Log-RPQM models, and Fig.~\ref{fig12a} for the PolyLog-glue-RPQM models. Focusing first on the basic RQM framework in Fig.~\ref{fig11a}, these critical fluctuation contours for the two-flavor model span $61.6$, $37.1$, and $20.9\text{ MeV}$ along the $T$ axis, while their respective $\mu$-axis extents are $63.7$, $34.6$, and $18.2\text{ MeV}$. In generalizing from the two-flavor to the (2+1)-flavor RQM model, these boundaries undergo a significant reduction: the corresponding contours shrink to $T$-axis spreads of $38.4$, $23.3$, and $13.1\text{ MeV}$, alongside $\mu$-axis extents of $45.0$, $25.5$, and $13.7\text{ MeV}$. A comprehensive breakdown detailing how these widths are distributed relative to the critical coordinates (i.e., above and below $T_{\cep}$, and left and right of $\mu_{\cep}$) is provided in Table~\ref{tab:table4}. Comparing the two frameworks reveals that the two-flavor contours have substantially more extension overall. While the (2+1)-flavor contours undergo a large reduction in both size and broadness, their $T$ and $\mu$ spreads remain nearly equivalent. These contours shrink proportionally in their sizes (on the $T$ and $\mu$ axes) as well as widths, and are not pinched or compressed—a geometrical deformation that, as will be seen later, is characteristic of the Log-RPQM model. Ultimately, this confirms that considering the (2+1)-flavor chiral symmetry breaking (spontaneous and explicit) scenario with the addition of the third flavor fundamentally dampens the critical fluctuations, resulting in a significantly reduced size and width for the critical regions enveloping the CEP. However, this dampening effect is offset by a distinct theoretical advantage: the relative position of the (2+1)-flavor CEP itself shifts to a more favorable location, higher on the $T$ axis and lower on the $\mu$ axis. Furthermore, it is relevant to recall that because the TCP lies well inside the $R_{q}=2$ contour of the two-flavor RQM model, it actively influences the critical fluctuations near the CEP. In contrast, this interplay is entirely absent in the (2+1)-flavor framework, as its TCP is located significantly far from the contour boundary.

For the two-flavor case, the critical fluctuation contours of the Log-RPQM model for $R_q = 2, 3,$ and $5$ (shown in Fig.~\ref{fig11b}) span 79.0, 55.1, and 33.6 MeV along the $T$-axis, with respective $\mu$-axis extents of 82.1, 46.8, and 24.2 MeV. Transitioning from the two-flavor to the $(2+1)$-flavor Log-RPQM model, the $T$-axis spreads shrink significantly to 46.9, 29.8, and 18.8 MeV, whereas their $\mu$-axis spans experience a more moderate reduction to 71.4, 42.6, and 23.8 MeV. Unlike the standard RQM model scenarios in Fig.~\ref{fig11a}, the $(2+1)$-flavor Log-RPQM contours in Fig.~\ref{fig11b} exhibit a large reduction in their $T$-axis spreads while their $\mu$-axis spans are only moderately affected due to the influence of the logarithmic Polyakov-loop potential.   Furthermore, the contours appear pinched as their widths are noticeably compressed by the Polyakov-loop potential. This compression and pinching become even more pronounced when the effects of the third flavor combine with the Polyakov-loop potential in the absence of quark back-reaction. To the left of their respective critical end points (CEPs) in Fig.~\ref{fig11b}, the $(2+1)$-flavor contours lie noticeably below those of the two-flavor Log-RPQM model. This behavior can be traced back to the locations of their CEPs: in the $(2+1)$-flavor Log-RPQM model, the CEP is located at $(\mu_{\cep}, T_{\cep}) = (252.00, 94.60)\text{ MeV}$, which is shifted by $\Delta\mu = 18.58\text{ MeV}$ to lower values and $\Delta T = 2.15\text{ MeV}$ to higher values compared to the two-flavor CEP at $(270.58, 92.45)\text{ MeV}$. Because the curvature of the $(2+1)$-flavor phase boundary at its CEP is smaller than that of the two-flavor case (see Figs.~\ref{fig10a} and \ref{fig10b}), the $(2+1)$-flavor contours are not lifted along the $T$-axis and instead incline toward the $\mu$-axis, placing them lower than their two-flavor counterparts. Finally, while the tricritical point (TCP) lies well inside the $R_q = 2$ contour for the two-flavor Log-RPQM model-thereby influencing critical fluctuations around the CEP-the third flavor generally suppresses critical fluctuations along both the $T$ and $\mu$ axes. In the $(2+1)$-flavor Log-RPQM model, the TCP is pushed right to the boundary of the $R_q = 2$ contour, rendering its effect on the critical fluctuations around the CEP marginal.

Examining the plots of the PolyLog-glue-RPQM formulation in Fig.~\ref{fig12a}, the critical fluctuation contours exhibit distinct geometric shifts under the influence of the third flavor. In the two-flavor regime, the contours for $R_q = 2, 3,$ and $5$ cover $T$-axis intervals of 86.7, 53.2, and 35.4 MeV, corresponding to $\mu$-axis widths of 80.8, 46.2, and 25.8 MeV. Expanding to the $(2+1)$-flavor system induces a drastic compression along the $T$-axis, narrowing these spans to 53.5, 35.9, and 21.7 MeV, while the $\mu$-axis extents experience a softer contraction to 65.0, 40.2, and 22.8 MeV. This substantial $T$-axis suppression sharply contrasts with standard RQM predictions (Fig.~\ref{fig11a}) and somewhat mirrors the Log-RPQM dynamics. However, a broader structural distinction emerges between the two frameworks: across both the two- and $(2+1)$-flavor regimes, the PolyLog-glue-RPQM model consistently yields larger $T$-axis spreads but narrower $\mu$-axis spans compared to the Log-RPQM model. Furthermore, while both models experience $\mu$-axis shrinkage upon introducing the third flavor, this contraction is notably more in the PolyLog-glue-RPQM framework.

It is worth emphasizing that the contour compression in the PolyLog-glue-RPQM model is significantly moderated by the effect of quark back-reaction. Consequently, these contours suffer much less geometric pinching than those in the Log-RPQM framework, yielding noticeably broader shapes. These structural differences are visually corroborated in Fig.~\ref{fig12b}, which compares the $R_q = 2, 3,$ and $5$ susceptibility contours of both models for the two-flavor case (the corresponding numerical spans for the Log-RPQM model are detailed in the preceding paragraph and Table~\ref{tab:table4}). The vertical ordering of these contours is directly tied to the locations of their respective critical end points (CEPs) and the local curvature of the phase boundary. Because the Log-RPQM CEP is situated at a higher $T$ and lower $\mu$—where the phase boundary curvature differs from that at the PolyLog-glue-RPQM CEP—its contours lie below those of the PolyLog-glue-RPQM model in Fig.~\ref{fig12b}. A similar hierarchy between the two models is observed for the $(2+1)$-flavor case, as recently reported in Fig.~8(b) of Ref.~\cite{Akakn}. Finally, this curvature argument also explains the internal ordering within the PolyLog-glue-RPQM model itself (Fig.~\ref{fig12a}): the smaller, $(2+1)$-flavor contours lie below their two-flavor counterparts because the phase boundary curvature is smaller at the $(2+1)$-flavor CEP (located at a higher $T$) than at the two-flavor CEP (located at a lower $T$). Furthermore, the relative position of the tricritical point (TCP) dictates its impact on these dynamics. In the two-flavor PolyLog-glue-RPQM model also, critical fluctuations around the CEP are noticeably influenced by the TCP, which resides well within the $R_q = 2$ contour. In contrast, the TCP in the $(2+1)$-flavor Log-RPQM model lies precisely on the boundary of the $R_q = 2$ contour, resulting in a minimal effect on the critical fluctuations near the CEP.

\begin{figure*}[htb]
\subfigure[]{
\label{fig13a} 
\begin{minipage}[b]{0.47\textwidth}
\centering \includegraphics[width=\linewidth]{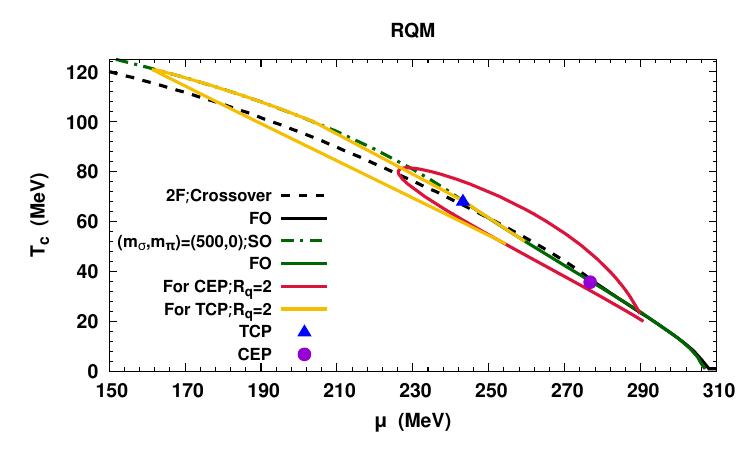}
\end{minipage}}
\hfill
\subfigure[]{
\label{fig13b}
\begin{minipage}[b]{0.49\textwidth}
\centering \includegraphics[width=\linewidth]{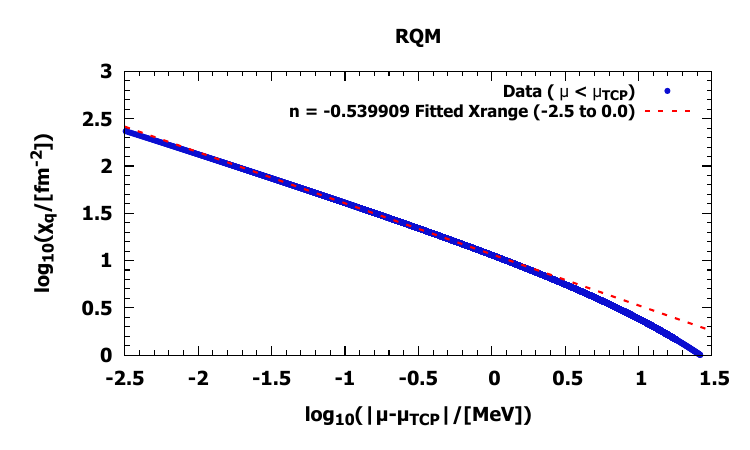}
\end{minipage}}

\caption{The left panel (a) presents an enlarged view of the RQM phase diagrams for the physical-point and chirallimit parameter sets. It illustrates the overlap between the normalized quark number susceptibility contours ($R_q=2$) that envelop the TCP and the CEP. The right panel (b) plots the logarithm of $\chi_{q}(\mu,T)$ as a function of the logarithm of $(\mu-\mu_{\tcp})$ close to the TCP in the RQM. Here, $\mu_{\tcp}$ is approached from the lower-$\mu$ side to extract the critical exponent governing the power-law divergence of the quark number susceptibility at the TCP.}
\label{fig:mini:fig13} 
\end{figure*}

In the previously described results for the two-flavor RQM, Log-RPQM, and PolyLog-glue-RPQM models, the tricritical point (TCP) lies well inside the $R_{q}=2$ quark number susceptibility contour for each respective model scenario. To determine in detail whether the critical region around the CEP overlaps with the one around the TCP, we plotted the $R_{q}=2$ contours around both the TCP and CEP in Fig.~\ref{fig13a} for the RQM model (the RPQM models are expected to yield similar results). This plot displays an enlarged view of the contours, spanning a range of 0–124 MeV on the temperature axis and 150–310 MeV on the chemical potential axis. The $R_q =2$ contour enveloping the TCP at $(\mu_{\tcp}, T_{\tcp}) = (243.19, 68.25)$ MeV extends up to 121 MeV and 51 MeV on the upper and lower bounds of the temperature axis, respectively. Along the $\mu$ axis, the contour spans from 161.17 MeV on the left to 259.51 MeV on the right of the TCP. The value $R_{q}=2$ on both the lower and upper branches of the contour is observed only up to $\mu=259.51$ MeV. Beyond this point, only the lower branch of the contour is present, extending even beyond the CEP (not shown) into the first-order region.This feature arises because the second-order chiral transition of the $O(4)$ universality class produces a very sharp susceptibility in the chiral limit $m_{\pi}=0$, resulting in a truncated critical region around the TCP in the chiral-symmetry restored phase. Employing the Proper-Time Renormalization Group (PTRG) approach to account for the effects of vacuum and thermal quantum fluctuations of quarks and mesons in the QM model, Ref.~\cite{Schaefer:2006ds} demonstrated in Fig. 19 of their work that the critical region of the TCP does not overlap with that of the CEP, as it lies in a distant region of the phase diagram. In contrast to this previous finding, the critical regions ($R_q=2$ contours) of the TCP and CEP show a noticeable overlap in our Fig.~\ref{fig13a}.

The critical exponent characterizing the power-law divergence of the quark number susceptibility at the TCP has been determined by plotting $\log \chi_{q}$, computed in the chiral limit, against $\vert{}\mu-\mu_{\tcp}\vert{}$. In our calculations, the TCP in the $\mu$-$T$ plane is approached along a path parallel to the $\mu$ axis, moving from lower values of $\mu$ toward $\mu \rightarrow \mu_{\tcp}$. Note that the first-order transition line and the second-order $O(4)$ critical line at the TCP are asymptotically parallel \cite{Schaefer:2006ds}. The critical exponent, obtained from the slope of a linear logarithmic fit of $\chi_{q}$ (expression is given in the following subsection) near the TCP, is $n = 0.5399 \pm 0.0001$, as shown in Fig.~\ref{fig13b}. Mean-field theory predicts a critical exponent of $\gamma=1/2$ for the TCP, which is also expected from its Gaussian fixed-point structure. The value of the exponent obtained from our calculation, $\gamma = n = 0.5399 \pm 0.0001$, is consistent with this theoretical expectation and matches exactly with its value $\gamma =0.54$ found in the PTRG approach in Ref.~\cite{Schaefer:2006ds}.

\subsection{Critical exponents.}
\label{secIIIC}

\begin{figure*}[htb]
\subfigure[]{
\label{fig14a} 
\begin{minipage}[b]{0.48\textwidth}
\centering \includegraphics[width=\linewidth]{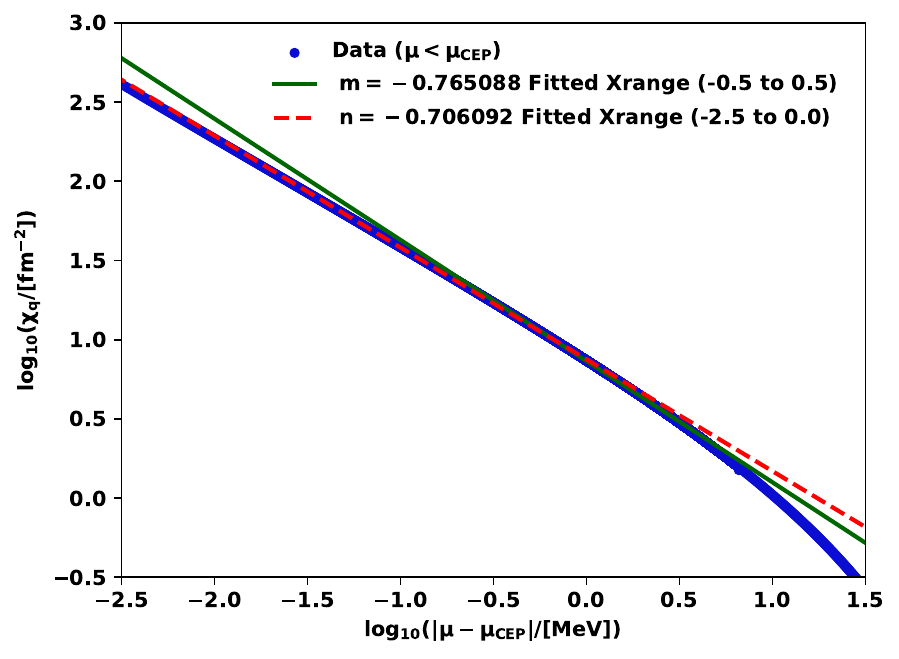}
\end{minipage}}
\hfill
\subfigure[]{
\label{fig14b}
\begin{minipage}[b]{0.48\textwidth}
\centering \includegraphics[width=\linewidth]{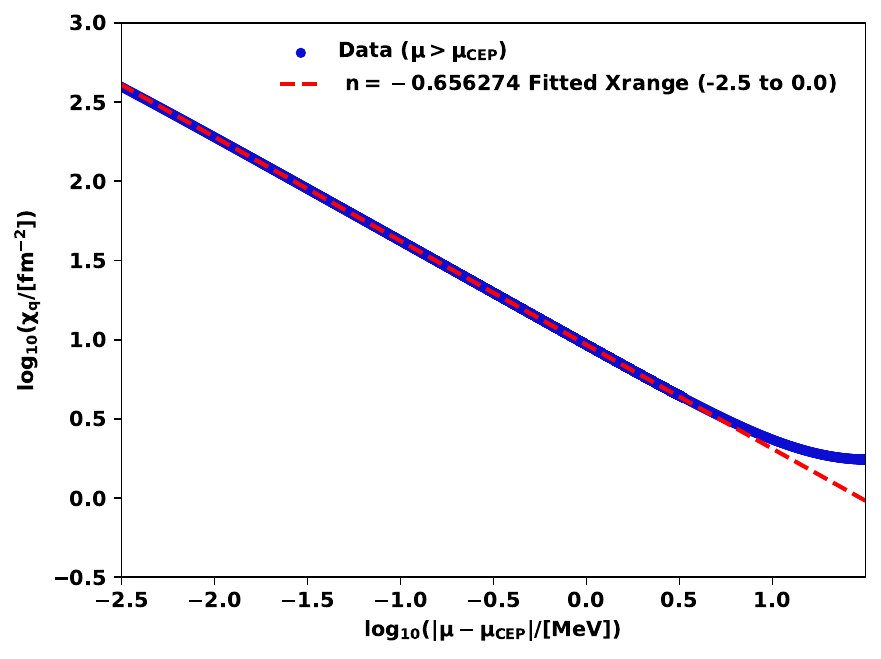}
\end{minipage}}

\caption{The left panel (a) and right panel (b) plot the logarithm of $\chi_{q}(\mu,T)$ as a function of the logarithm of $(\mu-\mu_{\cep})$ close to the CEP in the RQM. The critical end point $\mu_{\cep}$ is approached from the lower-$\mu$ and higher-$\mu$ sides, respectively, to extract the critical exponent governing the power-law divergence of the quark number susceptibility at the CEP.}

\label{fig:mini:fig14} 
\end{figure*}

The peak of the quark number susceptibility ($\chi_{q}$) grows at higher chemical 
potentials ($\mu>0$), reflecting the increasing strength of the chiral crossover 
transition as $\mu$ increases. At the critical end point (CEP), $\chi_{q}$ diverges 
according to a power law, and the exponent governing that divergence depends on the 
direction of approach to the CEP in the $\mu$--$T$ plane 
\cite{Hatta,Fuji,Schaefer:2006ds,schafwag12,vkkr12,Griffiths1970}. Within the mean-field 
approximation (MFA), $\chi_{q}$ is expected to diverge with exponent $\gamma=1$ along a 
path asymptotically parallel to the first-order transition line, while along any path 
not parallel to it the divergence scales with $\epsilon=1-1/\delta$. Since $\delta=3$ in 
the MFA, $\epsilon=2/3$, so $\gamma>\epsilon$: the enhancement of $\chi_{q}$ is stronger 
along the direction parallel to the first-order line. This directional asymmetry 
produces the elongated critical-region shapes seen in the phase diagrams of 
Figs.~(\ref{fig:mini:fig4}), (\ref{fig:mini:fig5}), (\ref{fig:mini:fig8}), 
(\ref{fig:mini:fig9}), and (\ref{fig:mini:fig10}). The critical exponents of $\chi_{q}$ in the two-flavor QM, PQM, QMVT, and Log-PQMVT 
models were evaluated in Ref.~\cite{vkkr12}; earlier, Ref.~\cite{Schaefer:2006ds} 
computed the exponents for the QM model using the s-MFA as well as fluctuation-corrected 
PTRG methods. Following this approach, we 
fix $T=T_{\cep}$ and approach $\mu_{\cep}$ along the $\mu$-axis from both below and 
above, extracting the critical exponent from the logarithmic fit
\begin{equation}
\log \chi_{q} = -n \log |\mu -\mu_{\cep} | + c\ ,
\end{equation}
where the slope $n$ gives the critical exponent $\epsilon$, while the intercept $c$ on 
the $y$-axis is independent of $\mu$. The critical exponents for the s-MFA QM and 
Log-PQM models, as well as the e-MFA QMVT and Log-PQMVT models, calculated in 
Ref.~\cite{vkkr12}, are presented in Table~\ref{tab:table5}. These values serve as a 
reference for comparison with the critical exponents characterizing the power-law 
divergence of $\chi_{q} (\mu, T)$, which we compute in the present work for the RQM, 
Log-RPQM, and PolyLog-glue-RPQM models.

\begin{table}
\caption{ Critical exponents of the
  quark-number susceptibility in the QM,PQM,QMVT, PQMVT, RQM,Log-RPQM and PolyLog-Glue RPQM models
  for two different paths parallel to the chemical potential axis 
  approaching the $\mu_{\cep}$ from the lower $\mu<\mu_{\cep}$ and 
  higher $\mu>\mu_{\cep}$ side. The last row gives critical exponent for the power law divergence of quark-number susceptibility (in chiral limit) at the TCP of RQM model.}
\label{tab:table5}
\begin{tabular}{|c|c|c|}
\hline
Model &$\mu-\mu_{\cep} < 0$& $\mu-\mu_{\cep} > 0$ \\ \hline \hline
QM	  & $0.6379 \pm 0.0002$ &$0.6648 \pm 0.0001$\\

PQM		& $0.6309 \pm 0.0001$ &$0.6668 \pm 0.0001$\\
RQM	&  $0.706\pm 0.0001$ &$0.656 \pm 0.00001$ \\

Log-RPQM	&  $0.718\pm 0.0003$ &$0.655 \pm 0.0001$ \\

PolyLog-Glue RPQM	&  $0.756\pm 0.0002$ &$0.689 \pm 0.0003$ \\

QMVT	&  $0.720\pm 0.00005$ &$0.6938 \pm 0.0002$ \\

Log-PQMVT	& $0.725 \pm 0.0002$ &$0.6886 \pm 0.0004$\\ 
\hline
TCP Exponent  & $\mu-\mu_{\tcp} < 0$&$...$ \\ \hline \hline
RQM	&  $0.5399\pm 0.0001$&$...$  \\
\hline

\end{tabular}
\end{table}

The logarithm of $\chi_{q} (\mu, T)$ is plotted in Fig.~\ref{fig14a} as a function of the 
logarithm of $|\mu-\mu_{\cep}|$ close to the CEP of RQM model.~The plot shows 
scaling over several orders of magnitude. Approaching the $\mu_{\cep}$ from the lower 
$\mu$ side yields the critical exponent of $\epsilon=n =0.706 \pm 0.0001$.~This value in 
RQM model is slightly larger than mean field critical exponent $\epsilon=2/3$ but 
closer to the QMVT model critical exponents $\epsilon=0.72$ reported in 
Ref.~\cite{vkkr12} (see Table~\ref{tab:table5}).~Similar to the QMVT model result in 
Ref.~\cite{vkkr12} and in our present QMVT model calculation (with slightly different 
$f_{\pi}=92.4$ MeV and $g=3.25$ than the values taken in \cite{vkkr12} as $f_{\pi}=93.0$ 
MeV and $g=3.33$),~one finds the TCP in the chiral limit ($m_{\pi}=0$) phase diagram of 
the RQM model and this TCP lies very well inside the $R_{q}=2$ susceptibility contour 
enveloping the CEP of the RQM model phase diagram in Fig.~\ref{fig4a}.~This slightly 
larger critical exponent may be due to the modification of criticality around the CEP 
caused by the presence of the TCP in its proximity~\cite{vkkr12}. The several orders of 
magnitude scaling in the data starts earlier when $\log |\mu- \mu_{\cep} | < 0.5$.~Here 
it is relevant to point out that the work in Ref.\cite{Schaefer:2006ds}, using the 
PTRG framework to account for the effect of quantum fluctuations in the QM model, finds the critical exponent $\epsilon=0.74$.~They find a crossover kind of phenomenon in the critical exponents, occurring in the range $-0.5<\log |\mu- \mu_{\cep} |<.5$, through which the slope of $0.77$ for data points with $\log |\mu- \mu_{\cep} |>.5$ changes to $0.74$ for data points in the range $-2.5<\log |\mu- \mu_{\cep} |<-.5$.~It is argued in Ref.~\cite{Schaefer:2006ds} that this change in exponents could be interpreted as a crossover between different universality classes.~A similar crossover phenomenon between different universality classes was witnessed in Ref.~\cite{Hatta}. Though our work does not find an analogous crossing behavior between universality classes,~the data points in Fig.~\ref{fig14a} show a bending trend, and when fitted in the smaller range $-0.5<\log |\mu- \mu_{\cep} |<.5$, they yield a higher slope $\epsilon=m = 0.765\pm 0.0002$. Such a trend might result from the phenomenon that critical fluctuations around the CEP are somehow affected by the TCP. Further, when $\mu_{\cep}$ is approached from the higher $\mu>\mu_{\cep}$ side, Fig.~\ref{fig14b} gives the critical exponent as $\epsilon=n=0.656 \pm 0.0001$, which aligns well with the mean-field exponent $\epsilon=2/3$, whereas the analogous critical exponent for the QMVT model is $\epsilon=n=0.6938 \pm 0.00002$, as shown in Table~\ref{tab:table5}.

\begin{figure*}[htb]
\subfigure[]{
\label{fig15a} 
\begin{minipage}[b]{0.48\textwidth}
\centering \includegraphics[width=\linewidth]{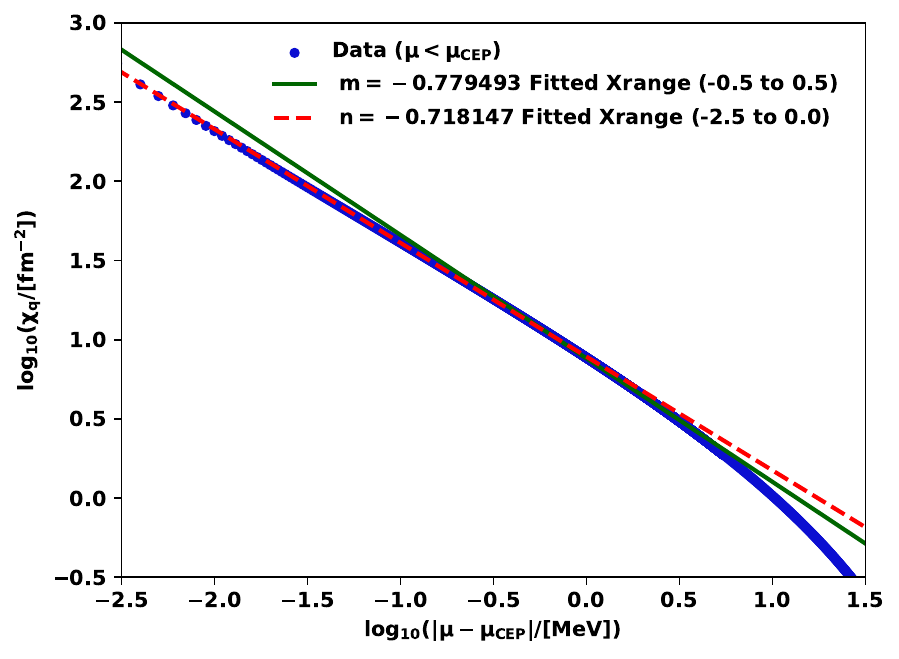}
\end{minipage}}
\hfill
\subfigure[]{
\label{fig15b}
\begin{minipage}[b]{0.48\textwidth}
\centering \includegraphics[width=\linewidth]{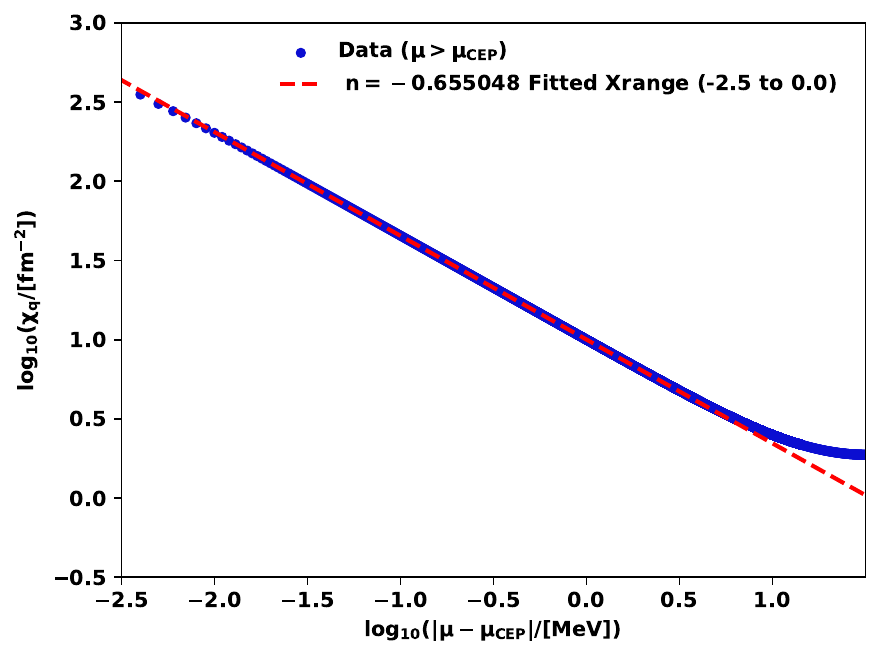}
\end{minipage}}

\caption{The left panel (a) and right panel (b) plot the logarithm of $\chi_{q}(\mu,T)$ as a function of the logarithm of $(\mu-\mu_{\cep})$ close to the CEP in the Log-RPQM. The critical end point $\mu_{\cep}$ is approached from the lower-$\mu$ and higher-$\mu$ sides, respectively, to extract the critical exponent governing the power-law divergence of the quark number susceptibility at the CEP.}

\label{fig:mini:fig15} 
\end{figure*}

\begin{figure*}[htb]
\subfigure[]{
\label{fig16a} 
\begin{minipage}[b]{0.48\textwidth}
\centering \includegraphics[width=\linewidth]{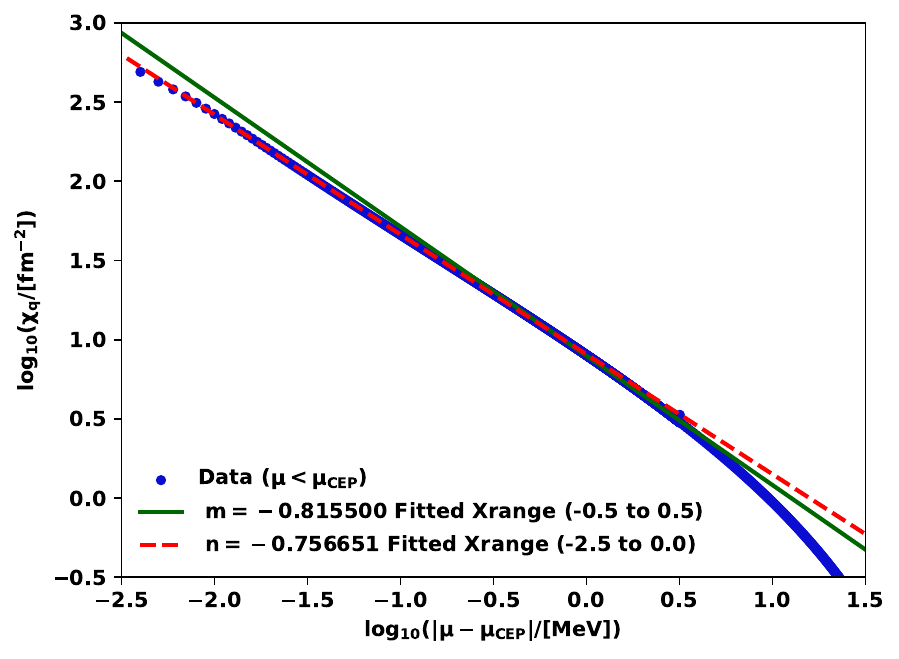}
\end{minipage}}
\hfill
\subfigure[]{
\label{fig16b}
\begin{minipage}[b]{0.48\textwidth}
\centering \includegraphics[width=\linewidth]{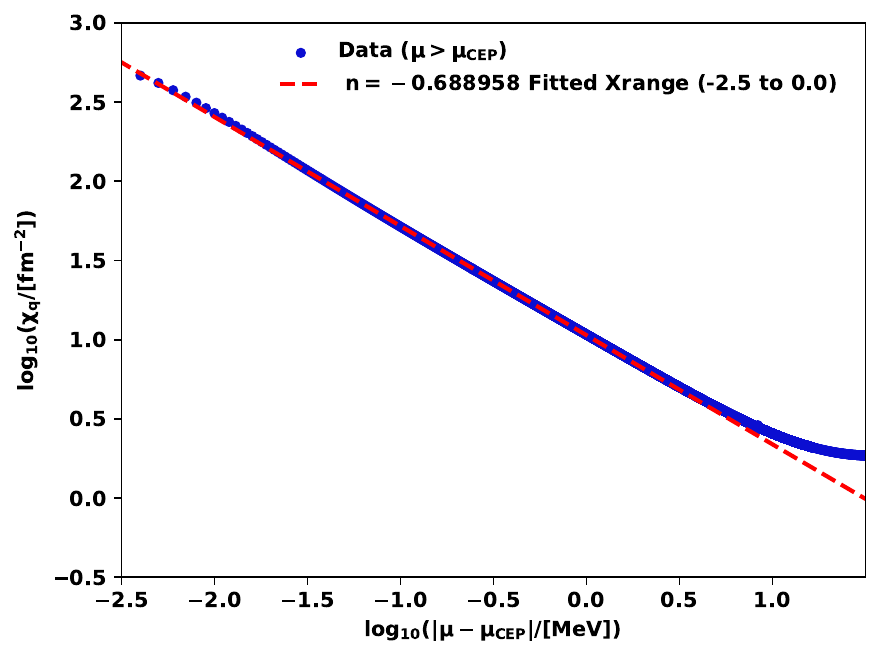}
\end{minipage}}

\caption{The left panel (a) and right panel (b) plot the logarithm of $\chi_{q}(\mu,T)$ as a function of the logarithm of $(\mu-\mu_{\cep})$ close to the CEP in the PolyLog-glue RPQM. The critical end point $\mu_{\cep}$ is approached from the lower-$\mu$ and higher-$\mu$ sides, respectively, to extract the critical exponent governing the power-law divergence of the quark number susceptibility at the CEP.}

\label{fig:mini:fig16} 
\end{figure*} 

One finds equivalent critical exponents when the logarithmic Polyakov-loop potential 
without quark back-reaction is integrated into the Log-RPQM model. The 
$\log \chi_{q} (\mu, T)$ versus $\log |\mu-\mu_{\cep}|$ plot for the Log-RPQM model in 
Fig.~\ref{fig15a} shows that when $\mu_{\cep}$ is approached from the lower $\mu$ side, 
the critical exponent is $\epsilon=n=0.718 \pm 0.0003$. Similar to the RQM model 
(Fig.~\ref{fig14a}), the data points for the Log-RPQM model in Fig.~\ref{fig15a} exhibit 
a bending trend, and when fitted in the small range 
$-0.5<\log |\mu- \mu_{\cep} |<.5$, they yield a higher slope 
$\epsilon=m = 0.779\pm 0.0002$.~Almost the same pattern is found in Fig.~8(a) of 
Ref.~\cite{vkkr12} for the Log-PQMVT model, where the critical exponent is 
$\epsilon=n=0.725 \pm 0.0002$ (see Table~\ref{tab:table5}). Approaching $\mu_{\cep}$ 
from the higher $\mu$ side in the Log-RPQM model, Fig.~\ref{fig15b} gives the critical 
exponent $\epsilon=n=0.655 \pm 0.0001$, whereas Table~\ref{tab:table5} shows the 
analogous critical exponent for the Log-PQMVT model to be $\epsilon=n=0.6886 \pm 0.00004$.
 
When the effect of quark back-reaction is taken into account through the PolyLog-glue 
form of the Polyakov-loop potential, in the PolyLog-glue-RPQM model, the 
$\log \chi_{q} (\mu, T)$ versus $\log |\mu-\mu_{\cep}|$ plot in Fig.~\ref{fig16a} yields 
a relatively larger critical exponent of $\epsilon=n=0.756 \pm 0.0002$ when $\mu_{\cep}$ 
is approached from the lower $\mu$ side. The data points for the PolyLog-glue-RPQM 
model in Fig.~\ref{fig16b} exhibit a stronger bending trend, and when fitted in the 
small range $-0.5<\log |\mu- \mu_{\cep} |<.5$, one finds a higher slope 
$\epsilon=m = 0.815\pm 0.0002$. A larger critical exponent, 
$\epsilon=n=0.689 \pm 0.0003$, is found in Fig.~\ref{fig16b} when $\mu_{\cep}$ is 
reached from the higher $\mu$ side, also in the PolyLog-glue-RPQM model.

Near the CEP, the singular part of $\chi_{q}$ falls into the three-dimensional Ising 
($Z(2)$) universality class, consistent with the expectation that the QCD critical point 
shares the universality class of the liquid-gas critical point. However, since no 
exact symmetry protects $(T,\mu)$ as the intrinsic scaling axes, the physical directions 
$\delta T$ and $\delta\mu$ around the CEP are generically non-trivial linear 
combinations of the two Ising scaling fields: the reduced-temperature-like field $r$ and 
the ordering (magnetic) field $h$ \citep{Son,Hatta}. Consequently, the effective 
critical exponent governing the divergence of $\chi_{q}$ depends on the direction of 
approach to the CEP relative to these mixed axes. Along a path aligned with the 
$r$-direction, $\chi_{q}$ diverges with exponent $\gamma$ ($\gamma=1$ in mean field; 
$\gamma\simeq1.24$ in the static 3D Ising class), whereas along a path aligned with the 
$h$-direction, $\chi_{q}\sim h^{-(\delta-1)/\delta}$, giving an effective exponent of 
$2/3\simeq0.67$ in mean field ($\delta=3$) and $\simeq0.79$ using the static 3D Ising 
value ($\delta\simeq4.8$). Since our models are treated at mean-field level, augmented 
only by fermionic vacuum fluctuations and without RG improvement where, apart from the 
quark one-loop vacuum fluctuation, quantum fluctuations of mesons are also included, 
these mean-field values provide the natural reference scale for the exponents extracted 
numerically.

Approaching the CEP along the $\mu$ direction at fixed $T$ predominantly probes the 
field-like ($h$) scaling direction, since the mixing angle between the Ising axes and 
the $(T,\mu)$ plane rotates the $\mu$-direction away from the pure temperature-like 
axis. We find that the effective critical exponent characterizing the divergence of 
$\chi_{q}$ as $\mu\to\mu_{\cep}$ lies in the range $0.66$--$0.76$ across the RQM, 
Log-RPQM, and PolyLog-glue-RPQM models studied here. This range brackets the mean-field 
field-like exponent $(\delta-1)/\delta=2/3\simeq0.67$, with the modest excess above this 
value reflecting the residual admixture of the temperature-like direction at finite 
mixing angle, together with subleading, non-universal contributions from the regular 
background term $\chi_{q}^{\rm reg}$. This is consistent with the expectation that 
mean-field models correctly reproduce the qualitative Ising-like scaling structure near 
the CEP, while the precise universal 3D Ising exponents are recovered only once 
renormalization-group (fluctuation) effects are included beyond mean field.

It is worth recalling that the RQM, Log-RPQM, and PolyLog-glue-RPQM models employed here retain only the quark one-loop vacuum and thermal quantum fluctuations, while the quantum 
fluctuations of the meson fields are not resummed as they are treated as mean field; this is precisely the distinction that identifies these models as extended mean-field (e-MFA) constructions, in the same sense used for the QM and PQM models with vacuum quark loops \cite{vac,guptiw,vkkr12,TranAnd,schafwag12,vkkt13,fix1,Adhiand1,Adhiand2,Adhiand3,asmuAnd,Rai,RaiTiw22,
raiti23,vkkr23,skrvkt24}. Functional renormalization group studies that restore the mesonic quantum fluctuations on top of the quark loop have shown that this additional resummation shrinks the size of the critical region around the CEP and shifts the CEP itself toward lower temperature and larger chemical potential relative to mean-field-type treatments \cite{THerbst2}. Since the effective critical exponent extracted along a fixed-$T$ path depends sensitively on the local geometry of the critical region and on the mixing angle between the physical $(T,\mu)$ axes and the intrinsic Ising scaling axes $(r,h)$, such a shrinking/reshaping of the critical region is expected to alter the numerically fitted exponents. This is borne out quantitatively in FRG studies of the quark-meson model, where the inclusion of meson quantum fluctuations shifts the $O(4)$/Ising-type exponents substantially away from their mean-field values and toward the corresponding lattice values \cite{Braun:2009}. We therefore expect that the critical exponents obtained here within the e-MFA RQM, Log-RPQM, and PolyLog-glue-RPQM models may differ somewhat from those of the fully fluctuation-resummed FRG-based QM and PQM models, precisely because meson quantum fluctuations - absent in our treatment by construction - are known to modify both the size of the critical region and the universal scaling behavior in its vicinity.

\section{Summary}
\label{sec:IV}

In standard mean-field studies, neglecting quark one-loop vacuum fluctuations in the QM model produced small and strongly pinched critical regions within Polyakov-loop augmented PQM models \cite{Schaefer:2006ds,vkkr12,schafwag12,Akakn}. In contrast, the present work not only explicitly incorporates these vacuum fluctuations but also addresses how distinct treatments of quark one-loop divergences impact the morphologies of the critical regions enveloping the CEP. Because matching counter-terms in the on-shell and $\overline{\text{MS}}$ schemes yields different effective potentials and renormalized parameters compared to evaluating the effective potential in the $\overline{\text{MS}}$ scheme while determining model parameters via meson curvature masses, we systematically compared the effects of these differently treated vacuum corrections across four two-flavor frameworks: the on-shell renormalized QM (RQM), its logarithmic Polyakov-loop extended Log-RPQM, the curvature-mass-parametrized QM with a vacuum term (QMVT), and its Polyakov-loop augmented Log-PQMVT model. To quantify the CEP's proximity to the TCP, we computed phase diagrams in both the chiral limit ($m_\pi = 0$) and at the physical point when $m_\sigma = 500\text{ MeV}$. The chemical potential dependence of the meson masses ($m_\sigma$, $m_\pi$) at $T=T_{\cep}$ and $T=T_{\cep}\pm 15$ was also analyzed. Since the quark number susceptibility $\chi_q \sim (\partial n_q/\partial x)^2/m_{\sigma,c}^2 \propto \chi_{s} = 1/m_{\sigma,c}^2$ \citep{Hatta,redlsaki} and the scalar susceptibility $\chi_{s}$ diverge at the CEP when the effective potential flattens ($m_{\sigma,c} \to 0$), we quantified the size and shape of the critical regions by plotting contours of the enhanced quark number susceptibility normalized by its free quark gas value ($R_{q}=\frac{\chi_{q}(\mu,T)}{\chi_{free}}$) and the normalized scalar susceptibility ($R_s = m_{\sigma,c}^2(0,0)/m_{\sigma,c}^2(\mu,T)$).

To fulfill our second objective—definitively isolating the roles of the third (strange) flavor and the axial $U_{A}(1)$ anomaly in governing the topography of critical fluctuations enveloping the CEP—we comprehensively compared the constant $R_q$ contours from our two-flavor RQM and RPQM calculations with the corresponding $(2+1)$-flavor results recently published in Ref.~\cite{Akakn}. Specifically, using the logarithmic and PolyLog-glue forms of the Polyakov-loop potential—without and with quark back-reaction, respectively—yields the two-flavor Log-RPQM and PolyLog-glue-RPQM models employed for this systematic comparison. Finally, to address our third objective, we computed the critical exponents characterizing the power-law divergence of the quark number susceptibility as $\mu$ approaches $\mu_{\cep}$ from both directions ($\mu < \mu_{\cep}$ and $\mu > \mu_{\cep}$) at fixed $T_{\cep}$. The resulting exponents for the RQM, Log-RPQM, and PolyLog-glue-RPQM models were then compared against existing QM and QMVT model findings \cite{Schaefer:2006ds,vkkr12} and Log-PQMVT results \cite{vkkr12}, respectively. Furthermore, the critical exponent for the RQM model TCP was extracted by approaching $\mu \rightarrow \mu_{\tcp}$ from below ($\mu < \mu_{\tcp}$) at $T = T_{\tcp}$.

Due to the pronounced smoothing of the chiral transition in the QMVT model, its critical region contours (for $R_{q}=2, 3,$ and $5$) are broad, extended, and markedly larger than those in the s-MFA QM model. Furthermore, the large bending of the QMVT model phase boundary around its CEP at high-$\mu_{\cep}$ (and low-$T_{\cep}$) induces a distinct neck-like structure in these contours (particularly for $R_{q}=2$) for $\mu > \mu_{\cep}$. In contrast, the moderate softening of the chiral transition strength in the RQM model yields critical regions that are smaller than those of the QMVT model, though still significantly larger than those of the s-MFA QM model. Because the CEP of the RQM model occurs at a significantly lower $\mu_{\cep}$ with a gentler phase boundary curvature, its critical regions have no neck-like structure. Integrating a logarithmic Polyakov-loop potential without quark back-reaction modifies the fermionic determinant and shifts the CEP to higher temperatures. The Log-RPQM model situates its CEP at a higher $T$ and significantly lower chemical potential ($\mu$) than that of the Log-PQMVT model, resulting in a gentler phase boundary curvature. Consequently, its critical region contours are smaller in size and somewhat more compressed (pinched) in width, but the neck-like bending disappears. By contrast, the CEP of the Log-PQMVT model at high-$\mu$ (low $T$) induces sharp phase boundary bending; this geometrically stretches the critical regions to a larger size while sharply compressing their width, forming a large, pronounced neck-like structure below the CEP for $\mu > \mu_{\cep}$ and $T < T_{\cep}$. Driven by these fundamental differences in phase boundary curvature and CEP positioning, the $R_{q}=2, 3,$ and $5$ contours of the RQM and Log-RPQM models are situated vertically below those of the corresponding QMVT and Log-PQMVT models. When quark back-reaction is incorporated into the PolyLog-glue-RPQM model, these effects are noticeably smoothed. Compared to the Log-RPQM model, its critical regions expand along the $T$-axis, shrink slightly along the $\mu$-axis, and experience a mitigated pinching effect, resulting in notably broader contours due to the back-reaction's smoothing influence.

Because the pseudo-critical (critical for the chiral limit) temperatures $T_c^\chi$ ($T_c$) at $\mu=0$ for the two-flavor QMVT and Log-PQMVT models are noticeably larger than those of the RQM and Log-RPQM models, the phase boundaries of the latter lie noticeably lower in the $\mu-T$ plane. A proper comparison therefore requires defining the $\mu_{r}-T_{r}$ plane, where the reduced temperature is $T_r = T/T_c^\chi$, and the reduced chemical potential is $\mu_r = \mu/\mu_0$, with $\mu_0$ denoting the first-order phase transition value on the $\mu$-axis at $T=0$. In terms of reduced coordinates, the CEP and TCP of the RQM model are located at $(\mu_{r\cep}, T_{r\cep}) = (0.892, 0.245)$ and $(\mu_{r\tcp}, T_{r\tcp}) = (0.784, 0.468)$, placing them to the upper left of the corresponding QMVT coordinates at $(0.948, 0.180)$ and $(0.835, 0.426)$. Similarly, the Log-RPQM model positions its CEP at $(0.877, 0.548)$ and TCP at $(0.719, 0.809)$, which lie higher and further left than the corresponding Log-PQMVT coordinates at $(0.950, 0.399)$ and $(0.777, 0.721)$, respectively. Although the RQM and Log-RPQM models yield smaller overall critical regions, they possess the theoretical advantage that these elevated, leftward-shifted CEPs drive their critical region contours significantly higher on the $T_r$-axis and further left on the $\mu_r$-axis. For instance, the $R_q=2$ contour in the RQM model reaches up to close at $(\mu_r, T_r) = (0.729, 0.560)$, whereas the QMVT contour closes lower at $(0.782, 0.534)$. This topological advantage is identically mirrored in the logarithmic variants: the Log-RPQM contour extends markedly higher and to the left to close at $(0.659, 0.860)$, while the broader Log-PQMVT contour closes lower and to the right at $(0.719, 0.776)$.

The morphology of the $R_s = 10, 15,$ and $25$ scalar susceptibility contours around the CEP varies considerably across models. The QMVT contours are significantly broader than the RQM contours because their $\sigma$ mass variation begins at $m_{\sigma,c}(0,0) = 500$ MeV (where curvature and pole masses coincide), whereas the RQM variation begins at a lower $m_{\sigma,c}(0,0) = 417$ MeV. The stronger effect of the Polyakov-loop potential along the temperature axis causes a sharper $\sigma$ mass drop in the Log-PQMVT model, yielding contours of large size but considerably compressed in width. A moderate softening of the chiral transition combined with the smaller initial $m_{\sigma,c}(0,0) = 417$ MeV produces an even sharper drop in $m_{\sigma,c}(\mu,T)$ in the Log-RPQM model, whose contours become remarkably small, thin, and pinched, with the $R_s = 25$ contour disappearing entirely.

Considering the $(2+1)$-flavor chiral symmetry breaking scenarios in these models, one finds that the third flavor universally shifts the CEP to significantly lower chemical potentials and noticeably higher temperatures in the $\mu-T$ plane. The TCP similarly shifts to lower chemical potentials across all models, but its temperature response diverges: the TCP temperature rises only in the RQM model, whereas it drops in both the Log-RPQM and PolyLog-glue-RPQM models. Demonstrating this, the $(2+1)$-flavor CEP and TCP in the RQM model—located at $(\mu_{\cep}, T_{\cep})=(265.42, 38.71)\text{ MeV}$ and $(\mu_{\tcp}, T_{\tcp})=(220.15, 71.52)\text{ MeV}$—both shift to higher temperatures relative to their two-flavor counterparts at $(276.69, 35.68)\text{ MeV}$ and $(243.19, 68.25)\text{ MeV}$, respectively. In contrast, the $(2+1)$-flavor CEP and TCP reside at $(252.00, 94.60)\text{ MeV}$ and $(194.30, 126.53)\text{ MeV}$ in the Log-RPQM model, compared to their two-flavor coordinates of $(270.58, 92.45)\text{ MeV}$ and $(221.93, 136.46)\text{ MeV}$. Similarly, the CEP and TCP lie at $(252.70, 70.90)\text{ MeV}$ and $(202.20, 108.14)\text{ MeV}$ in the $(2+1)$-flavor PolyLog-glue-RPQM model, relative to their two-flavor positions of $(268.29, 68.26)\text{ MeV}$ and $(228.19, 112.17)\text{ MeV}$.

Because the pseudo-critical temperatures $T_{c}^{\chi}$ at $\mu=0$ are noticeably larger in the two-flavor models (e.g., $145.2\text{ MeV}$ versus $133.6\text{ MeV}$ in the $(2+1)$-flavor RQM model), comparing their phase boundaries requires the dimensionless $\mu_r-T_r$ plane. In terms of reduced variables, the $(2+1)$-flavor critical points universally migrate to higher $T_r$ and lower $\mu_r$ relative to their two-flavor counterparts. Specifically, for the RQM model, the $(2+1)$- and two-flavor CEPs are located at $(\mu_{r\cep}, T_{r\cep}) = (0.87, 0.289)$ and $(0.892, 0.245)$, while their TCPs lie at $(\mu_{r\tcp}, T_{r\tcp}) = (0.724, 0.535)$ and $(0.784, 0.468)$, respectively. For the Log-RPQM model, the corresponding CEPs reside at $(0.828, 0.635)$ and $(0.877, 0.548)$, and the TCPs at $(0.639, 0.849)$ and $(0.719, 0.809)$. Finally, in the PolyLog-glue-RPQM model, the CEPs are positioned at $(0.831, 0.450)$ and $(0.869, 0.388)$, with TCPs at $(0.665, 0.687)$ and $(0.739, 0.638)$, respectively, for the $(2+1)$- and two-flavor cases. Furthermore, while the two-flavor $R_q=2$ critical regions are consistently broader, their topological containment diverges. The broader two-flavor RQM contour—closing at $(\mu_r, T_r) = (0.729, 0.560)$—completely envelops its $(2+1)$-flavor counterpart, which closes at $(0.7556, 0.5089)$. However, the two-flavor boundaries—due to the effect of confinement physics encoded in the Polyakov-loop potential, with and without quark back-reaction in the Log-RPQM and PolyLog-glue-RPQM models, respectively—fail to fully encompass their $(2+1)$-flavor contours despite substantial overlap: the two-flavor contours close at $(0.659, 0.860)$ and $(0.666, 0.720)$, respectively, in the Log-RPQM and PolyLog-glue-RPQM models, whereas their corresponding $(2+1)$-flavor contours close at $(0.6443, 0.852)$ and $(0.6797, 0.6956)$.

Introducing the $(2+1)$-flavor chiral symmetry breaking (both spontaneous and explicit) scenario in the presence of the axial $U_A(1)$ anomaly fundamentally dampens critical fluctuations across all models, reducing the overall footprint of the critical regions. However, this morphological suppression is model-dependent. Recall that the axial $U_A(1)$ anomaly—represented by the 't Hooft cubic coupling $c$—becomes significantly stronger after renormalization in the RQM model. In the basic RQM framework, the $(2+1)$-flavor contours shrink proportionally along both the $T$ and $\mu$ axes relative to the dimensions of their two-flavor counterparts, maintaining a broad, unpinched geometry. In contrast, the $(2+1)$-flavor critical region contours in the Log-RPQM suffer a drastic reduction in their $T$-axis spread, but only a moderate reduction along the $\mu$-axis compared to the two-flavor case. Consequently, these $(2+1)$-flavor critical regions exhibit considerable width compression (pinching) driven by the Polyakov-loop potential in the absence of quark back-reaction. Notably, while the PolyLog-glue-RPQM framework shares this strong $T$-axis suppression with the Log-RPQM for the $(2+1)$-flavor critical regions, its $\mu$-axis contraction is more pronounced relative to the two-flavor case. Crucially, the inclusion of quark back-reaction significantly moderates this width compression and prevents severe pinching, yielding noticeably broader critical region shapes than those in the Log-RPQM model. Furthermore, the vertical ordering of these contours is governed by their respective CEP locations and the local curvature of the phase boundary. Because the Log-RPQM CEP is situated at a higher $T$ and lower $\mu$, local curvature differences force its contours to lie vertically below those of the PolyLog-glue-RPQM model in both the two- and $(2+1)$-flavor scenarios~\cite{Akakn}. This curvature dependence similarly dictates internal hierarchies: within the Log-RPQM and PolyLog-glue-RPQM frameworks, the smaller $(2+1)$-flavor contours fall below their two-flavor counterparts due to the reduced phase boundary curvature at the higher-$T$ $(2+1)$-flavor CEPs. It is worth pointing out that the dampening effect of the third flavor is somewhat offset by a distinct theoretical advantage: the relative position of the $(2+1)$-flavor CEP itself shifts to a more favorable location—higher on the $T$ axis and lower on the $\mu$ axis. Furthermore, a TCP in close proximity to the CEP will impact its critical fluctuations. The explicit overlap of the $R_{q}=2$ critical region around the TCP with the $R_{q}=2$ contours of the CEP was shown for the two-flavor RQM model. While the TCP resides deep within the $R_q=2$ contour of the CEP across all two-flavor models—influencing these fluctuations—the third flavor decisively breaks this interplay. By displacing the TCP outward relative to the shrinking contours around the CEP, its influence is entirely neutralized in the $(2+1)$-flavor RQM framework—where it lies far from the boundary—and rendered marginal in both Polyakov-loop augmented RPQM models, where it is relegated to the contour edge.

The critical exponent $\gamma=n = 0.5399 \pm 0.0001$, obtained when the TCP is approached from the lower-$\mu$ side ($\mu \rightarrow \mu_{\tcp}$ at fixed $T_{\tcp}$), matches the value of $\gamma = 0.54$ derived via the PTRG approach in Ref.~\cite{Schaefer:2006ds}, which incorporates the full quantum fluctuations of quarks and mesons. Furthermore, the critical exponents characterizing the power-law divergence of $\chi_{q}(\mu,T)$ near the CEP are $\gamma=n = 0.706 \pm 0.0001$ ($0.656 \pm 0.0001$), $0.718 \pm 0.0003$ ($0.655 \pm 0.0001$), and $0.756 \pm 0.0002$ ($0.689 \pm 0.0003$) for the RQM, Log-RPQM, and PolyLog-glue-RPQM models, respectively. These values correspond to the CEP being approached from the lower (higher) $\mu$ side—that is, $\mu \rightarrow \mu_{\cep}$ with $\mu < \mu_{\cep}$ ($\mu > \mu_{\cep}$)—at a fixed $T_{\cep}$.

\section*{Acknowledgments} 
Pooja Kumari acknowledges the financial support of UGC (CRET) University Fellowship.~We are very much thankful to Akanksha Tripathi for valuable discussions, insightful suggestions, and careful reading of the manuscript.   

\section*{Data Availability}
There are no publicly available research data or software supporting this manuscript. Requests for further information or data should be sent to the authors. 

\appendix

\section{RQM model parameter fixing and effective potential}
\label{appendA}
Upon including the quark one-loop vacuum correction, the tree-level parameters $\lambda$, $m^2$, and $h$ become inconsistent unless the on-shell renormalization scheme is used.
While dimensional regularization is used to evaluate divergent loop integrals in the on-shell scheme, its counter-terms are defined differently from those in the minimal subtraction scheme. Specifically, they are chosen to exactly cancel the loop contributions to the self-energies. Because the couplings are evaluated on-shell, the renormalized parameters are independent of the renormalization scale. Note that the fields and parameters defined in Eq.~(\ref{lag}) are bare quantities. The counter-terms $\delta m^2$, $\delta \lambda$, $\delta g^2$ and $\delta h$ for the parameters and 
counter-terms $\delta Z_{\sigma}$, $\delta Z_{\pi}$, $\delta Z_{\psi}$ and $\delta x$ for the wave functions/fields are introduced in the Lagrangian (\ref{lag}) where renormalized fields and couplings are defined as : 
\bqa
\sigma_b&=\sqrt{Z_\sigma}\sigma, \hspace{0.8cm}\bm{\pi}_b=\sqrt{Z_\pi}\bm{\pi},\hspace{0.8cm} \psi_b=\sqrt{Z_{\psi}}\psi,\\
g_b&=\sqrt{Z_g}g,\hspace{0.8cm}m^2_b=Z_{m^2}m^2, \hspace{0.8cm} \lambda_b=Z_{\lambda}\lambda,\\
h_b&=Z_h h, \hspace{0.8cm} x_b=\sqrt{Z_{x}} x.\hspace{2.6cm}
\eqa

Here, $Z_(\sigma,\bm{\pi},\psi, x)=1+\delta Z(\sigma,\bm{\pi},\psi, x)$ identify the field strength renormalization constants while $Z_(m,\lambda,g, h)=1+\delta Z(m,\lambda,g,h)$ denote the mass and coupling renormalization constants.

In the large-$N_c$ limit, one-loop corrections to the quark fields and masses vanish, as the $\pi$ and $\sigma$ loops responsible for renormalizing the quark propagators scale as $\mathcal{O}(N_c^0)$. Hence, the quark field renormalization constant is $Z_\psi = 1$, and the self-energy correction is $\delta m_q = 0$. Furthermore, because the $\bm{\pi}\bar{\psi}\psi$ vertex is of order $\mathcal{O}(N_c^0)$, one gets the relation $Z_{\psi} \sqrt{Z_g g^2} \sqrt{Z_\pi} \approx g \left( 1 + \frac{1}{2} \frac{\delta g^2}{g^2} + \frac{1}{2} \delta Z_\pi \right) = g$ which gives 
\bqa
\frac{\delta x^2}{x^2}&=&-\frac{\delta g^2}{g^2}=\delta Z_{\pi} \ .
\eqa
The self-energy expressions for sigma and pion are given as
\bqa
\label{sigself}
\nonumber
\Sigma_\sigma(p^2) &=&-8g^2N_c[A(m_q^2)-\frac{1}{2}(p^2-4m_q^2)B(p^2)]\\ 
&&+\frac{24\lambda g x N_c m_q}{m_\sigma^2}A(m^2_q),\\ \nonumber \\
\label{piself}
\nonumber
\Sigma_\pi(p^2)&=&-8g^2N_c[A(m_q^2)-\frac{1}{2}p^2B(p^2)]\\ 
&&+\frac{24\lambda g x N_c m_q}{3m_\sigma^2}A(m^2_q)
\eqa
The sigma or pion inverse propagator is written as
\bqa
p^2-m^2_{\sigma,\bm{\pi}}-i\Sigma_{\sigma,\bm{\pi}}(p^2)+counterterms.
\eqa 
The physical mass is equal to renormalized mass in the on-shell scheme.~Hence one writes
\bqa
\Sigma_{\sigma,\bm{\pi}}(p^2=m^2_{\sigma,\bm{\pi}})+counterterms=0.
\eqa
Requiring the residue of the on-shell meson propagator to be unity yields
\bqa
\frac{\partial}{\partial p^2}(\Sigma_{\sigma,\bm{\pi}}(p^2))\Big|_{p^2=m^2_{\sigma,\bm{\pi}}}+counterterms=0.
\eqa 
The one-point function in the large-$N_c$ limit is 
\bqa
\label{oneP}
\delta \Gamma^{(1)}=-8g^2N_c x A(m_q^2)+i\delta t,
\eqa
here $\delta t$ is tadpole counter-term. The vanishing of the one-point function, which is equivalent to the equation of motion, gives $t=h-m^2_\pi \ x$ at tree level. This must also vanish at the one-loop level, yielding the renormalization condition :
\bqa
\delta \Gamma^{(1)}=0
\eqa
The counterterms are 
\bqa
\label{ct1}
\Sigma_{\sigma}^{ct1}&=&i[\delta Z_\sigma(p^2-m^2_\sigma)-\delta m^2_\sigma] \;, \\ \nonumber \\
\label{ct2}
\Sigma_{\pi}^{ct1}&=&i[\delta Z_\pi(p^2-m^2_\pi)-\delta m^2_\pi] \;,\\ \nonumber \\
\label{ct3}
\Sigma_\sigma^{ct2}&=&3\Sigma_\pi^{ct2}=-\frac{i6\lambda x}{2m^2_\sigma}\delta t \;,\\ \nonumber \\
\label{ct4}
\delta t &=& \delta h - f_\pi \delta m^2_\pi -m^2_\pi \delta f_\pi,
\eqa
the tadpole contribution to the self-energies are canceled by the counter-term in Eq.(\ref{ct3}). Evaluating the self-energies and their derivatives at the physical mass determines the on-shell renormalization constants, yielding
\bqa
\delta m^2_\sigma &=& -i \Sigma_\sigma(m^2_\sigma)\; , \\ \nonumber \\
\delta m^2_\pi &=& -i \Sigma_\pi(m^2_\pi)\; , \\ \nonumber \\
\delta Z_\sigma &=& i\frac{\partial}{\partial p^2}\Sigma_\sigma(p^2)\Bigg|_{p^2=m^2_\sigma}\; , \\ \nonumber \\
\delta Z_\pi &=& i\frac{\partial}{\partial p^2}\Sigma_\pi(p^2)\Bigg|_{p^2=m^2_\pi}\; .
\eqa
Using Eq.~(\ref{sigself})-(\ref{oneP}), one finds
\bqa
\label{cntr1}
\delta m^2_\sigma &=&8ig^2N_c[A(m_q^2)-\frac{1}{2}(m^2_\sigma-4m_q^2)B(m^2_\sigma)]\ , 
\\ \nonumber \\ 
\delta m^2_\pi &=&8ig^2N_c[A(m_q^2)-\frac{1}{2}m^2_\pi B(m^2_\pi)] \\ \nonumber \\
\delta Z_\sigma &=&4ig^2N_c[B(m^2_\sigma)+(m^2_\sigma-4m_q^2)B^\prime(m^2_\sigma)]\ , \\ 
\nonumber \\
\delta Z_\pi &=&4ig^2N_c[B(m^2_\pi)+m^2_\pi B^\prime(m^2_\pi)]\ , \\ \nonumber \\
\label{cntr5}
\delta t &=&-8ig^2N_cf_\pi A(m^2_q). 
\eqa 
We use Eq. (~\ref{cntr1})--(\ref{cntr5}) to express the terms $\delta m^2$, $\delta \lambda$, $\delta g^2$, and $\delta h$. Since 
\bqa
\delta m^2&=&\frac{1}{2}(3\delta m^2_\pi-\delta m^2_\sigma),\\ \nonumber \\
\delta \lambda &=& \frac{\delta m^2_\sigma-\delta m^2_\pi}{2f^2_\pi}-\lambda \delta Z_\pi \, \\ \nonumber \\
\delta h &=& \delta t + f_\pi\delta m^2_\pi+\frac{1}{2}m^2_\pi f_\pi \delta Z_\pi \ . 
\eqa
The expression for the counterterms are the following.
\begin{widetext}
\bqa
\nonumber
\delta m^2_{\os}&=&8ig^2N_c\biggl[A(m_q^2)+\frac{1}{4}(m^2_\sigma-4m_q^2)B(m^2_\sigma)-\frac{3}{4}m^2_\pi B(m^2_\pi)\biggr] \\
&=&\delta m^2_{div}+\frac{4g^2N_c}{(4\pi)^2}\biggl[m^2\log\frac{\Lambda^2}{m^2_q}-2m_q^2-\frac{1}{2}(m^2_\sigma-4m_q^2)\mathcal{C}(m^2_\sigma)+\frac{3}{2}m^2_\pi \mathcal{C}(m^2_\pi)\biggr] \,  \\ \nonumber \\
\nonumber \\ \nonumber
\delta \lambda_{\os}&=&-\frac{2ig^2N_c}{f^2_\pi}(m^2_\sigma-4m_q^2)B(m^2_\sigma)+\frac{2ig^2N_c}{f^2_\pi}m^2_\pi B(m^2_\pi)-4i\lambda g^2N_c\biggl[B(m^2_\pi)+m^2_\pi B^\prime(m^2_\pi)\biggr]\\ \nonumber
&=&\delta \lambda_{div}+\frac{2g^2N_cm^2_\sigma}{(4\pi)^2f^2_\pi}\biggl[\biggl(1-\frac{4m_q^2}{m^2_\sigma}\biggr)\biggl\{\log\frac{\Lambda^2}{m^2_q}+\mathcal{C}(m^2_\sigma)\biggr\}+\log\frac{\Lambda^2}{m^2_q}+\mathcal{C}(m^2_\pi)+m^2_\pi \mathcal{C}^{\prime}(m^2_\pi)\biggr]\\
&&-\frac{2g^2N_c m^2_\pi}{(4\pi)^2 f^2_\pi}\biggl[2\log\frac{\Lambda^2}{m^2_q}+2\mathcal{C}(m^2_\pi)+m^2_\pi \mathcal{C}^{\prime}(m^2_\pi)\biggr]\ \\ \nonumber \\ \nonumber \\. 
\delta g^2_{\os}&=& -4ig^4N_c\biggl[B(m^2_\pi)+m^2_\pi B^{\prime}(m^2_\pi)\biggr]=\delta g^2_{div}+\frac{4g^4N_c}{(4\pi)^2}\biggl[\log\frac{\Lambda^2}{m^2_q}+\mathcal{C}(m^2_\pi)+m^2_\pi \mathcal{C}^{\prime}(m^2_\pi)\biggr]\ , \\ \nonumber \\ \nonumber \\
\delta h_{\os}&=&-2ig^2N_cm^2_\pi f_\pi\biggl[B(m^2_\pi)-m^2_\pi B^{\prime}(m^2_\pi)\biggr]=\delta h_{div}+\frac{2g^2N_cm^2_\pi f_\pi}{(4\pi)^2}\biggl[\log\frac{\Lambda^2}{m^2_q}+\mathcal{C}(m^2_\pi)-m^2_\pi \mathcal{C}^{\prime}(m^2_\pi)\biggr] \, \\ \nonumber \\ \nonumber \\
\delta Z_\sigma^{\os}&=&\delta Z_{\sigma, div}-\frac{4g^2N_c}{(4\pi)^2}\biggl[\log\frac{\Lambda^2}{m^2_q}+\mathcal{C}(m^2_\sigma)+(m^2_\sigma-4m^2_q) \mathcal{C}^{\prime}(m^2_\sigma)\biggr] \ , \\ \nonumber \\ \nonumber \\ 
\delta Z_\pi^{\os} &=&\delta Z_{\pi, div}-\frac{4g^2N_c}{(4\pi)^2}\biggl[\log\frac{\Lambda^2}{m^2_q}+\mathcal{C}(m^2_\pi)+m^2_\pi \mathcal{C}^{\prime}(m^2_\pi)\biggr] \ , \\ \nonumber \\ \nonumber \\
\delta x^2_{\os} &=& 4ig^2 x^2N_c\biggl[B(m^2_\pi)+m^2_\pi B^{\prime}(m^2_\pi)\biggr]=\delta x^2_{div}-\frac{4g^2 x^2N_c}{(4\pi)^2}\biggl[\log\frac{\Lambda^2}{m^2_q}+\mathcal{C}(m^2_\pi)+m^2_\pi \mathcal{C}^{\prime}(m^2_\pi)\biggr]\ , 
\eqa
where $A(m_{q}^2)$, $B(m^2)$, $B'(m^2)$, $\mathcal{C}(m^2)$ and $\mathcal{C}^{\prime}(m^2)$ are given in Appendix A.
\end{widetext}

The divergent part of the counterterms  are $\delta m^2_{div}=\frac{4m^2g^2N_c}{(4\pi)^2\epsilon}$,$\delta \lambda_{div}=\frac{8N_cg^2}{(4\pi)^2\epsilon}(\lambda-g^2)$, $\delta g^2_{div}=\frac{4g^4N_c}{(4\pi)^2\epsilon}$, $\delta Z_{\sigma,div}=Z_{\pi,div}=-\frac{4g^2N_c}{(4\pi)^2\epsilon}$, $\delta h_{div}=\frac{2g^2hN_c}{(4\pi)^2\epsilon}$ and $\delta x_{div}^2=-\frac{4g^2x^2N_c}{(4\pi)^2\epsilon}$.
The divergent part of the counterterms are the same in both the on-shell and $\overline {\text{MS}}$ schemes i.e. $\delta m^2_{div}=\delta m^2_{\overline{\text{\tiny{MS}}}}$, $\delta \lambda_{div}=\delta \lambda_{\overline{\text{\tiny{MS}}}}$, etc. Because the bare parameters are independent of the renormalization scheme, one can relate the renormalized parameters in the on-shell and $\overline{\text{MS}}$ schemes as follows : 
\begin{align}
m_{\overline{\mathrm{MS}}}^2 &= m^2 + \delta m_{\mathrm{OS}}^2 - \delta m_{\overline{\mathrm{MS}}}^2,\\
\lambda_{\overline{\mathrm{MS}}} &= \lambda + \delta\lambda_{\mathrm{OS}} - \delta\lambda_{\overline{\mathrm{MS}}},\\
g_{\overline{\mathrm{MS}}}^2 &= g^2 + \delta g^2_{\mathrm{OS}} - \delta g^2_{\overline{\mathrm{MS}}},\\
h_{\overline{\mathrm{MS}}} &= h + \delta h_{\mathrm{OS}} - \delta h_{\overline{\mathrm{MS}}},\\
x_{\overline{\mathrm{MS}}}^2 &= x^2 + \delta x^2_{\mathrm{OS}} - \delta x^2_{\overline{\mathrm{MS}}}.
\end{align}
The minimum of the effective potential lies at $x=f_{\pi}$.~The scale $\Lambda$ dependent parameters of the model in the $\overline{\mathrm{MS}}$  scheme can be written as, 
\begin{align}
m_{\overline{\mathrm{MS}}}^2(\Lambda)
&= m^2+m^2_{\mathrm{FIN}},\\  \nonumber \\
\lambda_{\overline{\mathrm{MS}}}(\Lambda)
&= \lambda + \lambda_{\mathrm{FIN}},\\ \nonumber \\
h_{\overline{\mathrm{MS}}}(\Lambda)
&= h+h_{\mathrm{FIN}},\\  \nonumber \\ 
\label{rung}
g_{\overline{\mathrm{MS}}}^2(\Lambda)
&= g^2
+\frac{4N_c g^4}{(4\pi)^2}
\Bigg[
\log\frac{\Lambda^2}{m_q^2}
+\mathcal{C}(m_\pi^2)
+m_\pi^2 \mathcal{C}^{\prime}(m_\pi^2)
\Bigg],
\\ \nonumber \\
\label{runx}
x_{\overline{\mathrm{MS}}}^2(\Lambda)
&= f_\pi^2
-\frac{4N_c g^2}{(4\pi)^2 }
\Bigg[
\log\frac{\Lambda^2}{m_q^2}
+\mathcal{C}(m_\pi^2)
+m_\pi^2 \mathcal{C}^{\prime}(m_\pi^2)
\Bigg].
\end{align}
\begin{widetext}
\begin{align}
m^2_{\mathrm{FIN}}&=\frac{2N_c g^2}{(4\pi)^2 f_\pi^2}
\Bigg[
2m^2
\log\frac{\Lambda^2}{m_q^2}-4m_q^2-\left(m_\sigma^2-4m_q^2\right)\mathcal{C}(m_\sigma^2)
+3m_\pi^2\mathcal{C}(m_\pi^2)
\Bigg],\\
\lambda_{\mathrm{FIN}}&=\frac{2N_c g^2}{(4\pi)^2 f_\pi^2}
\Bigg[\left(m_\sigma^2-4m_q^2\right)\biggl(\log\frac{\Lambda^2}{m_q^2}+\mathcal{C}(m_\sigma^2)\biggr)+m_\sigma^2\biggl(\log\frac{\Lambda^2}{m_q^2}+\mathcal{C}(m_\pi^2)+m_\pi^2 \mathcal{C}^{\prime}(m_\pi^2)\biggr)
\nonumber\\
&\qquad\qquad
-m_\pi^2\biggl(2\log\frac{\Lambda^2}{m_q^2}+2\mathcal{C}(m_\pi^2)+m_\pi^2 \mathcal{C}^{\prime}(m_\pi^2)\biggr)
\Bigg],\\
h_{\mathrm{FIN}}&=\frac{2N_c g^2 h}{(4\pi)^2 }
\Bigg[
\log\frac{\Lambda^2}{m_q^2}
+\mathcal{C}(m_\pi^2)
-m_\pi^2 \mathcal{C}^{\prime}(m_\pi^2)
\Bigg],
\end{align}
\end{widetext}

In the large-$N_c$ limit the parameters $\lambda_{\ms}$, $m^2_{\ms}$, $h_{\ms}$,  and $g^2_{\ms}$ are running with the scale $\Lambda$ and a set of simultaneous renormalization group equations are satisfied , which are written as follows.
\bqa
\label{diffpara1}
\dfrac{d\lambda_{\ms}(\Lambda)}{d\log(\Lambda)}&=&\dfrac{16N_c}{(4\pi)^2}\left[\lambda_{\ms}(\Lambda)g^2_{\ms}(\Lambda)-g^4_{\ms}(\Lambda)\right]\;, \\ \nonumber
\\ \nonumber \\
\dfrac{dm^2_{\ms}(\Lambda)}{d\log(\Lambda)}&=&\dfrac{8N_c}{(4\pi)^2}g^2_{\ms}(\Lambda)m^2_{\ms}(\Lambda)\;,
\\  \nonumber \\ \nonumber \\
\dfrac{d h_{\ms}(\Lambda)}{d\log(\Lambda)}&=&\dfrac{4N_c}{(4\pi)^2}g^2_{\ms}(\Lambda)h{\ms}(\Lambda)\;,
\\ \nonumber \\ \nonumber \\
\dfrac{d g^2_{\ms}}{d\log(\Lambda)}&=&\dfrac{8N_c}{(4\pi)^2}g^4_{\ms}(\Lambda)\;,
\\  \nonumber \\ \nonumber \\
\label{diffpara7}
\dfrac{d {x}^2_{\ms}}{d\log(\Lambda)}&=&-\dfrac{8N_c}{(4\pi)^2}g^2_{\ms}(\Lambda){x}^2_{\ms}(\Lambda)
\eqa   
The solutions of Eq.~(\ref{diffpara1})--(\ref{diffpara7}) are the following.
\begin{align}
m_{\overline{\mathrm{MS}}}^2(\Lambda)
&=\frac{m_0^2}{1-\dfrac{4g_0^2N_c}{(4\pi)^2}
\log\dfrac{\Lambda^2}{\Lambda_0^2}},
\label{eq:running-m}
\\ \nonumber \\
g_{\overline{\mathrm{MS}}}^2(\Lambda)
&=
\frac{g_0^2}
{1-\dfrac{4g_0^2N_c}{(4\pi)^2}
\log\dfrac{\Lambda^2}{\Lambda_0^2}}, 
\label{eq:rung} 
\end{align} \\
\bqa
\lambda_{\overline{\mathrm{MS}}}(\Lambda)
&=
\frac{\lambda_0
-\frac{8g_0^4N_c}{(4\pi)^2}
\log\frac{\Lambda^2}{\Lambda_0^2}}
{\left[
1-\dfrac{4g_0^2N_c}{(4\pi)^2}
\log\dfrac{\Lambda^2}{\Lambda_0^2}
\right]^2},
\label{eq:running-lambda}
\\ \nonumber \\ \nonumber \\
h_{\overline{\mathrm{MS}}}(\Lambda)
&=
\frac{h_0}
{1-\dfrac{2g_0^2N_c}{(4\pi)^2}
\log\dfrac{\Lambda^2}{\Lambda_0^2}}. 
\label{eq:running-h}
\eqa
\begin{equation}
\label{para07}
{x}^2_{\overline{\mathrm{MS}}}(\Lambda) = f_\pi^2 \left[ 1 - \frac{N_c g_0^2}{(4\pi)^2} \log\left(\frac{\Lambda^2}{\Lambda_0^2}\right) \right]
\end{equation}

Where the parameters $\lambda_{0}$,\ $g^2_0$,\ $m^2_0$ and $h_0$, are the $\Lambda$ dependent parameters at the value $\Lambda_0$, we consider $\Lambda_0$ to satisfy the relation
\bqa
\log\left(\frac{\Lambda^2_0}{m_q^2}\right)+\mathcal{C}(m^2_\pi)+m^2_\pi \mathcal{C}^{\prime}(m^2_\pi)&=&0\;.
\eqa
\subsection{Effective Potential}
The vacuum effective potential in the $\overline{\text{MS}}$ scheme is
\bqa
\label{omegarqm}
\Omega_{vac}({x}_{\ms})&=&U({x}_{\ms})+\Omega^{q,vac}_{\ms}+\delta U({x}_{\ms})\;, \text{where} \\ \nonumber  \\
\label{omegams1}
\nonumber 
U({x}_{\ms})&=&\frac{m_{\ms}^2(\Lambda)}{2}{x}^2_{\ms}+\frac{1}{4}\left(\lambda_{\ms}(\Lambda)\right){x}^4_{\ms}-h_{\ms}(\Lambda){x}_{\ms}\;, \\
\eqa
\begin{widetext}
\bqa
\nonumber
\delta U({x}_{\ms})&=&\frac{1}{2}\delta m^2_{\ms}{x}_{\ms}^2+\frac{1}{2}m^2_{\ms}\delta {x}^2_{\ms}+\frac{1}{4}\left(\delta \lambda_{\ms}\right){x}^4_{\ms}+\frac{1}{4}\left(\lambda_{\ms}\right)\delta {x}^4_{\ms}
-\delta h_{\ms} {x}_{\ms}-h_{\ms}\delta{x}_{\ms}\;.
\eqa

The order $\mathcal{O}(N^2_c)$ terms are dropped as these are two loop terms and we get
\bqa
\delta U({x}_{\ms})&=&-\frac{2N_cg^4_{\ms}{x}^4_{\ms}}{(4\pi)^2}\frac{1}{\epsilon}=-\frac{2N_c\Delta^4}{(4\pi)^2}\frac{1}{\epsilon}
\eqa
\bqa
\label{omegavac}
\Omega^{q,vac}_{\ms}&=&\frac{2N_cg^4_{\ms}{x}^4_{\ms}}{(4\pi)^2}\left[\frac{1}{\epsilon}+\frac{3}{2}+\ln\left(\frac{\Lambda^2}{g^2_{\ms}{x}^2_{\ms}}\right)\right]=\frac{2N_c\Delta^4}{(4\pi)^2}\left[\frac{1}{\epsilon}+\frac{3}{2}+\ln\left(\frac{\Lambda^2}{\Delta^2}\right)\right]\;,
\eqa \\
One can define the scale $\Lambda$ independent parameter $\Delta={g_{\ms}{x}_{\ms}}$ using the Eq.(\ref{rung}) and Eq. (\ref{runx}).It is instructive to write the Eq.(\ref{omegams1}) in terms of the scale independent $\Delta$ as \\ 
\bqa
\nonumber
U(\Delta)&=&\frac{m_{\ms}^2(\Lambda)}{2g^2_{\ms}(\Lambda)}\Delta^2+\frac{\lambda_{\ms}(\Lambda)}{4g^4_{\ms}(\Lambda)}\Delta^4-\frac{h_{\ms}(\Lambda)}{g_{\ms}(\Lambda)}\Delta\; \\
&=&\frac{m^2_0}{2g^2_0}\Delta^2+\frac{\lambda_{0}}{4g^4_0}\Delta^4-\frac{h_0}{g_0}\Delta
\eqa

\begin{align}
\label{OmegDelx}
\Omega_{\text{vac}}(\Delta) &= \frac{m^2_0 \Delta^2}{2g^2_0} + \frac{\lambda_0 \Delta^4}{4g^4_0} - \frac{h_0 \Delta}{g_0} + \frac{2N_c\Delta^4}{(4\pi)^2}\left[\frac{3}{2} + \ln\frac{\Lambda^2}{\Delta^2}\right]
\end{align}
\end{widetext}

\section{Meson Masses}
\label{appendB}

The masses of sigma and pion mesons are calculated by determining the curvature (double derivative) of the grand potential at the global minimum:
\bqa
m^2_{\sigma}(T,\mu)&=&\frac{\partial^2 \Omega_{{\rm RQM }/{\rm QMVT }}(T,\mu)}{\partial \sigma^2}\Bigg|_{\text{min}} \, \\
m^2_{\pi,i}(T,\mu)&=&\frac{\partial^2 \Omega_{{\rm RQM }/{\rm QMVT }}(T,\mu)}{\partial \pi_i^2}\Bigg|_{\text{min}}
\eqa

\begin{widetext} 
The curvature masses of the $\sigma$ and $\pi$ mesons in the RQM are obtained by evaluating the second derivatives of the grand thermodynamic potential in Eq.  (\ref{GrandRQM}) along the $\sigma$ and $\pi$ directions in field space, respectively
\bqa
\nonumber
\left.m_{\sigma}^{2}\right|_{\pi=0,\;\sigma=\bar{\sigma}=x}
={}& m^{2}+3{\lambda}{x}^{2}+\frac{8N_{c}g^{4}{x}^{2}}{(4\pi)^{2}}+\frac{24N_{c}g^{4}{x}^{2}}{(4\pi)^{2}}
\left[2\left(\ln f_\pi-\ln x\right)-C\!\left(m_{\pi}^{2}\right)-m_{\pi}^{2}C'\!\left(m_{\pi}^{2}\right)\right]
\\[2mm]
\nonumber
&+\frac{2N_{c}g^{2}}{\pi^{2}}\int p^{2}\,dp\left\{\left(frac{e^{-\beta E_q^{+}}}{1+e^{-\beta E_q^{+}}}+\frac{e^{-\beta E_q^{-}}} {1+e^{-\beta E_q^{-}}}\right)\frac{p^{2}}{E_q^{3}}\right.
\\[2mm]
&\hspace{35mm}\left.-\frac{g^{2}\sigma^{2}}{T E_q}\left(\frac{e^{-\beta E_q^{+}}} {\left(1+e^{-\beta E_q^{+}}\right)^{2}}+
\frac{e^{-\beta E_q^{-}}} {\left(1+e^{-\beta E_q^{-}}\right)^{2}}\right)\right\}.
\eqa
\bqa
\nonumber
m_\pi^2 |_{\pi=0,\;\sigma=\bar{\sigma}=x}={}&m^2 + {\lambda}x^2 +\frac{8N_c g^4x^2}{(4\pi)^2} +\frac{8N_cg^4x^2}{(4\pi)^2}\left[2\left(\ln f_\pi-\ln x\right)-C(m_\pi^2)-m_\pi^2 C'(m_\pi^2)\right] \\
&+\frac{2N_cg^2}{\pi^2}\int p^2\,dp\,\left[\frac{e^{-\beta E_q^+}}{1+e^{-\beta E_q^+}}+\frac{e^{-\beta E_q^-}} {1+e^{-\beta E_q^-}}\right]\frac{1}{E_q}.
\eqa
The QMVT model curvature masses of the $\sigma$ and $\pi$ mesons are calculated by finding the second derivatives of the grand thermodynamic potential in Eq.   (\ref{GrandQMVT}) along the $\sigma$ and $\pi$ directions in field space, respectively.
\bqa
\nonumber
\left.m_{\sigma}^{2}\right|_{\pi=0,\;\sigma=\bar{\sigma}=x}
= & m_r^2 + 3 \lambda_r x^2 - \frac{2N_c g^4}{8\pi^2}\left[6 \ln \left(\frac{x}{f_\pi} \right)^2 + 7\right] x^2  \\
\nonumber
& + \frac{2N_c g^2}{\pi^2}\int p^{2}\,dp\left\{\left(\frac{e^{-\beta E_q^{+}}} {1+e^{-\beta E_q^{+}}}+\frac{e^{-\beta E_q^{-}}}
{1+e^{-\beta E_q^{-}}}\right)\frac{p^{2}}{E_q^{3}}\right.
\\[2mm]
&\hspace{35mm}\left.-\frac{g^{2}\sigma^{2}}{T E_q}\left(\frac{e^{-\beta E_q^{+}}}{\left(1+e^{-\beta E_q^{+}}\right)^{2}}+
\frac{e^{-\beta E_q^{-}}}{\left(1+e^{-\beta E_q^{-}}\right)^{2}}\right)\right\}.
\eqa
\bqa
\nonumber
m_\pi^2|_{\pi=0,\;\sigma=\bar{\sigma}=x} =& m_r^2 + \lambda_r x^2 - \frac{2N_c g^4}{8\pi^2}\left[2 \ln \left(\frac{x}{f_\pi} \right)^2 + 1\right] x^2  \\
&+\frac{2N_cg^2}{\pi^2}\int p^2\,dp\,\left[\frac{e^{-\beta E_q^+}}{1+e^{-\beta E_q^+}}+\frac{e^{-\beta E_q^-}} {1+e^{-\beta E_q^-}}\right]
\frac{1}{E_q}.
\eqa
\end{widetext}

\section{INTEGRALS}
\label{appendC}
The divergent loop integrals are regularized by encorporating dimensional regularization.
\bqa
\int_p=\left(\frac{e^{\gamma_E}\Lambda^2}{4\pi}\right)^\epsilon\int \frac{d^dp}{(2\pi)^d}\;,
\eqa
where $d=4-2\epsilon$ , $\gamma_E$ is the Euler-Mascheroni constant, and $\Lambda$ is renormalization scale associated with the $\overline{\text{MS}}$.
\bqa
\nonumber
\mathcal{A}(m^2_q)&=&\int_p \frac{1}{p^2-m^2_q}=\frac{i m^2_q}{(4\pi)^2}\left[\frac{1}{\epsilon}+1\right. \\
\nonumber
&&\left.+\log(4\pi e^{-\gamma_E})+\log\left(\frac{\Lambda^2}{m^2_q}\right)\right]\;
\eqa
we rewrite this after redefining $\Lambda^2\longrightarrow \Lambda^2\frac{e^{\gamma_E}}{4\pi}$.
\bqa
\label{aint1}
\mathcal{A}(m^2_q)&=&\frac{i m^2_q}{(4\pi)^2}\left[\frac{1}{\epsilon}+1+\log\left(\frac{\Lambda^2}{m^2_q}\right)\right]
\\
\label{bint1}
\nonumber
\mathcal{B}(p^2)&=&\int_k \frac{1}{(k^2-m^2_q)[(k+p)^2-m^2_q)]} \\
&=&\frac{i}{(4\pi)^2}\left[\frac{1}{\epsilon}+\log\left(\frac{\Lambda^2}{m^2_q}\right)+C(p^2)\right]
\\
\label{bprimeint1}
\mathcal{B}^\prime(p^2)&=&\frac{i}{(4\pi)^2}C^\prime(p^2)
\eqa
\begin{equation}
\label{eq:cp1}
\mathcal{C}(p^2)=2-2\sqrt{\dfrac{4 m^2_f}{p^2}-1}\arctan\left(\dfrac{1}{\sqrt{\dfrac{4 m^2_f}{p^2}-1}}\right)\\
\end{equation}
\begin{equation}
\label{eq:cp1prime}
 \mathcal{C}^{\prime}(p^2)=\frac{4 m^2_f}{p^4\sqrt{\dfrac{4 m^2_f}{p^2}-1}}\arctan\left(\dfrac{1}{\sqrt{\dfrac{4 m^2_f}{p^2}-1}}\right)-\frac{1}{p^2}\
\end{equation}



\bibliographystyle{apsrmp4-1}

\end{document}